\documentclass[11pt]{article}
\pdfoutput=1 
\usepackage{mathtools}
\usepackage{booktabs}
\usepackage[english]{babel}
\usepackage{amsmath,amssymb,amsbsy,amstext, amsthm, simplewick, amsfonts}
\usepackage{graphicx}
\usepackage[small]{caption}
\usepackage{siunitx}
\usepackage{upgreek}
\usepackage{framed}
\usepackage{wrapfig}
\usepackage{multirow}
\usepackage{bbm}
\usepackage[numbers,sort&compress]{natbib}
\usepackage[svgnames,dvipsnames,x11names]{xcolor}
\usepackage[utf8x]{inputenc}
\usepackage{selinput}
\usepackage{bm}
\usepackage{float}
\usepackage{geometry}
\usepackage{yfonts}
\usepackage{caption}
\usepackage{subcaption}
\usepackage{sidecap}
\usepackage{longtable}
\usepackage{anyfontsize}

\usepackage{genyoungtabtikz}
\Yboxdim{11pt} 
\Ylinethick{.75pt}
\usepackage{epstopdf}
\usepackage{cancel}
\usepackage{tcolorbox}

\usepackage{tocloft}

\newcommand{\ba}{\begin{eqnarray}}
\newcommand{\ea}{\end{eqnarray}}
\def\xyma{\xymatrix@M.7em}
\def\xymas{\xymatrix@M.1em}

\newcommand{\Comment}[1]{{}}
\definecolor{darkblue}{rgb}{0.15,0.35,0.55}
\definecolor{reddish}{rgb}{0.65, 0.2, 0.2}
\definecolor{darkgreen}{RGB}{50,150,0}
\definecolor{greyish2}{rgb}{.96,.96,.96}
\usepackage[linktocpage=true]{hyperref}
\hypersetup{
colorlinks=true,
citecolor=darkblue,
linkcolor=reddish,
urlcolor=darkblue,
pdfauthor={},
pdftitle={},
pdfsubject={}
}

\DeclareFontFamily{OT1}{rsfs10}{}
\DeclareFontShape{OT1}{rsfs10}{m}{n}{ <-> rsfs10 }{}
\DeclareMathAlphabet{\mathscript}{OT1}{rsfs10}{m}{n}

\def\gsim{ \lower .75ex \hbox{$\sim$} \llap{\raise .27ex \hbox{$>$}} }
\def\lsim{ \lower .75ex \hbox{$\sim$} \llap{\raise .27ex \hbox{$<$}} }
\def\be{\begin{equation}}
\def\ee{\end{equation}}
\def\bea{\begin{eqnarray}}
\def\eea{\end{eqnarray}}

\newcommand{\rd}{{\rm d}}

\newcommand{\tr}{\text{tr}}

\newcommand{\D}{{\rm d}}

\newcommand{\hypergeom}[2]{% #1 presubscript, #2 postsubscript
  \mathbin{_{#1}{\sf F}_{#2}} }

\usepackage{latexsym,amsmath,amssymb,epsfig}

\definecolor{greyish}{rgb}{.90,.90,.90}
\definecolor{greyish2}{rgb}{.96,.96,.96}
\usepackage{xcolor,colortbl}
\usepackage{tcolorbox}

\usepackage[all]{xy}

\usepackage{ytableau}

\title{}
\author{}

\numberwithin{equation}{section}

\begin{document}

\renewcommand{\thefootnote}{\fnsymbol{footnote}}

%\clearpage
%\thispagestyle{empty}
%
~
\begin{center}
{\fontsize{21.5}{18} \bf{Static response and Love numbers\\[7pt] 
of Schwarzschild black holes}}
\end{center} 

\vspace{.15truecm}

\begin{center}
{\fontsize{13.5}{18}\selectfont
Lam Hui,${}^{\rm a}$ Austin Joyce,${}^{\rm b}$ Riccardo Penco,${}^{\rm c,d}$\\[4.5pt]
Luca Santoni,${}^{\rm a}$ and
Adam R. Solomon${}^{\rm c,d}$
}
\end{center}
\vspace{.4truecm}

 \centerline{{\it ${}^{\rm a}$Center for Theoretical Physics, Department of Physics,}}
 \centerline{{\it Columbia University, New York, NY 10027}} 
 
  \vspace{.3cm}
 
 \centerline{{\it ${}^{\rm b}$Delta-Institute for Theoretical  Physics,}}
 \centerline{{\it University of Amsterdam, Amsterdam, 1098 XH, The Netherlands}}
 
  \vspace{.3cm}

\centerline{{\it ${}^{\rm c}$Department of Physics, Carnegie Mellon University, Pittsburgh, PA 15213}}

  \vspace{.25cm}

\centerline{{\it ${}^{\rm d}$McWilliams Center for Cosmology, Department of Physics,}}
\centerline{{\it Carnegie Mellon University, Pittsburgh, PA 15213}}
 \vspace{.25cm}

\vspace{.3cm}
\begin{abstract}
\noindent
We derive the quadratic action for the physical degrees of freedom of massless spin-0, spin-1, and spin-2 perturbations on a Schwarzschild--(A)dS background in arbitrary dimensions. We then use these results to compute the static response of asymptotically flat Schwarzschild black holes to external fields. Our analysis reproduces known facts about black hole Love numbers---in particular that they vanish for all types of perturbation in four spacetime dimensions---but also leads to new results. For instance, we find that neutral Schwarzschild black holes polarize in the presence of an electromagnetic background in any number of spacetime dimensions except four. Moreover, we calculate for the first time black hole Love numbers for vector-type gravitational perturbations in higher dimensions and find that they generically do not vanish. Along the way, we shed some light on an apparent discrepancy between previous results in the literature, and clarify some aspects of the matching between perturbative calculations of static response on a Schwarzschild background and the point-particle effective theory\hspace{.01cm}\raisebox{-0.05ex}{\includegraphics[scale=.0045]{cow.pdf}}

\end{abstract}

\newpage

\setcounter{tocdepth}{2}
\tableofcontents
\newpage
\renewcommand*{\thefootnote}{\arabic{footnote}}
\setcounter{footnote}{0}

\section{Introduction}

Despite their deeply mysterious nature, black holes appear to be extremely simple to an external observer. In fact, the uniqueness of the Kerr--Newman solution~\cite{Kerr:1963ud,Newman:1965my}, as encapsulated in the famous no-hair theorems~\cite{Israel:1967wq,Carter:1971zc,Bekenstein:1971hc}, indicates that black holes are amongst the simplest objects, described completely by their mass, charge, and spin.\footnote{The classic no-hair theorems can be extended to include more exotic forms of matter~\cite{Bekenstein:1995un,Hui:2012qt} with static boundary conditions. Interestingly, black holes can support time-dependent scalar profiles with nontrivial boundary conditions~\cite{Jacobson:1999vr,Horbatsch:2011ye,Hui:2019aqm,Clough:2019jpm} as well as superradiant~\cite{Penrose:1969pc,1971JETPL..14..180Z,1972JETP...35.1085Z,1972BAPS...17..472M,Bardeen:1972fi,Press:1972zz,Starobinsky:1973aij,Teukolsky:1974yv} clouds; see~\cite{Arvanitaki:2010sy,Endlich:2016jgc,Baumann:2019eav} for recent discussions.
}
Though black hole solutions can be described by a small number of parameters, if we think of black holes as ordinary objects in their own right, it is natural to ask how they respond to external stimuli, for example an impulse or a tidal perturbation. The responses to these external perturbations are also intrinsic quantities that characterize a black hole, and it is therefore worthwhile to understand their properties.
The question of how a black hole responds is far from academic, as these effects contribute to the form of gravitational wave emission, and hence are in principle measurable \cite{Cardoso:2017cfl}.

The study of weakly perturbed black holes has a long history, starting with the investigations of the Schwarzschild geometry by Regge and Wheeler~\cite{Regge:1957td} and Zerilli~\cite{Zerilli:1970se,Zerilli:1971wd}. This analysis was extended to the Kerr background by Teukolsky~\cite{Teukolsky:1972my,Teukolsky:1973ha}. As a result, wave equations for linearized massless particles are now known in all black hole backgrounds in a form that can be separated into ordinary differential equations, amenable to study.\footnote{Most of the developments in the study of black hole perturbations are reviewed nicely in Chandrasekhar's book~\cite{Chandrasekhar:1985kt}. It is more difficult to separate the wave equation for massive particles with spin. However, a separable ansatz for even parity spin-1 modes was found in~\cite{Krtous:2018bvk}, and separable equations for some of the odd parity modes are discussed in~\cite{Dolan:2018dqv}.}
These equations allow us to characterize the response of a black hole to small perturbations. The most-studied quantities of this type are the quasi-normal mode frequencies, which describe how a black hole relaxes back to the Kerr solution after being disturbed~\cite{Chandrasekhar:1975zza,Leaver:1985ax}; in particular these frequencies govern the ring-down phase of a binary black hole merger. See~\cite{Nollert:1999ji,Kokkotas:1999bd,Berti:2009kk} for exhaustive reviews. In this article, we focus on a different (but related) probe, which is the linear response of a black hole to a (static) external field. More specifically, we will consider the response of a black hole to background gravitational, electromagnetic, and scalar field profiles.

The response of an object to a long-wavelength tidal gravitational field is encoded in the so-called Love numbers~\cite{Love}, which can be thought of as measures of the deformability or rigidity of the object. In an orbiting binary system, the Love numbers of the constituents affect the gravitational wave signal at subleading post-Newtonian order~\cite{Flanagan:2007ix}, which means that gravitational waves can serve as a probe of the Love numbers of both black holes and other astrophysical objects, from neutron stars~\cite{Hinderer:2007mb} to more exotic compact objects~\cite{Cardoso:2017cfl,Chirenti:2020bas,Brustein:2020tpg}.

Astrophysical Love numbers were first defined and studied for spherically symmetric objects in~\cite{Damour:2009vw,Binnington:2009bb}, and display some interesting properties. 
In particular, the Love numbers of a Schwarzschild black hole are exactly zero in four spacetime dimensions~\cite{Damour:2009vw,Binnington:2009bb,Fang:2005qq,Kol:2011vg,Chakrabarti:2013lua,Gurlebeck:2015xpa}, indicating that these black holes are totally rigid, in a sense. This property remains mysterious: black hole Love numbers do not generically vanish in higher dimensions~\cite{Kol:2011vg,Cardoso:2019vof}, with anti-de Sitter asymptotics~\cite{Emparan:2017qxd}, in the presence of higher-curvature terms~\cite{Cardoso:2018ptl}, or in many alternative theories of gravity~\cite{Cardoso:2017cfl}, indicating that their vanishing is a special feature of general relativity in four dimensions.\footnote{The (conservative) static response of Kerr black holes also vanishes in four dimensions. This was first studied for axisymmetric perturbations of slowly spinning black holes in~\cite{Pani:2015hfa,Pani:2015nua,Landry:2015zfa,Landry:2015cva,Landry:2017piv,Poisson:2020mdi}. More recently~\cite{LeTiec:2020spy,LeTiec:2020bos,Chia:2020yla} studied general perturbations for arbitrary spin, finding that the conservative response vanishes. Kerr black holes do however display dissipative tidal heating as a consequence of their rotation. Since the present paper appeared, the issue has been investigated further in~\cite{Goldberger:2020fot,Charalambous:2021mea}.
}

From an effective field theory viewpoint, the vanishing of Love numbers corresponds to a fine-tuning of parameters. Specifically, at distances large compared to the Schwarzschild radius of the black hole, it can be described as a point particle in effective field theory~\cite{Goldberger:2004jt,Goldberger:2005cd}. The Love numbers arise as particular Wilson coefficients in this effective theory~\cite{Porto:2016zng}, and it is important to understand whether there is some deeper explanation (e.g., an underlying symmetry) for their being fine-tuned away.

With an eye toward shedding light on these issues, 
in this paper we take a unified approach to the computation of black hole responses in the presence of static background fields. Specifically, we compute Love numbers and their analogues for massless fields of spins 0, 1, and 2 around Schwarzschild black holes in arbitrary dimensions. This allows us to reproduce known facts about black hole Love numbers---in particular that they vanish for all types of perturbation in four spacetime dimensions---as well as to derive new results. For example, we show that uncharged Schwarzschild black holes acquire a polarization in the presence of an electromagnetic background in higher dimensions, but not in four dimensions. Additionally, we compute the black hole Love numbers for vector-type gravitational perturbations in higher dimensions, which to the best of our knowledge had not been previously computed, finding that they generically do not vanish.

The notion that black holes have properties similar to ordinary objects suggests that we should think of them in a similar way to how we describe other objects. In particular, viewed from very far away, black holes appear to be point particles, much like any other object. We can systematically correct this description and write down an effective field theory that describes the response of the black hole to external fields. In order to make contact with this description of black hole physics, we match our results from the full general relativity calculations to computations performed in this effective theory, in order to see the correspondence between operators in the effective theory and black hole response. This provides the sharpest definition of black hole Love numbers---they are just the coefficients of particular operators in the effective theory that encode the static response. 

As a technical byproduct of our computation we decompose the linearized action for massless particles of spin-0, spin-1, and spin-2 in general Schwarzschild--(anti-)de Sitter backgrounds.
We present fully gauge-fixed actions---with all auxiliary fields integrated out---for each of these cases in general dimension. In the higher-dimensional cases this was previously done at the equation of motion level: for the spin-1 case in~\cite{LopezOrtega:2006vn,Rosa:2011my,Avery:2016zce} and for the spin-2 case in~\cite{Kodama:2000fa,Kodama:2003jz,Ishibashi:2003ap}. Our results are completely consistent with these previous works, however the action-based approach that we take allows for a more transparent analysis of the symmetries, which we pursue further in~\cite{uspaper2}.

As we have already alluded to, one of our motivations for these computations is to shed light on the mysterious vanishing of Love numbers for asymptotically flat four-dimensional black holes. One of the insights from our explicit calculations is that it is not only tidal Love numbers that behave specially in $D=4$: electromagnetic and scalar responses {\it also} vanish only in four dimensions. Though a complete explanation remains elusive, this suggests that there might be some symmetry or other underlying structure of four-dimensional black holes.\footnote{A possible connection to Carroll symmetry in the context of the membrane paradigm was suggested in~\cite{Penna:2018gfx}.} We study the symmetries of four-dimensional black hole perturbations and their consequences in a related work~\cite{uspaper2}.

\vspace{-.5cm}
\paragraph{Outline:} We begin by briefly reviewing the Schwarzschild--(anti-)de Sitter geometry in Section~\ref{sec:background}. We then decompose the actions for massless spin-0, spin-1 and spin-2 fields into spherical harmonics and integrate out all auxiliary fields to obtain actions for the physical degrees of freedom in Section~\ref{sec:actions}. In Section~\ref{sec:lovenums} we turn to the computation of central interest and compute the static response of a black hole to massless perturbations of all types in general dimensions. In Section~\ref{sec:EFT}, we consider the point particle EFT that describes the black hole's response to external fields and match the Love numbers to operator coefficients in this effective theory. This has been done in some cases before, but we systematically treat all static responses. We then conclude in Section~\ref{sec:conclude}. Some technical details relevant for the computations are contained in the appendices. In Appendix~\ref{app:sphericalharmonics}, we review properties of (hyper)spherical harmonics relevant for the decomposition of perturbations in general dimension. The computation of Love numbers utilizes, in an essential way, properties of hypergeometric functions, so we provide a brief introduction to the theory of the hypergeometric equation in Appendix~\ref{App:hyperG}.

Some of the results we describe in this paper have appeared in the literature before. Tidal Love numbers have been computed for four-dimensional Schwarzschild black holes in~\cite{Damour:2009vw,Binnington:2009bb}. In higher dimensions, Love numbers for tensor- and scalar-type perturbations were computed in~\cite{Kol:2011vg,Cardoso:2019vof}, along with spin-0 responses. Here we compute the spin-2 vector-type Love numbers in general dimension for the first time, to our knowledge. Additionally, we explain an apparent discrepancy between~\cite{Kol:2011vg} and~\cite{Cardoso:2019vof}.  Four-dimensional electromagnetic susceptibilities were computed in~\cite{Damour:2009va,dos2016relativistic}, and here we extend these computations to general dimension. Despite the fact that some of the results we present overlap with previous work, we hope that our unified and systematic treatment of all cases will be illuminating.

\vspace{-.5cm}
\paragraph{Conventions:} Throughout we use the mostly-plus metric signature,  denote the spacetime dimension by $D$, denote the spatial dimensionality by $d=D-1$, and focus on the case where $D \geq 4$. In many cases we decompose fields into $(D-2)$-spherical harmonics, where we denote the angular momentum by $L$. We often utilize the black hole tortoise coordinate defined by~\eqref{eq:tortoise}, denoted as $r_\star$. When reporting response coefficients, like Love numbers, we list the dimensionless numbers. Units can be restored by multiplying by $r_s^{2L+D-3}$. Throughout we denote $D$-dimensional spacetime indices using Greek letters, e.g., $\mu,\nu, \rho,\cdots$, denote $d$-dimensional spatial indices by Latin letters from the beginning of the alphabet, e.g., $a,b,c,\cdots$, and we denote angular indices on the $(D-2)$-dimensional sphere using Latin indices from the middle of the alphabet, e.g., $i,j,k,\cdots$.

\newpage
\section{The Schwarzschild--(anti-)de Sitter geometry}
\label{sec:background}

We begin by reviewing some aspects of the Schwarzschild--(anti-)de Sitter (S(A)dS) geometry, which will serve as the setting for our investigations. For simplicity, in this paper we will focus on the case where the horizon has a spherical topology. We will eventually specialize to asymptotically flat black holes, but we leave the cosmological constant general for the time being as it does not complicate the calculation of the actions for perturbations.

The Schwarzschild--(A)dS geometry in $D$ spacetime dimensions is described by the line element
\be
\rd s^2 = -f(r) \rd t^2 + \frac{1}{f(r)}\rd r^2+r^2 \rd\Omega_{S^{D-2}}^2,
\label{eq:sphericalmetric}
\ee
where $\rd\Omega_{D-2}^2$ is the line element on the $(D-2)$-sphere,\footnote{\label{fn:spheelineelement}The line element on the $n$-sphere can be defined recursively:
$
\rd\Omega_{S^n}^2 = \rd \theta_n^2 + \sin^2\theta_n \rd\Omega_{S^{n-1}}^2,
$
where the line element on the circle is just $\rd\Omega_{S^1}^2 = \rd\theta_1^2$. The coordinate $\theta_1$ ranges from $0$ to $2\pi$, whereas all other angles $\theta_i$ range from $0$ to $\pi$.} and where the function $f(r)$ is defined as
\be
f_{\rm S(A)dS}(r)  = 1-\left(\frac{r_s}{r}\right)^{D-3}-\frac{2\Lambda r^2}{(D-1)(D-2)}.
\label{eq:sadsf}
\ee
The line element~\eqref{eq:sphericalmetric} solves the vacuum Einstein equation: $R_{\mu\nu}-2\Lambda/(D-2) g_{\mu\nu} = 0$. This spacetime is a solution for either sign of the cosmological constant, and corresponds to Schwarzschild--anti-de Sitter for negative $\Lambda$ and Schwarzschild--de Sitter for positive $\Lambda$.\footnote{Recall the conversion between the cosmological constant and Hubble parameter is $\Lambda = (D-1)(D-2)H^2/2$. This can be analytically continued to AdS by sending $H^2 \mapsto -\ell^{-2}$, where $\ell$ is the AdS length.
}

The parameter $r_s$ appearing in \eqref{eq:sadsf} is the Schwarzschild radius of the asymptotically flat black hole in the limit $\Lambda = 0$.\footnote{The Schwarzschild radius is related to the asymptotically flat black hole mass parameter by
\begin{equation*}
GM = \frac{(D-2)}{16\pi} \left(\frac{2\pi^\frac{D-1}{2}}{\Gamma\left[\frac{D-1}{2}\right]}\right)r_s^{D-3}.
\end{equation*}
In what follows, we will work primarily in terms of the Schwarzschild radius.
}
At finite $\Lambda$, the position of the horizon shifts, and depends on whether $\Lambda$ is positive or negative. For negative $\Lambda$, corresponding to Schwarzschild-anti-de Sitter black holes, the horizon sits at the single positive real root of the polynomial $f(r) = 0$. For $\Lambda >0$, there are two positive real roots, corresponding to the de Sitter cosmological horizon and the black hole horizon. In the Nariai limit~\cite{1950SRToh..34..160N,10026018884}, the two horizons coincide. We will not require explicit expressions for the locations of the horizons, but they can be found, for example, in~\cite{Tangherlini:1963bw,Cardoso:2004uz}.
In many cases it will be convenient to introduce the so-called tortoise radial coordinate, defined through
\be
\rd r_\star = \frac{1}{f(r)} \rd r.
\label{eq:tortoise}
\ee
The benefits of this coordinate are that the black hole horizon is pushed to $r_\star\to -\infty$, and the causal structure of the $(t, r_\star)$ subspace becomes particularly transparent. We will also find in the next section that $r_\star$ appears naturally when canonically normalizing perturbations.

The S(A)dS spacetime is both static and rotationally symmetric. It possesses $(D-1)(D-2)/2+1$ isometries which form the group ${\mathbb R}\times {\rm SO}(D-1)$. One of these isometries corresponds to time translations, reflecting the time independence of the geometry, while the rest are the rotational symmetries of surfaces of constant radius at fixed time. In addition to these continuous symmetries, the S(A)dS background is invariant under discrete parity transformations, which are typically taken to act by mapping points on a constant-radius sphere to their antipode.

In order to maximally utilize the ${\rm SO}(D-1)$ symmetry, it is convenient to decompose fields propagating in the spacetime into spherical harmonics. The symmetries of the background guarantee a certain degree of decoupling of linearized degrees of freedom once so expanded. For scalar fields, we will see that modes with different angular momentum decouple as a consequence of the background rotational symmetries. For spin-1 and spin-2 fields, there is the additional complication that we must introduce vector and tensor spherical harmonics in order to decompose them. Once again, though, rotational invariance of the background guarantees that modes decomposed into different types of spherical harmonics decouple from each other.\footnote{This can be understood as a consequence of the fact that scalar, vector, and tensor spherical harmonics have different spherical laplacian eigenvalues in general dimension. See Appendix~\ref{app:sphericalharmonics} for a review.} In the case of four dimensions, this coincides with the decoupling of modes of different parity.
In addition to the spherical harmonic decomposition, in many cases we will also leverage the time-translation symmetry by working in the frequency domain. In the following section we describe in detail  the decomposition of massless spin-0, spin-1, and spin-2 fields on a static black hole background.

\newpage
\section{Linearized fields in the black hole background}
\label{sec:actions}
We are interested in studying the dynamics of massless fields propagating in a Schwarzschild--(anti-)de Sitter black hole background. To facilitate this, in this section we describe the decomposition of fields into spherical harmonics, and derive the action for the physical degrees of freedom carried by gauge fields.

\subsection{Spin-0 field}
\label{sec:scalar}
We begin by considering the simplest case of interest: a 
free real scalar field propagating in the geometry defined by~\eqref{eq:sphericalmetric}. The dynamics is governed by the action
\be
S = \int\rd^Dx\sqrt{-g}\left(-\frac{1}{2}(\partial\phi)^2-\frac{m^2}{2}\phi^2\right).
\label{eq:complexscalar}
\ee
Motivated by the rotational invariance of the background, we decompose the scalar into spherical harmonics~as
\be
\phi(x) = \sum_{L,M} \Psi(t,r)r^\frac{2-D}{2}Y_L^M(\theta),
\label{scalardecomp}
\ee
where we have pulled out an additional factor of $r^{(2-D)/2}$ for later convenience and the coefficient functions $\Psi$ carry implicit $L,M$ labels that we suppress for notational simplicity. The functions $Y_L^M(\theta)$ are the $(D-2)$-hyperspherical harmonics discussed in Appendix~\ref{sec:scalarsphericalharmonics}, which have the spherical laplacian eigenvalue
\be
\Delta_{S^{D-2}}Y_L^M(\theta) = -L(L+D-3)Y_L^M(\theta),
\ee
and where $M$ is a multi-index cataloging the magnetic quantum numbers $\lvert m_1\rvert\leq m_2\leq \cdots \leq m_{D-3}\leq L\equiv m_{D-2}.$

Inserting the decomposition \eqref{scalardecomp} into the action~\eqref{eq:complexscalar}, integrating by parts, and using the completeness relation for the spherical harmonics~\eqref{eq:sharmcompleteness} to integrate over the angular variables, we eventually find
\be
S = \sum_{L, M }\int\rd t\rd r\, \left[\frac{1}{2f}|\partial_t\Psi|^2 -\frac{f}{2}|\partial_r\Psi|^2 -\frac{1}{2f}V_0(r)|\Psi|^2
\right],
\ee
where $[\Psi_L^{m_1, m_2, \dots, m_{D-3}} ]^* = (-)^{m_1}\Psi_L^{-m_1, m_2, \dots, m_{D-3}}$ because $\phi$ is real, and we have defined the scalar potential as
\be
V_0(r) \equiv f\frac{L(L+D-3)}{r^2} +ff'\frac{D-2}{2r}
+f^2\frac{(D-2)(D-4)}{4r^2}+ fm^2.
\label{eq:scalarpot}
\ee
Note that the potential depends only on $L$ and not on $M$ due to the spherical symmetry of the background, essentially because the magnetic quantum numbers are defined with respect to (arbitrary) reference axes which can be changed by ${\rm SO}(D-1)$ rotations. 

In order to canonically normalize the action, we trade $r$ for the tortoise coordinate $r_\star$ defined in \eqref{eq:tortoise}. After these simplifications, the action is 
\begin{tcolorbox}[colframe=white,arc=0pt,colback=greyish2]
\be
S = \sum_{L, M }\int\rd t\rd r_\star\left[\frac{1}{2} |\dot\Psi|^2 -\frac{1}{2}\left|\frac{\partial\Psi}{\partial r_\star}\right|^2 -\frac{1}{2}V_0(r)|\Psi|^2
\right].
\label{eq:scalaraction}
\ee
\end{tcolorbox}
\vspace{-5pt}
\noindent The frequency-space equation of motion following from the action~\eqref{eq:scalaraction} takes the form of a Schr\"odinger equation
\be
\frac{\rd^2\Psi(r_\star)}{\rd r_\star^2} + \Big(\omega^2 - V_0(r)\Big)\Psi(r_\star)= 0,
\label{eq:s0schrod}
\ee
where the potential is given in \eqref{eq:scalarpot}. In terms of the tortoise coordinate, the potential $V_0(r)$ is rather simple, it is essentially featureless except for a positive bump near $r \simeq 3r_s/2$.\footnote{The exact position of the maximum depends on the angular momentum number $L$. When $L\to \infty$, the location of the maximum is usually referred to as the {\it light ring}. }

As advertised earlier, the quadratic action \eqref{eq:scalaraction} is just a sum of an infinite number of terms, one for each set of angular momentum numbers:
\be
S = \sum_{L,M} S_{(L,M)}.
\ee
For this reason, when discussing spin-1 and spin-2 perturbations below we will omit the sum over $L$ and $M$ and focus directly on the action $S_{(L,M)}$ for some fixed values of the angular momentum. In particular, we will choose $M$ such that $m_1 =0$, in which case the perturbations are just real functions.

\subsection{Massless spin-1 field}
\label{sec:masslesspin1action}
We will now turn our attention to the dynamics of massless spin-1 fields in a S(A)dS background. We describe the decomposition of the field into spherical harmonics, and describe how to handle the gauge redundancies in order to derive an action for the physical degrees of freedom. 
The equations of motion for a massless spin-1 field in a black hole background have been previously discussed in~\cite{LopezOrtega:2006vn,Rosa:2011my,Avery:2016zce}.

\subsubsection{Decomposition of perturbations}

A massless vector field in a Schwarzschild background is described by the usual Maxwell action
\be
S = \int\rd^Dx \sqrt{-g}\left(-\frac{1}{4}F_{\mu\nu}^2\right),
\label{eq:maxwell}
\ee
where the field strength tensor is given by $F_{\mu\nu}\equiv \partial_\mu A_\nu-\partial_\nu A_\mu$ in terms of the gauge potential $A_\mu$.
In order to exploit the rotational invariance of the background, it is
useful to first split $A_\mu$ as
\be
A_\mu = \left(
\begin{array}{c}
a_0\\
a_r\\
\partial_ia^{(L)}+a_i^{(T)}
\end{array}
\right),
\label{eq:vectorsplit}
\ee
where the ``$i$'' subscript runs over the angular coordinates. The quantities $a_0$, $a_r$, and the longitudinal component $a^{(L)}$ transform as scalars under SO$(D-1)$, while the transverse component $a^{(T)}_i$ transforms as a vector, and is such that $\nabla^i a_i^{(T)} = 0$.
Under the gauge transformation $\delta A_\mu  = \partial_\mu\Lambda$, these individual components transform as
\begin{subequations}
\begin{align}
\delta a_0 &= \dot\Lambda,\\
\delta a_r &= \Lambda',\\
\delta a^{(L)} &= \Lambda, \\
\delta a_i^{(T)} &= 0.
\end{align}
\label{eq:spin1gaugetransformations}%
\end{subequations}
Note in particular that the vector $a_i^{(T)}$ is gauge invariant by itself.

Next we decompose the variables introduced in \eqref{eq:vectorsplit} into spherical harmonics. In this case we need both scalar and vector spherical harmonics:
\begin{subequations}
\begin{align}
a_0(t,\vec x) &=  \sum_{L,M} a_0(t,r)Y_L^M, \\
a_r(t,\vec x) &=  \sum_{L,M}a_r(t,r)Y_L^M ,\\
a^{(L)}(t,\vec x) &=  \sum_{L,M} a^{(L)}(t,r)Y_L^M, \\
a^{(T)}_i(t,\vec x) &= \sum_{L,M} a^{(T)}(t,r)Y_{i}^{(T)}{}_L^M.
\end{align}
\label{eq:vectdecomposition}%
\end{subequations}
In these expressions, 
we have suppressed $L,M$ labels on the coefficient functions of the spherical harmonic decomposition.\footnote{We use the same multi-index $M$ label to denote the magnetic quantum numbers of all types of spherical harmonics. However, since the dimension of the representation space is different for different types of harmonics (except on the 2-sphere), the magnetic quantum numbers range over different values. We only ever require the completeness properties of the harmonics, so this subtlety will not be important. Note also that $L=0$ is somewhat degenerate: for this value, $Y_i^{(T)}$ does not exist, so there is no vector-type perturbation. Furthermore, the derivative $\partial_i a^{(L)}$ annihilates the $L=0$ component of the $a^{(L)}$ decomposition. There is a single gauge-invariant combination $a_0'-\dot a_r$ that is well-defined for $L=0$, but this combination does not propagate: shifting its value instead corresponds to a shift of the monopole electric charge at the origin, so the $L=0$ mode should more properly be thought of as a shift of the background charge.} 
The variables $a_0, a_r, a^{(L)}$ have the same transformation properties under rotations---which is why they can all be expanded in regular spherical harmonics---and so are allowed to mix at linear level. On the contrary, $a_i^{(T)}$ admits an expansion in vector spherical harmonics (reviewed in Appendix~\ref{app:spin1sphericalharmonics}), and therefore decouples from the other fields.\footnote{Note that the decomposition into spherical harmonics obscures the counting of propagating degrees of freedom. However, this counting can be recovered by matching onto a plane wave basis for the field $A_\mu$, which is most simply done in the $L\to\infty$ limit~\cite{Thorne:1980ru}.
}

Inserting the decomposition~\eqref{eq:vectdecomposition} into the action~\eqref{eq:maxwell}, we obtain
\be
\begin{aligned}
S = \int\rd t\rd r\, r^{D-4}\Bigg(&\frac{1}{2f}\dot a_{(T)}^2 -\frac{f}{2} a'_{(T)}{}^2 -\frac{(L+1)(L+D-4)}{2r^{2}}a_{(T)}^2\\
&+\frac{r^2}{2} \dot a_r^2-\frac{L ( L + D-3)}{2} f a_r^2+\frac{L(L+D-3)}{2f}\dot a_{(L)}^2-\frac{L(L+D-3)}{2}f a_{(L)}'{}^2\\
&+\frac{r^2}{2}a_0'{}^2+\frac{L(L+D-3)}{2f}a_0^2-r^2a_0'\dot a_r+L(L+D-3)f a_r a_{(L)}'-\frac{L(L+D-3)}{f}a_0\dot a_{(L)}
\Bigg).
\label{eq:masslessspin1action}
\end{aligned}
\ee
As anticipated $a^{(T)}$ decouples and forms its own (parity odd) sector, while the other degrees of freedom mix. We therefore want to isolate the true degrees of freedom in this (even parity) sector.

\subsubsection{Spin-1 vector mode}
We first consider the vector mode, $a^{(T)}$. These modes are odd under parity transformations and are gauge invariant.
In order to canonically normalize the action, we define the variable
\be
\Psi_V \equiv r^\frac{D-4}{2} a^{(T)},
\ee
and transform to the 
 tortoise coordinate $\rd r_\star = f^{-1} \rd r$. With these redefinitions and after some integrations by parts, the action for $\Psi_V$ takes the form
\begin{tcolorbox}[colframe=white,arc=0pt,colback=greyish2]
\be
S = \int\rd t\rd r_\star \Bigg(\frac{1}{2}\dot\Psi_V^2-\frac{1}{2}\left(\frac{\partial\Psi_V}{\partial r_\star}\right)^2-\left[\frac{(L+1)(L+D-4)}{2r^2}f+\frac{(D-4)[(D-6)f^2+2rff']}{8r^2}\right]\Psi_V^2\Bigg).
\label{eq:spin1odd}
\ee
\end{tcolorbox}
\vspace{-5pt}
\noindent
The equation of motion for $\Psi_V$ in frequency space is the Schr\"odinger-like equation
\be
\frac{\rd^2\Psi_V}{\rd r_\star^2}+\left(\omega^2 - f(r)\frac{(L+1)(L+D-4)}{r^2}-f(r)\frac{(D-4)[(D-6)f+2rf']}{4r^2}\right)\Psi_V = 0.
\label{eq:masslessBmodespin10}
\ee
We next turn to the scalar (parity even) sector.

\subsubsection{Spin-1 scalar mode}
In the scalar sector there is also a single degree of freedom. Only one combination of the variables is physical. However, the mixing between modes makes it difficult to isolate this physical degree of freedom. In order to facilitate this, we will integrate in an auxiliary field. Before doing this, it is convenient to utilize our gauge freedom to choose the gauge $a_{(L)} = 0$ so that the action for the even modes simplifies,\footnote{\label{foot:actiongauge}Not all gauge choices are suitable at the level of the action \cite{Lagos:2013aua,Motohashi:2016prk}. Under a gauge transformation \eqref{eq:spin1gaugetransformations}, the action transforms as $\delta S =0 = \int\rd^Dx\left(\mathcal{E}_0\dot\Lambda+\mathcal E_r\Lambda'+\mathcal E_{(L)}\Lambda\right)$, where $\mathcal{E}_i=0$ is the equation of motion for $a_i$, implying the (off-shell) Noether identity $\dot{\mathcal{E}}_0+\mathcal{E}_r'-\mathcal{E}_{(L)}=0$. While our gauge choice eliminates $a_{(L)}$, its equation of motion $\mathcal{E}_{(L)}$ can be recovered from the Noether identity, implying this gauge is suitable for use in the action.}
\be
S = \int\rd t\rd r\, r^{D-4}\Bigg(
\frac{r^2}{2} \dot a_r^2-\frac{L ( L + D-3)}{2} f a_r^2+\frac{r^2}{2}a_0'{}^2-r^2a_0'\dot a_r+\frac{L(L+D-3)}{2f}a_0^2
\Bigg).
\label{eq:evenspin1modes}
\ee
We then introduce the auxiliary field $\Psi_S$ as
\be
S = \int\rd t\rd r\, \Bigg(
\sqrt{L(L+D-3)}r^\frac{D-4}{2}\Psi_S\left(a_0'-\dot a_r\right)-\frac{L(L+D-3)}{2r^2}\Psi_S^2+\frac{L(L+D-3)}{2 r^{4-D}}\left[\frac{1}{f}a_0^2- f\,a_r^2\right]
\Bigg),
\ee
where the precise normalization has been chosen for later convenience.
The equation of motion for $\Psi_S$ is
\be
\Psi_S = \frac{r^{D/2}}{\sqrt{L(L+D-3)}}\left(a_0'-\dot a_r\right).
\ee
Using the gauge transformations~\eqref{eq:spin1gaugetransformations} it is straightforward to check that $\Psi_S$ is gauge invariant.
If we substitute this expression back into the action, we recover \eqref{eq:evenspin1modes}, so the two actions are equivalent. The benefit of introducing $\Psi_S$ is that we can now integrate out both $a_0$ and $a_r$ to obtain an action purely for $\Psi_S$ by using the equations of motion
\begin{align}
\label{eq:a0intermsofpsi}
a_ 0 &= \frac{r^\frac{2-D}{2}f}{2\sqrt{L(L+D-3)}}\left[ (D-4)\Psi_S+2r\Psi_S'\right],\\
a_r &=  \frac{r^\frac{4-D}{2}}{\sqrt{L(L+D-3)}f}\dot \Psi_S.
\end{align}
The resulting action for $\Psi_S$ is given by
\be
S = \int\rd t\rd r \left(\frac{1}{2f}\dot \Psi_S^2-\frac{1}{2} f\,\Psi_S'^2 - \left[\frac{L(L+D-3)}{2r^2}+\frac{(D-4)\big[(D-2)f-2rf'\big]}{8r^2} \right]\Psi_S^2
\right),
\ee
where we have integrated by parts. We again transform to the tortoise coordinate so that the action is canonically normalized:
\begin{tcolorbox}[colframe=white,arc=0pt,colback=greyish2]
\be
S = \int\rd t\rd r_\star \left(\frac{1}{2}\dot \Psi_S^2-\frac{1}{2} \left(\frac{\partial\Psi_S}{\partial r_\star}\right)^2 - \left[\frac{L(L+D-3)}{2r^2}f+\frac{(D-4)\big[(D-2)f^2-2rff'\big]}{8r^2} \right]\Psi_S^2
\right).
\label{eq:spin1even}
\ee
\end{tcolorbox}
\vspace{-5pt}
\noindent
In these variables, the equation of motion for $\Psi_S$ also takes a Schr\"odinger-like form in frequency space
\be
\frac{\rd^2\Psi_S}{\rd r_\star^2}+\left(\omega^2 - f(r)\frac{L(L+D-3)}{r^2}-f(r)\frac{(D-4)[(D-2)f-2rf']}{4r^2}\right)\Psi_S = 0.
\label{eq:masslessEmodespin10}
\ee
The sum of the actions~\eqref{eq:spin1odd} and~\eqref{eq:spin1even} therefore parameterizes the dynamics of a massless spin-1 field in a Schwarzschild background in arbitrary dimensions.

Notice that the variables $\Psi_V$ and $\Psi_S$ are gauge invariant, so that the natural expectation is that they are related to the usual electric and magnetic fields. This is indeed the case; for static field profiles, the electric field is built from $\Psi_S$ and the magnetic field is comprised of $\Psi_V$. However, in dynamical situations the electric and magnetic fields each contain an admixture of $\Psi_S$ and $\Psi_V$, which is easily verified by computing the Faraday tensor, $F_{\mu\nu}$.

\subsubsection{Electric-magnetic duality in $D=4$}

Though our focus is on black hole perturbations in general dimension, it is worth briefly remarking on an interesting aspect that is special to $D=4$. Consider the Maxwell action, which is a sum of~\eqref{eq:spin1odd} and~\eqref{eq:spin1even} in four dimensions:
\be
\begin{aligned}
S_{\rm em}= \int\rd t\rd r_\star \Bigg(&\frac{1}{2}\dot\Psi_V^2-\frac{1}{2}\left(\frac{\partial\Psi_V}{\partial r_\star}\right)^2-\frac{L(L+1)}{2r^2}f\,\Psi_V^2+\frac{1}{2}\dot \Psi_S^2-\frac{1}{2} \left(\frac{\partial\Psi_S}{\partial r_\star}\right)^2 - \frac{L(L+1)}{2r^2}f\,\Psi_S^2
\Bigg)\,.
\end{aligned}
\label{eq:spin1actionemdual}
\ee
The fields $\Psi_V$ and $\Psi_S$ continue to decouple in $D=4$, now as a consequence of their different parity transformation properties. Notice that the action~\eqref{eq:spin1actionemdual} is completely symmetric in $\Psi_S$ and $\Psi_V$. This can be made more manifest by grouping the two potentials into an SO$(2)$ vector:
\be
\vec \Psi \equiv \left(
\begin{array}{c}
\Psi_S\\
\Psi_V
\end{array}
\right),
\ee
in terms of which the action can be written in a manifestly rotationally invariant form:
\be
S_{\rm em}= \int\rd t\rd r_\star \Bigg(\frac{1}{2}\dot{\vec{\Psi}}^2-\frac{1}{2}\bigg(\frac{\partial\vec{\Psi}}{\partial r_\star}\bigg)^2-\frac{L(L+1)}{2r^2}f\,\vec{\Psi}^2
\Bigg)\,.
\ee
The SO$(2)$ transformations that rotate the potentials into each other are nothing other than electric-magnetic duality.\footnote{There is some confusion in the literature about whether electric-magnetic duality is a symmetry of the source-free Maxwell action. Electric-magnetic duality can indeed be made an off-shell symmetry of the action, but it acts (spatially) non-locally~\cite{Deser:1976iy}. This is consistent with the action in our case because the decomposition into spherical harmonics is effectively spatially nonlocal. The subtlety is that the transformations of the gauge potential $A$ that are a symmetry of the action only act to interchange the Faraday tensor $F$ with its dual $\star F$ on-shell.
}

Since the electric-magnetic duality symmetry is continuous, it has a corresponding Noether current, which takes the form
\be
J_\mu  = \Psi_S\partial_\mu \Psi_V - \Psi_V\partial_\mu\Psi_S,
\ee
where the $\mu$ index runs over $(t,r_\star)$. This current is clearly conserved on-shell because the two potentials obey the same equation of motion. In~\cite{uspaper2} we explore the consequences of this symmetry for spin-1 perturbations in a black hole background.

\subsection{Massless spin-2 field}
\label{sec:spin2decomp}

Finally, we consider the dynamics of a massless spin-2 field in a fixed Schwarzschild--(A)dS geometry. This describes the linearized gravitational perturbations around an (un-charged, un-spinning) black hole solution. In the four-dimensional context, this was first worked out with asymptotically flat boundary conditions by Regge and Wheeler~\cite{Regge:1957td} and by Zerilli~\cite{Zerilli:1970se,Zerilli:1971wd} and in a gauge-invariant formalism by Moncrief, Cunningham, and Price~\cite{Moncrief:1974am,Cunningham:1978zfa,Cunningham:1979px}. In higher dimensions, the equations of motion for these perturbations were derived in~\cite{Kodama:2000fa,Kodama:2003jz,Ishibashi:2003ap}.
For the generalization to massive and partially massless spin-$2$ fields on a  Schwarzschild--(A)dS spacetime in $D$-dimensions, see~\cite{Rosen:2020crj}.

\subsubsection{Decomposition of perturbations}

Our starting point is 
the action for the graviton linearized around an Einstein background (a spacetime satisfying $R_{\mu\nu} = 2\Lambda/(D-2) g_{\mu\nu}$)\footnote{We have canonically normalized the graviton fluctuation. In terms of the Einstein--Hilbert action, this corresponds to expanding $M_{\rm Pl}^{D-2}(R-2\Lambda)/2$ by defining the metric as $g_{\mu\nu} = \bar{g}_{\mu\nu}+2h_{\mu\nu}/M_{\rm Pl}^{(D-2)/2}$, with $\bar{g}_{\mu\nu}$ the Einstein background metric.}
\be
S=\int {\rm d}^Dx\sqrt{-g}\left[ -\frac{1}{2}\nabla_\lambda h_{\mu\nu} \nabla^\lambda h^{\mu\nu}+\nabla_\lambda h_{\mu\nu} \nabla^\nu h^{\mu\lambda}-\nabla_\mu h\nabla_\nu h^{\mu\nu} +\frac{1}{2} \nabla_\mu h\nabla^\mu h
+\frac{2\Lambda}{D-2}\left( h^{\mu\nu}h_{\mu\nu}-\frac{1}{2} h^2\right)\right]. \label{eq:graviton-action}
\ee
We now want to decompose the metric perturbation $h_{\mu\nu}$ into pieces that transform nicely under the SO$(D-1)$ rotational symmetry. We therefore split $h_{\mu\nu}$ as\footnote{The notation for spin-2 perturbations is intended to mirror that of~\cite{Franciolini:2018uyq}. There are, however, a couple small convention differences that we record here. In four dimensions, we can replace the unit-normalized vector spherical harmonic in terms of a gradient of a scalar harmonic as $Y_i^{(T)}{}_L^M \xrightarrow{D\to 4} \epsilon^j_{~i}\nabla_j Y_L^M/\sqrt{L(L+1)}$. This normalization factor is not included in~\cite{Franciolini:2018uyq}, which shifts the definition of the fields multiplying these harmonics by the same factor. Further, there is a sign difference between the definition of the variable $h_2$ here and the one defined there. Finally, the definitions of ${\cal K}$ and $G$ differ by the subtraction of the trace from the harmonic multiplying $G$ here, which was not done in~\cite{Franciolini:2018uyq}.}
\begin{subequations}
\begin{align}
h_{tt} &= \sum_{L,M}  f(r)H_0(t,r) Y_L^M, \\
h_{tr} &= \sum_{L,M}  H_1(t,r)Y_L^M,\\
h_{rr} &= \sum_{L,M}   f(r)^{-1}H_2(t,r)Y_L^M,\\
h_{ti} &= \sum_{L,M}\left[{\cal H}_0(t,r)\nabla_i Y_L^M +h_0(t,r)Y_i^{(T)}{}_L^M \right],\\
h_{ri} &=\sum_{L,M}\left[ {\cal H}_1(t,r)\nabla_i Y_L^M+ h_1(t,r)Y_i^{(T)}{}_L^M\right], \\
h_{ij} &= \sum_{L,M}r^2\left[{\cal K}(t,r)\gamma_{ij} Y_L^M +G(t,r)\nabla_{(i}\nabla_{j)_T}Y_L^M +h_2(t,r)\nabla_{(i}Y_{j)}^{(T)}{}_L^M +h_T(t,r) Y^{(TT)}_{ij}{}_L^M
\right].
\end{align}
\label{eq:hdecomp}%
\end{subequations}
In this expression $(\cdots)_T$ denotes the trace-free symmetrized part of the enclosed indices.
In order to decompose the spin-2 field $h_{\mu\nu}$ we have had to introduce three distinct types of spherical harmonics: tensor harmonics, $Y_{ij}^{(TT)}$; (transverse) vector harmonics, $Y_i^{(T)}$; and scalar spherical harmonics, $Y$. See Appendix~\ref{app:sphericalharmonics} for a review of their properties. There are correspondingly three different sectors of perturbations: one consists of the perturbations proportional to (derivatives of) scalar harmonics:
\be
H_0, \quad H_1, \quad H_2, \quad {\cal H}_0, \quad {\cal H}_1, \quad G, \quad {\cal K}  \qquad\qquad ({\rm scalar~perturbations}),
\ee
which we will refer to as scalar perturbations for simplicity. Another sector includes perturbations proportional to (derivatives of) vector harmonics:
\be
\hspace{1.6cm}h_0, \quad h_1, \quad h_2  \hspace{4cm}({\rm vector~perturbations}).
\ee
Finally there is a single perturbation proportional to a tensor harmonic:
\be
\hspace{2.5cm}h_T, \hspace{4.8cm}({\rm tensor~perturbation}).
\ee
Since these perturbations all multiply different kinds of spherical harmonics (which have different SO$(D-1)$ Casimir eigenvalues), the three sectors decouple at the linear level. 
 
In $D=4$, the $h_T$ perturbation is absent and in this case the eigenvalues of the scalar and vector spherical harmonics happen to coincide for the two-sphere but the scalar and vector modes continue to decouple because they have different parity eigenvalues (exactly as for spin-1 perturbations). In this case, the scalar (vector) modes correspond to the usual even (odd) modes. It is worth stressing the difference in terminology from the more familiar decomposition in Cartesian coordinates: there, the gravitational wave degrees of freedom are often referred to as tensor modes; in our decomposition in spherical coordinates, the same degrees of freedom live in the scalar and vector sectors.\footnote{In analogy with the spin 1 case, the $L=0$ and $L=1$ perturbations of spin 2 fields must be treated separately and turn out not to propagate. The $L=0$ mode corresponds to a shift of the black hole mass parameter in the static limit~\cite{Regge:1957td}, as a consequence of Birkhoff's theorem, while the $L=1$ mode corresponds to shifting the background to that of a slowly spinning black hole at leading order in the spin expansion~\cite{Regge:1957td,Kobayashi:2012kh}. (Note that the Kerr metric to leading order in spin is simply the Schwarzschild metric with a single odd perturbation, $h_0^L\propto (r_s/r)\delta^L_1$.) See \cite{Martel:2005ir} for a nice discussion of these modes.}

A massless spin-2 field enjoys gauge invariance under linearized 
diffeomorphisms
\be
\delta h_{\mu\nu} = \nabla_\mu\xi_\nu+\nabla_\nu\xi_\mu.
\ee
In order to see how the various fields shift under this transformation, 
we split the diffeomorphism parameter, $\xi$ in a similar way to what we did for the spin-1 field:
\begin{subequations}
\begin{align}
\xi_t &= \sum_{L,M} f(r)\xi_0(t,r)Y_L^M,\\
\xi_r &=\sum_{L,M}f^{-1} \xi_1(t,r)Y_L^M,\\
\xi_i &= \sum_{L,M} \xi_S(t,r)\nabla_i Y_L^M +\xi_V(t,r)Y_i^{(T)}{}_L^M.
\end{align}
\end{subequations}
From this we can determine how the individual variables shift under a diffeomorphism:
\begin{subequations}
\begin{align}
\delta H_0 &= 2\dot\xi_0-\frac{f'}{f}\xi_1,& \delta h_0 &= \dot\xi_V, & \delta h_T &=0,\\
\delta H_1 &=  f^{-1} \dot\xi_1+ f \xi_0', & \delta h_1 &= -2 r^{-1} \xi_V+\xi_V',\\
\delta H_2 &= 2\xi_1'-\frac{f'}{f}\xi_1,& \delta h_2 &= 2r^{-2}\xi_V,\\
\label{eq:h2gauge}
\delta{\cal H}_0 &= \dot\xi_S+f \xi_0,\\
\delta{\cal H}_1 &=f^{-1}\xi_1-2r^{-1}\xi_S,\\
\delta{\cal K} &= 2 r^{-1}\xi_1-\frac{2L(L+D-3)}{(D-2)r^2}\xi_S,\\
\delta G &= 2 r^{-2} \xi_S.
\end{align}
\end{subequations}
Notice that the mode multiplying the tensor spherical harmonic $Y_{ij}^{(TT)}$ is gauge invariant. 
It is convenient to discuss this mode first. 

\subsubsection{Spin-2 tensor sector}
\label{sec:tensorspin2}
Inserting the decomposition~\eqref{eq:hdecomp} into the linearized Einstein--Hilbert action, the $h_T$ mode decouples from all the other modes, so that its action is given  (up to a total derivative) by
\be
S_T = \int\rd t\rd r\, r^{D-2}\left(\frac{1}{2f}\dot h_T^2-\frac{1}{2}fh_T'{}^2  +f\left[\frac{D-3}{r^2}+\frac{6-2D-L(L+D-3)}{2 f r^2}+\frac{f'}{rf}+\frac{2\Lambda}{(D-2)f}\right]h_T^2
\right).
\ee
The equation of motion following from this action was derived in~\cite{Gibbons:2002pq,Kodama:2003jz}. We can express this action in terms of a canonically normalized variable by making the field redefinition
\be
\Psi_T \equiv r^\frac{D-2}{2} h_T,
\ee
and adopting the tortoise coordinate $\rd r_\star = f^{-1} \rd r$. The action for $\Psi_T$ is then (once again, after some integration by parts)
\begin{tcolorbox}[colframe=white,arc=0pt,colback=greyish2]
\be
S_T = \int\rd t\rd r_\star \Bigg(\frac{1}{2}\dot\Psi_T^2-\frac{1}{2}\left(\frac{\partial\Psi_T}{\partial r_\star}\right)^2-
\frac{1}{2}V_T(r)\Psi_T^2\Bigg),
\ee
\end{tcolorbox}
\vspace{-5pt}
\noindent
where the potential is
\be
V_T(r) = f\frac{L(L+D-3)+2(D-3)}{r^2}+f^2\frac{D(D-14)+32}{4r^2} +ff'\frac{D-6}{2r}- \frac{4\Lambda f}{D-2}.
\label{eq:tensorpotential}
\ee
As in the other cases, in frequency space the equation of motion for the radial degree of freedom takes the form of a Schr\"odinger equation
\be
\frac{\rd^2\Psi_T}{\rd r_\star^2} + \Big(\omega^2 - V_T(r)\Big)\Psi_T = 0,
\label{eq:tensorequation2}
\ee
with the potential given in \eqref{eq:tensorpotential}.

\subsubsection{Spin-2 vector sector (odd parity)}
\label{sec:vectorspin2}
We next move on to the vector perturbations. Since this sector coincides with the odd sector $D=4$, we will often refer to it as such.
The variables in this sector are $h_0, h_1, h_2$. However, we expect that there should only be one physical combination of these degrees of freedom, so we will have to fix a gauge and integrate out auxiliary variables.

We choose to work in the so-called Regge--Wheeler gauge~\cite{Regge:1957td}, defined by the condition
\be
h_2 = 0.
\ee
It is clear from \eqref{eq:h2gauge} that we have enough freedom to reach this gauge by choosing $\xi_V$ appropriately. The remaining degrees of freedom are $h_0$ and $h_1$, and their action is given (up to integrations by parts) by 
\be
\begin{aligned}
S_{\rm RW}=\int\rd t\rd r\,r^{D-4}\bigg[&\dot h^2_1+h'_0{}^2+\frac{4}{r}h_0 \dot h_1-2 h_0' \dot h_1+\frac{2(D-3)f+2rf'-(L+1)(D-4+L)}{r^2}f\,h_1^2\\
&+\frac{(L+1)(D-4+L)-2r f'}{r^2 f}h_0^2+\frac{4\Lambda}{D-2}\left(f h_1^2-f^{-1}h_0^2\right)
\bigg].
\end{aligned}
\label{eq:h0h1action}
\ee
In this choice of variables, it is somewhat difficult to isolate the physical degree of freedom because neither $h_0$ nor $h_1$ is obviously auxiliary. It is therefore useful to integrate in an additional auxiliary field $Q$ in a similar manner to~\cite{DeFelice:2011ka,Franciolini:2018uyq,Franciolini:2018aad}, so that our action becomes
\be
\begin{aligned}
S_{\rm RW}=\int\rd t\rd r\,r^{D-4}\bigg[&2Q\left(\dot h_1+\frac{2}{r}h_0-h_0'\right)-Q^2+\frac{2(D-3)f+2rf'-(L+1)(D-4+L)}{r^2}f\,h_1^2\\
&+\frac{(L+1)(D-4+L)-2(D-3)f-2r f'}{r^2 f}h_0^2+\frac{4\Lambda}{D-2}\left(f h_1^2-f^{-1}h_0^2\right)
\bigg].
\end{aligned}
\label{eq:Qaction}
\ee
The organizing principle is to make the derivatives of $h_0$ and $h_1$ appear as a perfect square and then to introduce $Q$ in such a way that integrating it out reproduces the original action.

The actions~\eqref{eq:Qaction} and~\eqref{eq:h0h1action} are equivalent, but we can now integrate out $h_0$ and $h_1$ to get an action only for $Q$. Their equations of motion set
\begin{align}
h_0 &=-\frac{r f}{(L-1)(D-2+L)}\left[(D-2)Q+rQ'\right],\label{h0odd}\\
h_1&= -\frac{r^2}{(L-1)(D-2+L)f}\dot Q.
\end{align}

We can then substitute these equations back into the action. In simplifying the resulting expression, it is helpful to use the background equations of motion, which imply
\be
f'' +\frac{(D-2) f'}{r}+\frac{4\Lambda}{D-2}=0.
\label{eq:feinsteineq}
\ee
The action for $Q$ then takes the form  (after several integrations by parts)
\be
S_{\rm RW}=\int\rd t\rd r \frac{r^{D-2}}{(L-1)(D-2+L)}\bigg[ \frac{1}{f}\dot Q^2-f  Q'{}^2-\left(\frac{(L+1)(D-4+L)-(D-4)f}{r^2}-\frac{Df'}{r}-\frac{4\Lambda }{D-2}\right)Q^2
\bigg].
\ee
It is again useful to write things in terms of a canonically normalized Schr\"odinger variable. We define
\be
\Psi_{\rm RW} \equiv \left(\frac{2r^{D-2}}{(L-1)(D-2+L)}\right)^{1/2}Q,
\label{Qvarle}
\ee
and transform to the tortoise coordinate. After integrations by parts the action takes the form
\begin{tcolorbox}[colframe=white,arc=0pt,colback=greyish2]
\be
S_{\rm RW}=\int\rd t\rd r_\star\Bigg( \frac{1}{2}\dot \Psi_{\rm RW}^2-\frac{1}{2}\left(\frac{\partial\Psi_{\rm RW}}{\partial r_\star}\right)^2-\frac{1}{2}V_{\rm RW}(r)\Psi_{\rm RW}^2
\Bigg),
\label{eq:RWaction}
\ee
\end{tcolorbox}
\vspace{-5pt}
\noindent
with the Regge--Wheeler potential 
\be
V_{\rm RW}(r) = f\frac{(L+1)(D-4+L)}{r^2}+f^2\frac{(D-4)(D-6)}{4r^2}-ff'\frac{(D+2)}{2r}-\frac{4\Lambda f }{D-2}\,.
\ee
The Schr\"odinger equation following from the action~\eqref{eq:RWaction} is
\be
\frac{\rd^2\Psi_{\rm RW}}{\rd r_\star^2} + \Big(\omega^2 - V_{\rm RW}(r)\Big)\Psi_{\rm RW} = 0,
\label{eq:ddimRWeq}
\ee
which is precisely
the $D$-dimensional Regge--Wheeler equation~\cite{Kodama:2003jz}. Note, however, that $\Psi_{\rm RW}$ is not quite the usual Regge--Wheeler variable: it is more properly called the Cunningham--Price--Moncrief variable~\cite{Cunningham:1978zfa,Martel:2005ir}. The usual Regge--Wheeler variable is (up to a numerical factor) the time derivative of what we have called $\Psi_{\rm RW}$. However, since both of these variables satisfy the same (Regge--Wheeler) equation, we will slightly abuse terminology and refer to $\Psi_{\rm RW}$ as the Regge--Wheeler variable.

\subsubsection{Spin-2 scalar sector (even parity)}
\label{sec:scalarspin2}
Finally we consider the scalar sector. In $D=4$, this coincides with the even parity sector. In this case, the relevant degrees of freedom are the variables $H_0, H_1, H_2, {\cal H}_0, {\cal H}_1, G, {\cal K}$. 

It is convenient to fix a gauge where\footnote{As discussed in Footnote~\ref{foot:actiongauge}, only a subset of gauge choices can be imposed at the level of the action without losing information contained in the equations of motion. Both this gauge and the Regge--Wheeler gauge in the vector/odd sector can be used in the action, as shown explicitly in \cite{Motohashi:2016prk}.}
\be
{\cal H}_0 = {\cal K} = G = 0,
\label{eq:evengauge}
\ee
so that the residual degrees of freedom are $H_0,H_1,H_2, {\cal H}_1$. It is clear that we have enough freedom to fix this gauge: setting $G =0$ uses the $\xi_S$ freedom, setting ${\cal K} = 0$ uses up the $\xi_1$ freedom and setting ${\cal H}_0 = 0$ fixes $\xi_0$ gauge transformations. With the gauge choice~\eqref{eq:evengauge}, the action becomes\footnote{In deriving this form of the action we have integrated by parts freely and used the background equation~\eqref{eq:feinsteineq}.}
\be
\begin{aligned}
S_{\rm Z} = L(L+D-3) \int\rd t\rd r\,r^{D-4}\Bigg[&\dot{\cal H}_1^2+\frac{2f}{r^2}\left((D-3)f+\frac{2\Lambda r^2}{D-2}+rf'\right){\cal H}_1^2+\frac{(D-2)(rf'+(D-3)f)}{2L(L+D-3)}H_2^2\\
&\hspace{-5cm}+H_0\Bigg( \bigg[f'+\frac{2(D-3)f}{r}\bigg]{\cal H}_1+2f{\cal H}_1' - \left(1+\frac{(D-3)(D-2)f+(D-2)rf'}{L(L+D-3)}\right)H_2-\frac{(D-2)rf H_2'}{L(L+D-3)}\Bigg)\\
&~~~~-\frac{(2(D-3)f+rf')}{r}H_2{\cal H}_1+ H_1\left(H_1-2 \dot{\cal H}_1+\frac{2(D-2)r}{L(L+D-3)}\dot H_2\right)
\Bigg].
\end{aligned}
\label{actionevenD}
\ee
Though this expression is fairly complicated, we
expect that there should be a single physical degree of freedom, and we would like to isolate it.

It is reasonably clear from the action that $H_1$ is auxiliary; it can be integrated out via its equation of motion, which sets
\be
H_1 = \dot{\cal H}_1-\frac{(D-2) r}{L(L+D-3)}\dot H_2.
\ee
The variable $H_2$ is also auxiliary, but in order to integrate it out we first trade off ${\cal H}_1$ for another variable $\mathcal{V}$ defined by
\be
 {\cal V} = {\cal H}_1 - \frac{(D-2)r}{2L(L+D-3)} H_2.
\label{eq:Vaux}
\ee
This makes $H_2$ appear algebraically in the $H_0$ equation of motion (which is a constraint), so that it can be solved for in terms of ${\cal V}$:
\be
H_2 = \frac{2L(L+D-3)\Big[2f(r {\cal V}'+(D-3){\cal V})+r f' {\cal V}\Big]}{r\Big[2L(L+D-3)-2(D-2)f+(D-2)rf'\Big]}.
\label{eq:H2aux}
\ee
Substituting this solution back into the action eliminates both $H_2$ and $H_0$, because we have solved the constraint that the latter enforces. We are therefore left with an action 
 for only the variable ${\cal V}$. After integration by parts it can be written as
\be
S_{\rm Z} = \int \rd t\rd r\, r^{D-4}{\cal F}(r) \left(  \dot {\cal V}^2- f^2{\cal V}'{}^2 +{\cal N} (r){\cal V}^2\right)\,,
\label{actioncalV}
\ee
where the functions that appear are rather complicated:
\begin{subequations}
\begin{align}
{\cal F}(r)&\equiv \frac{8(D-2)L(L+D-3)f\Big[(D-3)L(L+D-3)-(D-3)(D-2)f-2\Lambda r^2-(D-2)rf'\Big]}{\Big[2L(L+D-3)-2(D-2)f+(D-2)r f'\Big]^2},\\
{\cal N}(r) &\equiv \frac{{\cal A}+{\cal B}+{\cal C}}{r^2(2L(L+D-3)^2-2(D-2)f+(D-2)rf')},\\\nonumber
\end{align}
\end{subequations}
with the following expressions appearing in ${\cal N}$:
\begin{subequations}
\begin{align}
&~{\cal A} = 2f\left[-L^2(L+D-3)^2+f\left(2L(L+D-3)+(D-4)(D-2)f+4(D-3)\Lambda r^2\right)\right],\\
&~{\cal B} = rf f'\left[3(D-4)L(L+D-3)+(D-2)(24+D(2D-13))f+4\Lambda r^2\right],\\
&~{\cal C} = r^2 f'{}^2\left[2L(L+D-3)+(D-2)(3D-10)f+(D-2)rf'\right].
\end{align}
\end{subequations}
We can canonically normalize the action by defining the Zerilli variable
\be
\Psi_Z \equiv \left(2f r^{D-4}{\cal F}\right)^{1/2}{\cal V},
\label{eq:Zvariable}
\ee
to finally obtain
\begin{tcolorbox}[colframe=white,arc=0pt,colback=greyish2]
\be
S_{\rm Z}=\int\rd t\rd r_\star\Bigg( \frac{1}{2}\dot \Psi_{\rm Z}^2-\frac{1}{2}\left(\frac{\partial\Psi_{\rm Z}}{\partial r_\star}\right)^2-\frac{1}{2}V_{\rm Z}(r)\Psi_{\rm Z}^2
\Bigg),
\label{eq:Zaction}
\ee
\end{tcolorbox}
\vspace{-5pt}
\noindent
with the Zerilli potential 
\be
V_{\rm Z}(r) = \frac{f\, \hat V_Z(r)}{4(D-2)r^2H(r)^2},
\ee
where we have defined the functions
\begin{align}
H(r) &\equiv 2L(L+D-3)-2(D-2)f+(D-2)rf'\\\nonumber
\hat V_Z(r) &\equiv 4(D-4)(D-2)^4 f^3-8(D-2)^2\Big[(D-2)(D-6)L(L+D-3)-8(D-3)\Lambda r^2\Big]f^2\\\nonumber
&~~~+4(D-2)\Big[(D-2)(D-12)L^2(L+D-3)^2-16(D-4)L(L+D-3)\Lambda r^2+32\Lambda^2 r^4\Big]f\\
&~~~+2(D-2)^3(D+2)r^3 f'{}^3-4(D-2)^2 r^2\Big[(D-6)L(L+D-3)-4\Lambda r^2\Big]f'{}^2\\\nonumber
&~~~-8(D-2)^2L^2(L+D-3)^2rf'+12(D-2)^5 rf^2f'+(D-2)^3(D(D+10)-32)r^2 ff'{}^2\\\nonumber
&~~~-4(D-2)^2\Big[(D-2)(3D-8)L(L+D-3)-8D\Lambda r^2\Big]rff'\\\nonumber
&~~~+16L^2(L+D-3)^2\Big[(D-2)L(L+D-3)-4\Lambda r^2\Big].
\end{align}
The corresponding equation of motion
\be
\frac{\rd^2\Psi_{\rm Z}}{\rd r_\star^2} + \Big(\omega^2 - V_{\rm Z}(r)\Big)\Psi_{\rm Z} = 0,
\label{eq:ddimZeq}
\ee
is the $D$-dimensional Zerilli equation, and agrees with the expression derived in~\cite{Kodama:2003jz}.
In $D=4$ with $\Lambda=0$, it agrees with the usual Zerilli variable \cite{Martel:2005ir}.

\subsubsection{Chandrasekhar's symmetry in $D=4$}
There is also a unique aspect of the theory of massless spin-2 perturbations in $D=4$. Much like the spin-1 case, a massless spin-2 in a black hole background exhibits a duality symmetry, though of a much more nontrivial form. This symmetry was first discovered by Chandrasekhar~\cite{Chandrasekhar:1985kt,1975RSPSA.343..289C}, motivated by the observation that the spectra of quasi-normal modes of the Regge--Wheeler and Zerilli equations are identical~\cite{Chandrasekhar:1975zza}.

In four dimensions, the action for the physical degrees of freedom simplifies dramatically. The Regge--Wheeler and Zerilli variables still decouple (now as a consequence of parity) and their combined action can be written as
\begin{align}
S&= \frac{1}{2}\int{\rm d} t {\rm d} r_\star\Bigg(\dot\Psi_\mathrm{RW}^2 - \left(\frac{\partial\Psi_\mathrm{RW}}{\partial r_\star}\right)^2 - V_\mathrm{RW}(r)\Psi_\mathrm{RW}^2+\dot\Psi_\mathrm{Z}^2 - \left(\frac{\partial\Psi_\mathrm{Z}}{\partial r_\star}\right)^2 - V_\mathrm{Z}(r)\Psi_\mathrm{Z}^2  \Bigg).\label{eq:EH-Sch}
\end{align}
The Regge--Wheeler and Zerilli potentials take the simplified form
\begin{align}
\label{eq:rwpot1}
V_{\rm RW}(r) &= f(r) \left(\frac{L(L+1)}{r^2}-\frac{3r_s}{r^3}\right),\\
V_{\rm Z}(r) &= f(r)\bigg(\frac{r_s}{r^3}+\frac{2\lambda}{3r^2}+\frac{8\lambda^2(2\lambda+3)-18\Lambda r_s^2}{3(2\lambda r+3r_s)^2}\bigg),
\label{eq:zpot2}
\end{align}
where the function $f(r)$ is given by
\be
f(r)  = 1-\frac{r_s}{r}-\frac{\Lambda r^2}{3},
\ee
and we define
\be
\lambda \equiv\frac{(L-1)(L+2)}{2}.
\ee
The action~\eqref{eq:EH-Sch} possesses a duality symmetry, which follows from Chandrasekhar's observation 
that both the Regge--Wheeler and Zerilli potentials can be derived from a single {\it superpotential}~\cite{Berti:2009kk},
\be
W(r) = \frac{3r_s(r_s-r)}{r^2(3r_s+2\lambda r)}-\frac{2\lambda(\lambda+1)}{3r_s}+\frac{\Lambda r_s \,r}{3r_s+2\lambda r}, \label{eq:superpot}
\ee
in the sense that 
\begin{align}
V_{\rm RW} &= W^2 + f(r)\frac{{\rm d} W}{{\rm d} r}+\beta, \\
V_{\rm Z} &= W^2 - f(r)\frac{{\rm d} W}{{\rm d} r}+\beta,
\end{align}
where we have defined the constant
\be
\qquad\beta \equiv -\frac{4\lambda^2(\lambda+1)^2}{9r_s^2}.
\ee
This relation between the potentials is ultimately responsible for the isospectrality of the even and odd sectors (the fact that they have same set of quasi-normal modes) when $\Lambda \geq 0$ \cite{Berti:2009kk}, and is a manifestation of the fact that the Regge--Wheeler and Zerilli potentials are partner potentials in the sense of supersymmetric quantum mechanics~\cite{Cooper:1994eh}.\footnote{An alternative perspective on the relation between the Regge--Wheeler and Zerilli potentials is that the Chandrasekhar symmetry is an example of a Darboux transformation between differential equations~\cite{Glampedakis:2017rar}. From this point of view, the distinguishing feature of $D=4$ is that a transformation can be found that preserves the boundary conditions of interest in physical situations~\cite{uspaper2}. This is also what goes wrong with AdS asymptotics: there the Chandrasekhar transformation does not preserve such boundary conditions, so the two sectors are not isospectral even in $D=4$.}
Rewriting the Regge--Wheeler and Zerilli potentials in terms of $W$, it is in fact straightforward to check that the action is invariant under the duality transformation 
\begin{subequations} \label{eq:psisym}
\begin{align} 
\delta\Psi_\mathrm{Z} &= \left(\frac{\partial}{\partial r_\star}-W(r)\right)\Psi_\mathrm{RW}, \\
\delta\Psi_\mathrm{RW} &= \left(\frac{\partial}{\partial r_\star}+W(r)\right)\Psi_\mathrm{Z}, 
\end{align}
\end{subequations}
which is a true off-shell symmetry, much as electric-magnetic duality is for the spin-1. Incidentally, because the Regge--Wheeler and Zerilli equations are linear, this implies that the right hand sides of eqs.~\eqref{eq:psisym} are solutions to the Zerilli and Regge--Wheeler equations respectively. The symmetry is continuous, and therefore it also gives rise to a conserved Noether current:
\begin{align}
J^t &=  -\Psi_\mathrm{Z}'\dot\Psi_\mathrm{RW} - \dot\Psi_\mathrm{Z}\Psi_\mathrm{RW}' + W\left(\Psi_\mathrm{RW}\dot\Psi_\mathrm{Z}-\Psi_\mathrm{Z}\dot\Psi_\mathrm{RW}\right), \label{eq:Jt}\\
J^{r_\star} &=\dot\Psi_\mathrm{Z}\dot\Psi_\mathrm{RW} + \Psi_\mathrm{Z}'\Psi_\mathrm{RW}' + W\left(\Psi_\mathrm{Z}\Psi_\mathrm{RW}'-\Psi_\mathrm{RW}\Psi_\mathrm{Z}'\right) -\left(W^2+\beta\right)\Psi_\mathrm{Z}\Psi_\mathrm{RW}, \label{eq:Jrs}
\end{align}
which obeys the conservation law
$\partial_tJ^t + \partial_{r_\star} J^{r_\star} = 0$.

This symmetry has a number of interesting consequences, which we explore in~\cite{uspaper2}. For instance, in addition to underlying isospectrality, the conservation of this current also is responsible for even and odd parity tidal Love numbers being equal in $D=4$, in certain cases. Importantly, this symmetry does not rely on auxiliary variables having been integrated out: it is possible to uplift the Chandrasekhar duality to a novel, off-shell symmetry of the Einstein--Hilbert action linearized around Schwarzschild \cite{uspaper2}.

\newpage
\section{Static solutions and response to external fields}
\label{sec:lovenums}
In the previous sections we have derived the actions governing linearized perturbations in a Schwarzschild--(A)dS background and studied their symmetries. We now turn to the study of solutions to their equations of motion in the Schwarzschild limit (with $\Lambda = 0$). A physically interesting first step, and the focus of the rest of this paper, is to study static solutions in a black hole background. These solutions capture the linear response of a black hole to an externally applied field. A particularly interesting special case is the black hole response to a gravitational tidal field. This response is encoded in the so-called Love numbers, which measure the induced gravitational perturbation away from spherical symmetry  due to the tidal forces, and could possibly be measured by next-generation gravitational wave experiments~\cite{Flanagan:2007ix,Hinderer:2007mb,Cardoso:2017cfl,Chirenti:2020bas}.

In this section, we solve the equations governing the dynamics of massless spin-$0$, spin-$1$ and spin-$2$ fields in the zero-frequency limit in a Schwarzschild background and compute the linear response coefficients as functions of $D$ and $L$ for  various spins.\footnote{In an abuse of terminology we will often refer to all of the response coefficients as ``Love numbers" despite the fact that, for example, the spin-$1$ responses are more properly called (electric) polarizabilities or (magnetic) susceptibilities.} Some of these cases have already been analyzed in the literature \cite{Damour:2009va,dos2016relativistic,Kol:2011vg}, while others, such as spin-$1$ fields and parity-odd Love numbers for spin-$2$ perturbations in general dimensions have not, to our knowledge, been computed before. 
We find that---similar to the case of parity-even perturbations---the Love numbers corresponding to spin-2, parity-odd perturbations vanish in $D=4$, but not in higher dimensions~\cite{Damour:2009vw,Binnington:2009bb,Fang:2005qq,Kol:2011vg,Chakrabarti:2013lua,Gurlebeck:2015xpa}. 
We also find a similar behavior for the spin-1 electric polarizability and magnetic susceptibility: they vanish in $D=4$, but not for higher-dimensional black holes.
We treat all of these cases in a unified manner, reproducing known results and studying those cases that had not been previously considered.
All of the relevant equations can be solved in the zero frequency limit in terms of hypergeometric functions. This allows for the simultaneous treatment of the various cases of interest using general properties of the hypergeometric equation. For the reader's convenience, we have compiled in Appendix~\ref{App:hyperG} a list of elementary properties of hypergeometric functions which we will use in the computations below.

\vspace{-.5cm}
\paragraph{An instructive example:}
Though the setting might be somewhat foreign, the problem that we are trying to solve in computing Love numbers is in fact quite familiar. Conceptually, it is precisely the same as computing the electric polarizability of a material. It is therefore helpful to review this computation in this simplified setting: the (dipole) polarizability of a conducting sphere in flat space and $D = 4$.

In the static limit $\Psi_S$ and the radial electric field $E_r$ are very simply related, up to normalization we have $\Psi_S= r^2 E_r$. Therefore the electric field outside such a conducting sphere is governed by equation~\eqref{eq:masslessEmodespin10}. Moreover, in flat space we have $f = 1$ and $r_\star = r$. Therefore, $E_r$ satisfies the following equation:\footnote{This equation is nothing more than (a derivative of) Gauss' law written in spherical coordinates.}
\be
E_r''+\frac{4}{r}E_R'- \frac{L(L+1)-2}{r^2}E_r = 0,
\label{eq:externalE}
\ee
where $(\phantom{c})'\equiv \rd (\phantom{c})/\rd r$. The most general solution to this differential equation is straightforward to write down:
\be
E_r = c_{L-1} \,r^{L-1}+c_{-L-2} \,r^{-L-2},
\ee
where $c_{L-1}$ and $c_{-L-2}$ are free parameters fixed by boundary conditions.
We will focus on the dipole case, $L=1$, so that the two independent solutions are a constant and  $1/r^3$. The goal is to understand the response of the sphere to a long-wavelength external electric field, so we imagine applying a constant electric field as $r\to \infty$, with angular structure given by an $L=1$ harmonic and with some particular normalization.
This forms one of the boundary conditions to the differential equation~\eqref{eq:externalE}. As a second boundary condition, we demand that the tangent electric field vanishes at the surface of the conductor, which we take to have radius $R$. For all but the $L=0$ mode, this implies that we should set $E_r$ to zero at the surface of the conductor.\footnote{This is just the standard boundary condition for a conductor, demanding that the electric field tangent to its surface vanishes. Since all modes with $L\geq 1$ have a nontrivial angular structure, their coefficients must consequently vanish. Notice that, much like the black hole case, this boundary condition does not require us to know anything about the internal structure of the conductor.} This implies the equation
\be
E_r(R) = c_0 +c_{-3} R^{-3} = 0,
\ee
which allows us to solve for $c_{-3}$ in terms of $c_{0}$ as $c_{-3}= -c_0 R^{3}$. We therefore see that the solution for $E_r$ is given by:
\be
E_r = c_0\left[ 1- \left(\frac{R}{r}\right)^3\right].
\ee
Once we have this solution, the idea is to read off the coefficient of the $r^{-3}$ term, which is the dipole induced by the long-wavelength background field.
More explicitly, the electric polarizability is the ratio of the coefficient of the $1/r^3$ dipole to the coefficient of the constant term and is given by
\be
\alpha_E = -R^{3}.
\ee
This implies that the polarizability of a conducting sphere is proportional to the volume of the sphere, which is a well-known fact.
The minus sign captures the fact that the charges deform in such a way as to counteract the applied electric field. Notice also that the overall normalization of the external field drops out because it is common to both parts of the solution, so that the polarizability can really be thought of as an intrinsic property of the object that is independent of the magnitude of the external field.

In the following we solve the analogous problem for fields in a Schwarzschild background geometry.

\subsection{Spin-0: Scalar field response}
\label{Sec:spin-0}

As a first example, we consider the induced scalar charge from a long-wavelength scalar field perturbation in a Schwarzschild background. This computation was first carried out in~\cite{Kol:2011vg}, but we review it here for completeness, and because it provides a simple illustrative example in preparation for the spin-1 and spin-2 cases.

Our starting point is the Schr\"odinger like equation for a real scalar field in eq.~\eqref{eq:s0schrod}, which has been decomposed into spherical harmonics. In the zero-frequency limit (and setting the mass to zero), it reduces to
\begin{equation}
f\Psi'' +f' \Psi'  - \left(\frac{L(L+D-3)}{r^2}+f'\frac{D-2}{2r}+f\frac{(D-4)(D-2)}{4r^2}\right)\Psi = 0 \, ,
\label{spin0}
\end{equation}
where we have reverted from the tortoise coordinate back to the original radial coordinate $r$. It is also convenient to introduce the dimensionless radial variable
\begin{equation}
x \equiv \left(\frac{r_s}{r} \right)^{D-3} ,
\label{x}
\end{equation}
such that the black hole horizon is now located at $x=1$, while spatial infinity corresponds to $x =0$. In addition to this coordinate change, we also perform the field redefinition
\begin{equation}
u(x) \equiv x^{-\frac{D+2 L-4}{2 (D-3)}}  \Psi(r(x)) \, ,
 \label{rescalingphi}
\end{equation}
which recasts eq.~\eqref{spin0} as a hypergeometric equation in the standard form (see Appendix~\ref{App:hyperG} for more details):
\begin{equation}
x(1-x) u''(x) + \big[c - (a+b+1)x\big] u'(x) - a \,  b \,  u(x) =0 \, ,
\label{hyper0}
\end{equation}
where the parameter values are given by
\begin{equation}
a = \hat L+1  \, ,
\qquad
b= \hat L +1 \, ,
\qquad
c= 2\hat L+2 \, ,\qquad{\rm with}\qquad\hat L \equiv \frac{L}{D-3} .
\label{abcS}
\end{equation}
Notice that the parameters $a$, $b$ and $c$ satisfy the condition $a+b-c=0$.
The benefit of these transformations is that the hypergeometric equation is extremely well-studied, and therefore the solutions of interest are readily available in the literature. 

The differential equation~\eqref{hyper0} is a second-order equation, so we require two boundary conditions to specify completely a solution. On physical grounds, the first requirement we will impose is that our solution be regular at the black hole horizon, i.e., at $x=1$. The second boundary condition fixes instead the normalization of the growing mode solution at radial infinity, i.e., at $x=0$. We can then read off the induced sub-leading fall-off at radial infinity, which captures the linear response to the externally applied field. Note that the overall normalization of the solution at infinity is formally a boundary condition, but it does not affect the ratio of the growing and decaying modes at infinity, which is ultimately what we are interested in.

As reviewed in Appendix~\ref{App:hyperG}, the form of the solutions to the hypergeometric  equation \eqref{hyper0} depends drastically on whether the numbers $a$, $b$, $c-a$, and $c-b$ take integer or non-integer values, i.e., on~whether $\hat L$ is integer, half-integer, or neither integer nor half-integer. Indeed, for some of these values solutions to the hypergeometric equation that would otherwise be linearly independent become degenerate, i.e., linearly dependent. When that is the case, new linearly independent solutions can be found, but their asymptotic structure is typically very different from the original solutions. As we will see, this phenomenon is at the heart of the differences in the computation of Love numbers in different dimensions.

In the following we enumerate each of the individual cases and describe both the linearly independent solutions, as well as the particular combination of such solutions that is regular at the black hole horizon. We then read off the Love numbers by expanding this regular solution near infinity.

\begin{itemize}

\item
{\bf $\hat L$ is  neither integer nor half-integer:}
In this case, all   the  parameters $a$, $b$, $c-a$, $c-b$, and $c$   are non-integer. The two linearly  independent solutions to \eqref{hyper0} are \cite{slavjanov2000special,Bateman:100233,beals_wong_2010}
\be
\def\arraystretch{.8}
u_1(x) = \hypergeom{2}{1}\left[\begin{array}{c}
\hat L+1,~~\hat L+1\\
2\hat L+2
\end{array}\Big\rvert \,x\,\right]~~~~{\rm and}~~~~
u_5(x) = x^{-2\hat L-1 }\hypergeom{2}{1}\left[\begin{array}{c}
-\hat L,~~-\hat L\\
-2\hat L
\end{array}\Big\rvert \,x\,\right]\,.
\label{spin0solhyper}
\ee
This basis of solution is particularly natural because it corresponds to the two linearly independent fall-offs near $x = 0$ (or, a bit more formally, the asymptotic form of these solutions are eigenfunctions of the dilation operator with two different eigenvalues). See also eq.~\eqref{hypergcase1} for more details. 
Using the identity \eqref{app:id1} and the asymptotic expansion at the horizon in eq.~\eqref{app:id1bis}, one finds that the particular linear combination that remains finite at the horizon ($x=1$) is given by,\footnote{See also eq.~\eqref{hypergcase1-regular} in the Appendix.}
\begin{equation}
\def\arraystretch{.8}
u (x) = A \left( \frac{\Gamma(-2\hat L-1)}{\Gamma(-\hat L)^2} \hypergeom{2}{1}\left[\begin{array}{c}
\hat L+1,~~\hat L+1\\
2\hat L+2
\end{array}\Big\rvert \,x\,\right]
+ \frac{\Gamma(2\hat L+1)}{\Gamma(\hat L+1)^2} x^{-2\hat L-1 }\hypergeom{2}{1}\left[\begin{array}{c}
-\hat L,~~-\hat L\\
-2\hat L
\end{array}\Big\rvert \,x\,\right]\right) \, ,
\label{reg0}
\end{equation}
with $A$ an overall normalization constant. Note that we can use the connection formula
\be
\def\arraystretch{.8}
\begin{aligned}
 \hypergeom{2}{1}\left[\begin{array}{c}
a,~~b\\
a+b-c+1
\end{array}\Big\rvert \,1-x\,\right]
= &~\frac{\Gamma(1-c)\Gamma(a+b-c+1)}{\Gamma(a-c+1)\Gamma(b-c+1)} \hypergeom{2}{1}\left[\begin{array}{c}
a,~~b\\
c
\end{array}\Big\rvert \,x\,\right]
\\
&\,+ \frac{\Gamma(c-1)\Gamma(a+b-c+1)}{\Gamma(a)\Gamma(b)} x^{1-c} \hypergeom{2}{1}\left[\begin{array}{c}
a-c+1,~~b-c+1\\
2-c
\end{array}\Big\rvert \,x\,\right] \, ,
\end{aligned}
\label{identity0}
\ee
to rewrite eq.~\eqref{reg0} in a more compact form as~\cite{Kol:2011vg}
\be
\def\arraystretch{.8}
u (x) = A
\hypergeom{2}{1}\left[\begin{array}{c}
\hat L+1,~~\hat L+1\\
1
\end{array}\Big\rvert \,1-x\,\right]\,,
\label{spin0regsolv2}
\ee
which is manifestly regular in the limit $x\to 1$, since hypergeometric functions are normalized in such a way that $\hypergeom{2}{1} \left(a,b;c;0\right)=1$.

In order to extract the Love numbers, we expand the solution \eqref{spin0regsolv2} around $x=0$ to find 
\begin{equation}
u (x\rightarrow0) \simeq A \left( \frac{\Gamma(-2\hat L-1)}{\Gamma(-\hat L)^2} 
+ \dots + \frac{\Gamma(2\hat L+1)}{\Gamma(\hat L+1)^2} x^{-2\hat L-1 } + \dots \right)  \, ,
\label{reg0Love}
\end{equation}
where we are keeping only the contributions corresponding to the two linearly independent fall-offs at infinity, which are the ones relevant for the calculation of the response.

In terms of the radial coordinate, eq.~\eqref{reg0Love} takes the form
\begin{equation}
u(r\rightarrow\infty) \simeq A \left(\frac{r}{r_s} \right)^{L+D-3}  \left(
 \frac{\Gamma(2\hat L+1)}{\Gamma(\hat L+1)^2} \left(\frac{r}{r_s} \right)^{L}
 + \dots +  \frac{\Gamma(-2\hat L-1)}{\Gamma(-\hat L)^2}  \left(\frac{r_s}{r} \right)^{L+D-3} +\dots\right)  \, ,
\label{reg0Lover}
\end{equation}
with the first term to be interpreted as an external tidal field with overall amplitude $A$, while the second term encodes the response of the system. As expected for weak perturbations, the response is linear in the magnitude of the external field.

The static response is then defined as the ratio between the coefficient of the induced $r^{-(L+D-3)}$ tail of the solution and the $r^L$ tidal component, measured in units of $r_s^{2L+D-3}$:
\begin{tcolorbox}[colframe=white,arc=0pt,colback=greyish2]
\begin{equation}
k = \frac{\Gamma(-2\hat L-1)}{\Gamma(-\hat L)^2}  \frac{\Gamma(\hat L+1)^2 }{  \Gamma(2\hat L+1)} 
=  \frac{2\hat L+1}{2\pi }
 \frac{\Gamma(\hat L+1)^4}{ \Gamma(2\hat L+2)^2  } \tan (\pi \hat L ) \,,
 \label{scalarLoven}
\end{equation}
\end{tcolorbox}
where we have used some of the identities in Appendix~\ref{App:hyperG} to simplify the final result, which agrees with the expression found in~\cite{Kol:2011vg}.

\item
{\bf $\hat L$ is  half-integer:} If $\hat L$ is  half-integer, the two solutions~\eqref{spin0solhyper} cease to be linearly independent. Translating $\hat L$ back into the parameters $a,b,c$ using~\eqref{abcS}, $\hat L$ being half-integral implies that  $a$, $b$, $c-a$ and $c-b$ are non-integer, while $c$  takes positive integer values. Using this information we can use standard results to find a new basis of solutions~\cite{Bateman:100233}---see also eq.~\eqref{halfinteger1},
\be
\def\arraystretch{.8}
u_1(x) = \hypergeom{2}{1}\left[\begin{array}{c}
\hat L+1,~~\hat L+1\\
2\hat L+2
\end{array}\Big\rvert \,x\,\right]~~~~{\rm and}~~~~
u_2(x) = \hypergeom{2}{1}\left[\begin{array}{c}
\hat L+1,~~\hat L+1\\
1
\end{array}\Big\rvert \,1-x\,\right]\,.
\label{half-integer1}
\ee
Note that the first solution $u_1(x)$ contains a logarithmic divergence of the form $\log(1-x)$ around $x=1$. On the other hand $u_2(x)$ is finite as $x\to 1$ and therefore it is the solution
describing the physical scalar perturbations around a Schwarzschild black hole.  In particular, since $c=1,2,3,\cdots$ and $a,b\neq c-1,c-2,\cdots, 0 ,-1,-2,\cdots$, one can infer its asymptotic expansion in the neighborhood of $x=0$ via the formula~\eqref{half-integer2} 
\be
\def\arraystretch{.8}
\begin{aligned}
 \hypergeom{2}{1}\left[\begin{array}{c}
a,~~b\\
1+a+b-c
\end{array}\Big\rvert \,1-x\,\right]&=\hypergeom{2}{1}\left[\begin{array}{c}
a,~~b\\
c
\end{array}\Big\rvert \,x\,\right] 
\log x - \sum_{n=1}^{c-1} \frac{(c-1)!(n-1)!}{(c-n-1)!(1-a)_n(1-b)_n}(-x)^{-n}
\\
&~~~\,+ \sum_{n=0}^{\infty}\frac{(a)_n(b)_n}{(c)_n n!} \big[\psi(a+n) + \psi(b+n) - \psi(1+n) - \psi(c+n)\big]  x^n \, ,
\label{half-integer20}
\end{aligned}
\ee
where $\psi(x)\equiv \Gamma'(x)/\Gamma(x)$ is the digamma function. This asymptotic expansion is of a drastically different form than eq.~\eqref{reg0Love}. In fact, keeping in~\eqref{half-integer20} only the leading term and the one that scales like $x^{-2 \hat L -1}$, as we did in eq.~\eqref{reg0Love}, and substituting in~\eqref{abcS} for $a,b,c$ one finds:
\begin{equation}
u_2(x) \simeq \log x + \dots 
+ (-1)^{2\hat{L}} (2\hat{L})!(2\hat{L}+1)! \frac{\Gamma(-\hat{L})^2}{\Gamma(\hat{L}+1)^2}x^{-2\hat{L}-1} +\cdots\, .
\label{eq:scalrexpdeg}
\end{equation}
An important difference compared to the case studied above is that~\eqref{eq:scalrexpdeg} does not consist only of powers of $x$, but contains also a logarithmic divergence as $x\to 0$. This logarithm can be understood as a classical running of the value of the induced response~\cite{Kol:2011vg}. In more detail, we can take the  ratio of the two fall-offs in~\eqref{eq:scalrexpdeg} to define the dimensionless response (in units of $r_s^{2L+D-3}$):
\begin{tcolorbox}[colframe=white,arc=0pt,colback=greyish2]
\vspace{-4pt}
\begin{equation}
k =  \frac{(-1)^{2\hat{L}}(D-3) \Gamma(\hat{L}+1)^2  }{(2\hat{L})!(2\hat{L}+1)! \Gamma(-\hat{L})^2} \log \left( \frac{r_0}{r} \right)
\qquad\qquad
\text{(half-integer $\hat{L}$)} .
\label{lovescalardiv}
\end{equation}
\end{tcolorbox}
Here we have only recorded the coefficient of the logarithmic term in the ratio of fall-offs in~\eqref{eq:scalrexpdeg}. This is because only these terms are unambiguous. The dependence on $r$---which we can think of as the distance at which we measure the response of the system---is an example of classical renormalization group (RG) running. The length scale $r_0$ is a renormalization scale to be fixed by experiments, but that on physical grounds we expect to be of $\mathcal{O}(r_s)$.\footnote{There is an inherent ambiguity in the quantity~\eqref{lovescalardiv} in that it depends on the distance/scale $r$ at which we choose to measure the response. However, if we fix the value at some particular distance $r_0$, the way in which it changes (or ``runs") with scale is unambiguous. In particular, the coefficient of the logarithmic term is unambiguous and independent of the calculational technique. 
In \cite{Kol:2011vg}, the Love numbers in this degenerate case are obtained from the general expression~\eqref{scalarLoven} by taking the limit of half-integer values for $\hat L$. This limit is singular, but it is possible to isolate a finite contribution by a suitable (classical) renormalization procedure that  removes the divergent piece.
As expected, the value~\eqref{lovescalardiv} has precisely the same logarithmic term as in~\cite{Kol:2011vg}, but differs in the finite terms. This is consistent, as one can adjust the scale $r_0$ so that the two quantities agree at some distance, and they will continue to agree at all scales thereafter.}

\item
{\bf $\hat L$ is an integer:} Lastly we consider the case where $\hat L$ is an integer. This is the case of interest for $D=4$ spacetime dimensions.
If $\hat L$ is  integer, we see from~\eqref{abcS} that  $a$, $b$ and $c$ are all integers as well. As reviewed in Appendix~\ref{App:hyperG}, in this situation, the two solutions~\eqref{spin0solhyper} are again degenerate. 
Two linearly independent solutions are instead given by\footnote{See also eq.~\eqref{case2indsols} in Appendix~\ref{App:hyperG} with $b-a \equiv l=0$.}
\begin{equation}
\def\arraystretch{.8}
u_1(x) = \hypergeom{2}{1}\left[\begin{array}{c}
\hat L+1,~~\hat L+1\\
2\hat L+2
\end{array}\Big\rvert \,x\,\right]~~~{\rm and}~~~
u_4(x)=(-x)^{-\hat L-1}\hypergeom{2}{1}\left[\begin{array}{c}
-\hat L,~~\hat L+1\\
1
\end{array}\Big\rvert \,\frac{1}{x}\,\right]\, .
\label{solsscalrdeg3}
\end{equation}
The solution $u_1(x)$ is logarithmically divergent at the horizon ($x=1$) while $u_4(x)$ is finite there. Thus, regularity of the scalar profile in the vicinity of the horizon forces to discard the first solution in~\eqref{solsscalrdeg3} and retain only the second one. For this special choice of parameters, the hypergeometric function $u_4$ is highly degenerate---in fact, it is just a polynomial:
\be
\def\arraystretch{.8}
u_4(x) = (-x)^{-\hat L-1 }\hypergeom{2}{1}\left[\begin{array}{c}
-\hat L,~~\hat L+1\\
1
\end{array}\Big\rvert \,\frac{1}{x}\,\right]
= (-x)^{-\hat L-1 }\sum_{n=0}^{\hat L} \frac{(-\hat L)_n(\hat L+1)_n}{ (n!)^2}x^{-n} \, ,
\label{solsscalrdeg32}
\ee
where we have used the identity~\eqref{case3appreg}.
Note that \eqref{solsscalrdeg32} contains only positive powers of $r$, which can be made more apparent by transforming back to the $r$ variable, and writing the solution for the original field $\phi(r)$:
\be
\phi(r) = (-1)^{-(\hat L+1)}r_s^\frac{D-2}{2}\sum_{n=0}^{\hat L} \frac{(-\hat L)_n(\hat L+1)_n}{ (n!)^2} \left(\frac{r}{r_s} \right)^{n(D-3)} \, .
\ee
As a result, this solution is pure growing mode (or tidal field) at infinity, with the highest power scaling as $r^{L}$, as expected, and subleading powers all the way down to $r^0$, where the series terminates.  In particular this implies that there is no subleading induced fall-off. This means that the Love numbers vanish. 
\begin{tcolorbox}[colframe=white,arc=0pt,colback=greyish2]
\vspace{-7pt}
\begin{equation}
k=0
\qquad\qquad\quad
\text{(integer $\hat{L}$)} \, .
\label{lovescalarzero}
\end{equation}
\end{tcolorbox}
This result can equivalently be obtained by starting from the generic case~\eqref{scalarLoven} and then taking the limit of integer $\hat L$~\cite{Kol:2011vg}.
\end{itemize}

\paragraph{Summary:} It is worth summarizing briefly the results in the scalar case. We have considered three possible cases: $\hat L$ generic, half-integer $\hat L$, and integer $\hat L$. The resulting susceptibilities in each of these cases are given by
\be
k_{\rm scalar} = 
\begin{cases}
~ \frac{2\hat L+1}{2\pi }
 \frac{\Gamma(\hat L+1)^4}{ \Gamma(2\hat L+2)^2  } \tan (\pi \hat L ) &\qquad\text{for generic $\hat L$}\,,\\
~ \frac{(-1)^{2\hat{L}} (D-3)\Gamma(\hat{L}+1)^2}{(2\hat{L})!(2\hat{L}+1)! \Gamma(-\hat{L})^2}\log \left( \frac{r_0}{r} \right) & \qquad\text{for half-integer $\hat L$}\,,\\
~0&\qquad\text{for integer $\hat L$}\,.
\end{cases}
\label{eq:scalarfieldLove}
\ee
In the most physical case, $D=4$, for all $L$ we have that $\hat L = L/(D-3)$ is an integer, so the scalar field Love numbers vanish.

\subsection{Spin-1: Electric/magnetic susceptibilities}
\label{sec:EMresponse}
We next consider the case where a black hole is immersed in a long-wavelength electric or magnetic field and compute its induced polarization or magnetization.
In four dimensions, these response coefficients have been computed in~\cite{Damour:2009va,dos2016relativistic}, here we generalize the results to arbitrary dimension. Similar to the scalar case, we find that the polarizability/susceptibility is nonzero in general dimensions, but happens to vanish in $D=4$, indicating that four-dimensional Schwarzschild black holes are not polarizable, while their higher-dimensional counterparts are.

\subsubsection{Electric polarizability}

In this section, we consider how a black hole responds to an externally applied massless spin-1 field. To study this case, we make use of the decomposition of the spin-1 equations performed in Section~\ref{sec:masslesspin1action}.
In contrast to the scalar field case, there are now two different types of external field that we can apply to the black hole, corresponding to the scalar and vector sectors discussed in section~\ref{sec:masslesspin1action}. 
We begin by considering the scalar (parity-even) sector.
The equation of motion~\eqref{eq:masslessEmodespin10}
in the zero frequency limit, reads 
\be
\frac{\rd^2\Psi_S}{\rd r_\star^2}- f (r)\left( \frac{L(L+D-3)}{r^2}+\frac{(D-4)[(D-2)f-2rf']}{4r^2}\right)\Psi_S = 0 \, .
\label{eq:masslessEmodespin1}
\ee
In the static limit, the electric field is built from $\Psi_S$, so we will refer to the static response as a polarizability of the black hole.

In order to recast \eqref{eq:masslessEmodespin1} in the standard form of a hypergeometric equation, we again define the radial variable  $x$ as in \eqref{x} and we make the field redefinition
\begin{equation}
u_S (x) =  x^{-\frac{D+2 L-4}{2 (D-3)}}  \Psi_S(r(x)) \, .
\end{equation}
After these manipulations, eq.~\eqref{eq:masslessEmodespin1} becomes
\begin{equation}
x(1-x)u_S''(x) + \big[c - (a+b+1)x\big] u_S'(x) - a \,  b \,  u_S(x) =0 \, ,
\label{hyperEq}
\end{equation}
where the parameters are given in terms of $\hat L = L/(D-3)$ as
\begin{equation}
a = \hat L  \, ,
\qquad\qquad
b= \hat L +2 \, ,
\qquad\qquad
c= 2\hat L+2 \, ,
\end{equation}
 which again satisfy the relation $a+b-c=0$.
 
 Much as in the scalar case considered previously, 
 the form of the two linearly independent solutions of~\eqref{hyperEq} depends on the value of $\hat L$ takes. In particular, there are three distinct possibilities: $(1)$  $\hat L$  is neither integer nor half-integer, $(2)$  $\hat L$ is  half-integer, and $(3)$  $\hat L$ is an integer. These correspond to a subset of all possible parameter values for $a,b,c$, discussed more generally in Appendix~\ref{App:hyperG}. 

We now proceed to enumerate the various cases and compute the relevant polarizabilities. Since the systematics of the calculations closely follow the scalar case of the previous section, we will not belabor the details. Instead, we will report the main equations and the final result.
 
\begin{itemize}

\item
{\bf $\hat L$ is  neither integer nor half-integer:} In this case, none of $a$, $b$, $c-a$, $c-b$ and $c$   is an integer. Thus, two linearly independent solutions for  $u_S(x)$ are given by
\be
\def\arraystretch{.8}
u_1(x) = \hypergeom{2}{1}\left[\begin{array}{c}
\hat L,~~\hat L+2\\
2\hat L+2
\end{array}\Big\rvert \,x\,\right]~~~~{\rm and}~~~~
u_5(x) = x^{-2\hat L-1 }\hypergeom{2}{1}\left[\begin{array}{c}
-1-\hat L,~~1-\hat L\\
-2\hat L
\end{array}\Big\rvert \,x\,\right]\,.
\ee
As for the scalar case, there is only one specific linear combination of these solutions that is regular at the horizon, $x=1$, which is given by eq.~\eqref{hypergcase1-regular} in  Appendix~\ref{App:hyperG}. We can then expand this solution at large radial distances (near $x=0$) and extract the polarizability from the ratio of the sub-leading to leading fall-off, in units of $r_s^{2L+D-3}$ they are given by
\begin{tcolorbox}[colframe=white,arc=0pt,colback=greyish2]
\vspace{-7pt}
\be
k_S  = \frac{\Gamma (-1-2\hat L)}{\Gamma (-\hat L-1) \Gamma (1-\hat L)}
\frac{\Gamma (\hat L) \Gamma (\hat L+2)}{\Gamma (1+2\hat L)}= -2^{-4\hat L-2}\frac{\hat L +1}{\hat L} \frac{ \Gamma (\hat L) \Gamma (\hat L+2)}{  \Gamma (\hat L+\tfrac{3}{2}) \Gamma (\hat L+\frac{1}{2})} \tan{(\pi \hat L)}\, ,
\label{E-Love}
\ee
\end{tcolorbox}
where in the last equation we have 
 used the Legendre duplication formula~\eqref{Legendredup} to simplify the expression, since $\hat L$ is not an integer or half-integer.

The quantities in eq.~\eqref{E-Love} represent the electric polarizability of a black hole (equivalently, its static susceptibility under spin-$1$ perturbations of the scalar type) in the general case.

\item
{\bf $\hat L$ is  half-integer:} 
If $\hat L$ is  half-integer,   the  numbers $a$, $b$, $c-a$ and $c-b$ are not integer-values, but $c$ takes positive integer values. In this degenerate case, the two linearly independent solutions to the hypergeometric equation~\eqref{hyperEq} are given by~\eqref{halfinteger1}. The solution that is regular at the black hole horizon is $u_2$, which can be written as~\eqref{half-integer2}.
Just as we found in the scalar case, the asymptotic expansion of this solution at large $r$ is not a pure polynomial in $r$.
In particular, the decaying component contains a logarithmic component, again leading to a classical running. The polarizability in this case
may be defined as (in units of $r_s^{2L+D-3}$)
\begin{tcolorbox}[colframe=white,arc=0pt,colback=greyish2]
\vspace{-4pt}
\begin{equation}
k_S =
 \frac{(-1)^{2\hat{L}}(D-3) \Gamma(\hat{L}+2)\Gamma(\hat{L}) }{(2\hat{L}+1)!(2\hat{L})!\Gamma(1-\hat{L})\Gamma(-\hat{L}-1)}\log \left(\frac{r_0}{r} \right)
\qquad\qquad
\text{(half-integer $\hat{L}$)} .
\end{equation}
\end{tcolorbox}
As before $r$ represents the distance at which the response is measured, and $r_0 \sim r_s$ is the renormalization scale.

\item
{\bf $\hat L$ is an integer:} Finally we consider the case where $\hat L$ is an integer. In this case, the solution that is regular at the horizon is given by\footnote{See also eq.~\eqref{case2indsols} in Appendix~\ref{App:hyperG} with $b-a \equiv l=2$.  }
\be
\def\arraystretch{.8}
u_4(z)=(-x)^{-\hat L-2}\hypergeom{2}{1}\left[\begin{array}{c}
1-\hat L,~~\hat L+2\\
3
\end{array}\Big\rvert \,\frac{1}{x}\,\right]\, .
\ee  
Since the first parameter is a non-positive integer and the bottom parameter is positive, we can use the  formula~\eqref{ana1} to rewrite this solution as a polynomial which contains only positive powers of $r$, much as in the scalar case.
As a result, the polarizabilities vanish,
\begin{tcolorbox}[colframe=white,arc=0pt,colback=greyish2]
\vspace{-7pt}
\be
k_S = 0
\qquad\qquad\quad
\text{(integer $\hat{L}$)} \, .
\ee
\end{tcolorbox}
This includes the $D=4$ case of greatest physical interest, so four-dimensional Schwarzschild black holes are not polarizable.
\end{itemize}

\subsubsection{Magnetic susceptibility}
\label{sec:odd-spin-1}

Next, we consider the response of a black hole to a magnetic field. In the static limit, this is equivalent to asking for the response to a  parity-odd massless spin-1 perturbation. These perturbations are governed by equation~\eqref{eq:masslessBmodespin10}
\begin{equation}
\frac{\D^2\Psi_V}{\D r_\star^2}-f\left(  \frac{(L+1)(L+D-4)}{r^2}+\frac{(D-4)[(D-6)f+2rf']}{4r^2}\right)\Psi_V = 0  \, ,
\label{eq:masslessBmodespin1}
\end{equation}
where we have taken the $\omega = 0$ (static) limit.
It is again possible to recast this equation in hypergeometric form. Relabeling the radial coordinate as $x = (r_s/r )^{D-3} $ and redefining the field through
\begin{equation}
  u_V (x) =  x^{-\frac{D+2 L-4}{2 (D-3)}}  \Psi_V(r(x)) \, ,
\end{equation}
we can put eq.~\eqref{eq:masslessBmodespin1} into the standard form
\begin{equation}
x(1-x) u_V''(x) + \big[c - (a+b+1)x\big] u_V'(x) - a \,  b \,  u_V(x) =0 \, ,
\label{hyperEqB}
\end{equation}
where now the parameters are given by 
\begin{equation}
a =   \hat L +1 - \frac{1}{D-3} \, ,
\qquad\qquad
b=\hat L +1 + \frac{1}{D-3}  \, ,
\qquad\qquad
c= 2\hat L+2 \, ,
\label{oddpars}
\end{equation}
where we have again defined $\hat L = L/(D-3)$.
Notice that these parameters~\eqref{oddpars} again satisfy the relation $a+b = c$ and they are all positive, $a,b,c>0$. However, they are no longer functions of $\hat L$ alone, so it is more convenient to distinguish among different possibilities based on the values of $a, b,$ and $c$:

\begin{itemize}

\item
{\bf $a$, $b$ and $c$ are non-integer:} We first consider the case where none of the parameters 
 $a$, $b$, $c-a$, $c-b$ and $c$   is an integer. In this case, a basis of  independent solutions is given by eq.~\eqref{hypergcase1}.
The linear combination that is regular at the horizon ($x=1$) can be read off from eq.~\eqref{hypergcase1-regular} with the parameters $a$, $b$ and $c$ given by~\eqref{oddpars}.  After expanding for large $r$, and taking the ratio of the subleading (induced) fall-off to the leading (applied) fall-off, one finds the following expression for the magnetic susceptibility\footnote{See also eq.~\eqref{generallove}.}  (in units of $r_s^{2L+D-3}$)

\begin{tcolorbox}[colframe=white,arc=0pt,colback=greyish2]
%\vspace{-7pt}
\be
k_V =(2\hat{L}+1)
 \frac{\Gamma(\hat L +1 - \frac{1}{D-3})^2\Gamma(\hat L +1 + \frac{1}{D-3})^2}{ \Gamma(2\hat{L}+2)^2  } \frac{\sin[\pi (\hat L  - \frac{1}{D-3} )] \sin[\pi(\hat L  + \frac{1}{D-3})]}{\pi \sin (2\pi\hat{L})}\,,
 \label{kVspin1m}
\ee
\end{tcolorbox}
which is valid in the generic case.

\item
{\bf $a$ and $b$  are non-integer, while $c$ is an integer:}  This case occurs  whenever $\hat L$ is integer in $D\geq5$ or when $\hat L$ is half-integer in $D>5$. The solution that is regular at the black hole horizon is given by $u_2$ in~\eqref{halfinteger1}, and its large-$r$ expansion is given in~\eqref{half-integer2}. From this we can extract the magnetic susceptibility in the usual way, again in units of $r_s^{2L+D-3}$
\begin{tcolorbox}[colframe=white,arc=0pt,colback=greyish2]
\vspace{-9pt}
\be
k_V= \frac{ (-1)^{2\hat{L}} (D-3)\Gamma(\hat{L}+1-\tfrac{1}{D-3})\Gamma(\hat{L}+1+\tfrac{1}{D-3}) }{(2\hat L+1)!(2\hat L)!\Gamma(-\hat{L}-\tfrac{1}{D-3})\Gamma(-\hat{L}+\tfrac{1}{D-3})}\log \left(\frac{r_0}{r}\right)
\qquad\quad\text{(integer $c$)}
 \, .
\ee
\end{tcolorbox}
Note that this case has a logarithmic running of the susceptibility, familiar from previous examples.

\item
{\bf $a$, $b$ and $c$ are integers:} We next consider the situation where all of the parameters are integers. This occurs for all $L$ in $D=4$ and whenever $\hat L$ is half-integer in $D=5$. As shown in Appendix~\ref{app:abcints}, the solution that is regular at the horizon is a pure polynomial in positive powers of $r$. In particular, there is no fall-off scaling as $r^{-L-D+3}$ at large distances and therefore the  magnetic susceptibility
\begin{tcolorbox}[colframe=white,arc=0pt,colback=greyish2]
\vspace{-7pt}
\begin{equation}
k_V=0 \,\qquad\qquad\quad \text{(integer $a,b,c$)} .
\end{equation}
\end{tcolorbox}
As stated before, this implies that four-dimensional black holes are not magnetizable.

\item
{\bf $a$ or  $b$ is an integer, while $c$ is non-integer:} 
This case  occurs in $D>5$ whenever $L=n(D-3)\pm1$, where $n$ is a positive integer. In this case again---as is shown explicitly in Appendix~\ref{app:aorbintcnonint}---imposing regularity of the solution at the horizon implies that the solution is a polynomial (see~\eqref{case4ex}). This implies that the magnetizability is also zero in this case.
\begin{tcolorbox}[colframe=white,arc=0pt,colback=greyish2]
\vspace{-7pt}
\begin{equation}
k_V=0 \,\qquad\qquad\quad \text{(integer $a$ or $b$)} .
\end{equation}
\end{tcolorbox}
Note that this case is one of the main differences  between the electric and magnetic sectors. Values of $\hat L= n\pm \tfrac{1}{D-3}$ in $D>5$ that are neither integer nor half-integer have non-vanishing electric polarizabilities---see eq.~\eqref{E-Love}---but have vanishing magnetic susceptibilities.

\end{itemize}

\paragraph{Summary:} We again briefly summarize the final results for spin-1 response. For the scalar (electric) sector we found
\be
k_S = 
\begin{cases}
~  -2^{-4\hat L-2}\frac{\hat L +1}{\hat L} \frac{ \Gamma (\hat L) \Gamma (\hat L+2)}{  \Gamma (\hat L+\tfrac{3}{2}) \Gamma (\hat L+\frac{1}{2})} \tan{(\pi \hat L)}&\qquad\text{for generic $\hat L$}\,,\\
~ \frac{(-1)^{2\hat{L}}(D-3) \Gamma(\hat{L}+2)\Gamma(\hat{L}) }{(2\hat{L}+1)!(2\hat{L})!\Gamma(1-\hat{L})\Gamma(-\hat{L}-1)}\log \left(\frac{r_0}{r} \right) & \qquad\text{for half-integer $\hat L$}\,,\\
~0&\qquad\text{for integer $\hat L$}\,.
\end{cases}
\label{eq:spin1Ssummary}
\ee
while for the vector (magnetic) sector we found:
\be
k_V = 
\begin{cases}
\vspace{.15cm}
~ (2\hat{L}+1)
 \frac{\Gamma(\hat L +1 - \frac{1}{D-3})^2\Gamma(\hat L  +1+ \frac{1}{D-3})^2}{ \Gamma(2\hat{L}+2)^2  } \frac{\sin[\pi (\hat L  - \frac{1}{D-3} )] \sin[\pi (\hat L + \frac{1}{D-3})]}{\pi \sin (2\pi\hat{L})} &\qquad\text{for generic $\hat L, D$}\,,\\
~\frac{ (-1)^{2\hat{L}}(D-3) \Gamma(\hat{L}+1-\tfrac{1}{D-3})\Gamma(\hat{L}+1+\tfrac{1}{D-3}) }{(2\hat L+1)!(2\hat L)!\Gamma(-\hat{L}-\tfrac{1}{D-3})\Gamma(-\hat{L}+\tfrac{1}{D-3})}\log \left(\frac{r_0}{r}\right)
 & \qquad\text{for integer $c$}\,,\\
~0&\qquad\text{for integer $a$ or $b$}\,,
\end{cases}
\label{eq:spin1Vsummary}
\ee
where the parameters $a,b,c$ are defined in~\eqref{oddpars}.
Interestingly, we find again in this case that $D=4$ is exceptional, in that four-dimensional Schwarzschild black holes can be neither polarized nor magnetized, while higher-dimensional black holes can.

\subsection{Spin-2: Love numbers}
\label{sec:Lovenumbers}
When placed in an external gravitational field, ordinary objects mechanically deform in response. In turn, this induces 
multipole moments in the gravitational field at infinity. The coefficients of these induced multipoles are the so-called Love numbers, which encode the deformability of objects due to an external tidal field.
In this section, we compute the Love numbers of all types for Schwarzschild black holes in all dimensions. These numbers are in a sense 
a measure of the rigidity of black holes. Much like the other linear response coefficients, they vanish for four-dimensional black holes, indicating infinite rigidity, suitably understood. 

In four dimensions, Schwarzschild black hole Love numbers were computed by~\cite{Binnington:2009bb,Damour:2009vw}, while in higher dimensions they were computed for scalar-type and tensor-type perturbations in~\cite{Kol:2011vg,Cardoso:2019vof}. Here we provide a unified computation of all cases, including the vector-type perturbations in general dimension.

\subsubsection{Spin-2 tensor perturbations}
We begin by discussing the tensor-type perturbations, for which the relevant action was derived in Sec.~\ref{sec:tensorspin2}. As mentioned previously, these modes only exist in $D>4$, and consequently the same is true for the corresponding Love numbers.
The zero-frequency limit of the equation of motion for tensor-type perturbations is obtained easily from~\eqref{eq:tensorequation2}
\begin{equation}
f\Psi_T'' +f' \Psi_T' - \left( \frac{L(L+D-3)+2(D-3)}{r^2}+f'\frac{D-6}{2r}+ f\frac{D(D-14)+32}{4r^2}  \right) \Psi_T = 0 \, .
\label{eqT}
\end{equation}
As was already noticed in \cite{Kol:2011vg}, this equation is equivalent to the scalar equation~\eqref{spin0} for $\Lambda = 0$. This is somewhat non-obvious, but substituting the explicit form of $f(r)$ in eq.~\eqref{eq:sadsf} with $\Lambda = 0$ into each equation, they turn out to be identical.
 Therefore, the solutions and Love numbers are exactly the same as in the scalar case, and are summarized in~\eqref{eq:scalarfieldLove}.

\subsubsection{Spin-2 vector-type perturbations (odd sector)}
We next consider the vector-type perturbations of a massless spin-2 field. In four dimensions this coincides with the odd (Regge--Wheeler) sector. To our knowledge the Love numbers corresponding to these modes have not previously been computed in general dimension. The details of the computation are somewhat similar to the scalar and spin-1 cases considered previously. 

The action for the vector-type modes was derived in Sec.~\ref{sec:vectorspin2}.
The zero-frequency limit of the $D$-dimensional Regge--Wheeler equation can be read off from eq.~\eqref{eq:ddimRWeq}:
\be
\frac{\rd^2\Psi_{\rm RW}}{\rd r_\star^2}  - f\left( \frac{(L+1)(D-4+L)}{r^2}+f\frac{(D-4)(D-6)}{4r^2}-f'\frac{(D+2)}{2r}\right)\Psi_{\rm RW} = 0.
\ee
Performing the coordinate change \eqref{x} and the field redefinition 
\begin{equation}
u_{\rm RW} (x) \equiv  x^{-\frac{L+D-3}{D-3}} \left(\frac{2r^{D-2}}{(L-1)(D-2+L)}\right)^{-1/2}  \Psi_{\rm RW} (r(x)) \, ,
\label{odduPsi}
\end{equation}
puts this equation
into the standard hypergeometric form
\be
x(1-x) u_{\rm RW}''(x) + \big[c - (a+b+1)x\big] u_{\rm RW}'(x) - a \,  b \,  u_{\rm RW}(x) =0,
\ee
where the parameters are given in terms of $L$ and $D$ by 
\be
a=\hat L - \frac{1}{D-3}\, ,
\qquad
b= \hat L +2 + \frac{1}{D-3}  \, ,
\qquad
c= 2\hat L +2 \, .
\label{abcoddspin2}
\ee
As before, these parameters satisfy $a+b=c $ and $a,b,c>0$. Note the striking resemblance to the parameters in the magnetic spin-1 sector~\eqref{oddpars}---they differ only by a shift of $a$ and $b$ by $1$. As a result, the structure of solutions is quite similar, and the presence of degeneracies in solutions will again depend on $D$ and $L$.

\begin{itemize}

\item
{\bf $a$, $b$ and $c$ are non-integer:} In this case, the solution that is regular at the black hole horizon is given by~\eqref{hypergcase1-regular}, with the parameters $a$, $b$ and $c$  in \eqref{abcoddspin2}. Expanding this solution around $r\to\infty$ and using
the expression \eqref{generallove} for the (dimensionless) Love numbers, we find
%\vspace{-7pt}
\begin{tcolorbox}[colframe=white,arc=0pt,colback=greyish2]
\vspace{-7pt}
\be
k_{\rm RW} = (2\hat L +1)
 \frac{\Gamma(\hat L +2 + \frac{1}{D-3})^2\Gamma(\hat L - \frac{1}{D-3})^2}{ \Gamma(2\hat L +2)^2  } \frac{\sin[\pi (\hat L + \frac{1}{D-3}) ] \sin[\pi (\hat L - \frac{1}{D-3}) ]}{\pi \sin (2\pi \hat L)} \, ,
\label{kRWodd2}
\ee
\end{tcolorbox}
in units of $r_s^{2L+D-3}$. This represents the vector-type Love number for generic $L,D$.
Note that, in $D=4$, odd Love numbers for black holes are usually defined at the level of the field component $h_0$. In these terms, using \eqref{h0odd}, \eqref{Qvarle}, \eqref{odduPsi} and \eqref{kRWodd2}, one finds that the tidal response associated with $h_0$ in general dimension is
\begin{equation}
k_0 = \frac{1-L}{L+D-2} k_{\rm RW} \, ,
\end{equation}
with $k_0$ defined by $h_0 = A_0 r^{L+1} (1+ k_0 r^{-2L-D+3})$, where $A_0 $ is some irrelevant overall factor.

\item
{\bf $a$ and $b$  are not integers, while $c$ is an integer:}
 In analogy with the spin-$1$ case,   this happens  whenever $\hat L$ is integer in $D\geq5$ or when $\hat L$ is half-integer in $D>5$. The solution that is regular at the horizon is $u_2$ in~\eqref{halfinteger1}. Its expansion at large radius again follows from~\eqref{half-integer2}, which we use to compute the Love number:
\begin{tcolorbox}[colframe=white,arc=0pt,colback=greyish2]
\vspace{-9pt}
\be
k_{\rm RW}=
\frac{ (-1)^{2\hat{L}}(D-3) \Gamma(\hat{L}-\tfrac{1}{D-3})\Gamma(\hat{L}+2+\tfrac{1}{D-3}) }{(2\hat L+1)!(2\hat L)!\Gamma(-\hat{L}-1-\tfrac{1}{D-3})\Gamma(-\hat{L}+1+\tfrac{1}{D-3})}\log\left(\frac{r_0}{r}\right)
\qquad\text{(integer $c$)}\, ,
\ee
\end{tcolorbox}
which is measured in units of $r_s^{2L+D-3}$.
Note that this is another case that displays a classical running of the response.

\item
{\bf $a$, $b$ and $c$ are integers:} The parameters of the hypergeometric function happen to all be integers in $D=4$ (for all $L$) and   whenever $\hat L$ is half-integer in $D=5$.
This corresponds to a degenerate case where the most general solution regular at the horizon is (see Appendix~\ref{app:abcints}) 
\be
u_{\rm RW}(r) = (-1)^{\hat L +2 + \frac{1}{D-3}}\left(\frac{r}{r_s} \right)^{ L +1+2(D-3 ) }\sum_{n=0}^{\hat L - \frac{1}{D-3}-1} \frac{(1 + \frac{1}{D-3}-\hat L)_n( 2 + \frac{1}{D-3}+\hat L )_n}{(3+\tfrac{2}{D-3})_n n!}\left(\frac{r}{r_s} \right)^{n(D-3)}.
\ee
Since this solution only has positive powers of $r$ appearing, the Love numbers vanish:
\begin{tcolorbox}[colframe=white,arc=0pt,colback=greyish2]
\vspace{-7pt}
\begin{equation}
k_{\rm RW}=0 \,\qquad\qquad\quad \text{(integer $a,b,c$)} .
\end{equation}
\end{tcolorbox}
In particular this implies that the odd Love numbers vanish in $D=4$.

\item
{\bf  $a$ or  $b$ is integer, $c$ is not an integer:} 
The last  case to be considered is when $L=n(D-3)\pm1$ with $n$ a positive integer in $D>5$. 
Similar to the spin-$1$ case discussed above---and as is shown more explicitly in App.~\ref{app:aorbintcnonint}---the only solutions to the hypergeometric equation in this case that are  regular at the horizon are again polynomials with only positive powers in $r$. This results in  vanishing  Love numbers:
 \begin{tcolorbox}[colframe=white,arc=0pt,colback=greyish2]
\vspace{-7pt}
\begin{equation}
k_{\rm RW}=0 \,\qquad\qquad\quad \text{(integer $a$ or $b$)} .
\end{equation}
\end{tcolorbox}

\end{itemize}

\subsubsection{Spin-2 scalar-type perturbations (even sector)}
Finally, we consider the case of scalar-type spin-2 perturbations. In $D=4$, this sector coincides with the parity even perturbations. We saw that the vector and tensor-type spin-2 perturbations behave substantially similar to the scalar and spin-1 cases considered previously. In particular, the relevant equations are of hypergeometric type so that the analysis has many common features. The scalar-type spin-2 perturbations, on the other hand, are described by a Heun equation. Heun equations are analogous to hypergeometric equations, but with four regular singular points, as opposed to three. Consequently, they are less well-understood, and harder to solve. Fortunately, in the case of interest the Heun equation can be transformed into a hypergeometric equation supplemented by a first order ordinary differential equation, as we will describe.

The calculation of  these scalar-type Love numbers for spin-$2$ fields in arbitrary dimensions was originally done in \cite{Kol:2011vg}, and has been repeated recently numerically in~\cite{Cardoso:2019vof}. One of our goals in this section is to resolve an apparent discrepancy between these two computations. For simplicity, we will focus on the case where the parameters are generic in order to avoid enumerating all the specific cases separately. If one is interested in one of the degenerate cases, they can be obtained from the generic answer by suitably taking the limit (as in~\cite{Kol:2011vg}), or can be obtained in a parallel manner to that described above and in Appendix~\ref{App:hyperG}.

Scalar-type spin-2 perturbations have the most complicated equation of motion of the cases considered here. The canonically normalized Zerilli variable is described by the action~\eqref{eq:Zaction}, but it is actually more convenient to first work with the non-canonically normalized variable ${\cal V}$, which has the action~\eqref{actioncalV}.\footnote{Note  that  our field ${\cal V}$ is related to the quantity $\Phi$ introduced in \cite{Kodama:2003jz} through the following relation,
\begin{equation}
{\cal V}
= -\frac{r^{2-\frac{D}{2}}}{\left(\frac{1}{r}\right)^{D-3}-1}  \left[2 (L-1) (D+L-2)+(D-2) (D-1) \left(\frac{1}{r}\right)^{D-3}\right]
\Phi(r) \, ,
\notag
\end{equation}
where we have set $r_s=1$ for simplicity.} As before we change radial coordinates to the $x$ variable $x\equiv (r_s/r)^{D-3}$ and then make the change of variables:
\be
 w(x) \equiv  x^{-j}   \left( L (D+L-3)-D+2+\frac{1}{2 }(D-2) (D-1)x\right)^{-l} {\cal V}(r(x)) \, .
 \label{frheun}
\ee
After this, the equation of motion for ${\cal V}$ in the zero-frequency limit takes the 
takes the standard form of  a Heun equation \cite{1971ApJ...166..197F,MR1392976,slavjanov2000special}:
\begin{equation}
w''(x) + \left( \frac{ \gamma}{x} + \frac{\delta}{x-1} + \frac{\epsilon}{x- a} \right) w '(x)
+ \frac{ \alpha \,\beta \,x -  q }{x(x-1)(x- a)} w(x) =0 \, ,
\label{heunxw}
\end{equation}
where the various parameters $\gamma$, $\delta$, $\epsilon$, $a$, $\alpha$, $\beta$, and $q$ depend on $j$ and $l$. 
Explicit expressions  for different choices  of $j$ and $l$, are listed in Table~\ref{table:heun}.
\bgroup
\def\arraystretch{1.75}% 
\begin{table}[tb]
\centering
{\small
\begin{tabular}{|c c || c c c c c|}
\hline
$j$   & $l$  &  $\gamma$   &  $\epsilon$  &  $\alpha$ &  $\beta$  &  $q$
\\
\hline \hline
 $-\hat L-\frac{D-2}{2 (D-3)}$  &  $-1$  & $-2\hat L$   & $-2$   & $-(\hat L+1) $  & $-(\hat L+1) $ &  $-\frac{2 (\hat{L}+1) \left[(-2 D^2+9 D-10) \hat{L}+(D-3)^2 \hat{L}^3-(D-2)^2\right]}{(D-1)(D-2)}$ 
  \\
 $-\hat L-\frac{D-2}{2 (D-3)}$    & 2  & $-2\hat L$    & $4$  & $-\hat L +2$   & $-\hat L +2$  &
 $ \frac{2\hat{L} \left[-\hat{L} \left((D-3) \hat{L}-D+2\right) \left((D-3) \hat{L}+2 D-5\right)-4 D+8\right]+2(D-2)^2}{(D-2) (D-1)} $
     \\
$\hat L+\frac{D-4}{2 (D-3)}$   & $-1$  & $2(\hat L+1)$  & $-2$  & $\hat L$  & $\hat L$   & 
$-\frac{2 \hat{L} \left[(D-3) \hat{L}-1\right] \left[\hat{L} \left((D-3) \hat{L}+3 D-8\right)+D-3\right]}{(D-2) (D-1)}$
     \\
$\hat L+\frac{D-4}{2 (D-3)}$     & $2$  & $2(\hat L+1)$  & $4$  & $\hat L+3$  & $\hat L+3$   & 
$6+\frac{2 \hat{L} \left[\hat{L} \left(-\hat{L}(D-3)^2 (\hat{L}+3)-(D-9) D-17\right)+D (3 D-8)+3\right]}{(D-2) (D-1)}$
    \\ \hline
\end{tabular}
}
\caption[Table caption text]{\small Values of the parameters entering the Heun equation~\eqref{heunxw} for different choices of $j$ and $l$  in the field redefinition \eqref{frheun}. Notice that  for all cases. The parameters satisfy the condition $\gamma+\delta+\epsilon=\alpha+\beta+1$. }
\label{table:heun}
\end{table}
\egroup

The Heun equation \eqref{heunxw} is a second order differential equation with four regular singular points, located at $0$, $1$, $a$ and $\infty$ \cite{MR1392976,slavjanov2000special}. It is solved by the so-called Heun function, which is defined analogously to the hypergeometric function as the series solution to the differential equation. Compared to hypergeometric functions, solutions to the Heun equation are considerably less studied. In particular, the connection formulas that relate solutions with given asymptotics near one of the singular points to solutions with asymptotics specified near a different singular point are not known in full generality.
This is the main obstruction to the analytic computation of Love numbers using the Heun function directly---in particular the large-distance extrapolation of the solution that is regular at the horizon is not straightforward to extract.

However, there are special cases where---after a field redefinition involving the field and its first derivative---the Heun equation can be recast in the form of a hypergeometric equation. 
In these situations, the Heun function admits a series representation in terms of a finite sum of hypergeometric functions~\cite{MR1392976,slavjanov2000special}. The most familiar example is provided by the Zerilli equation for parity-even spin-$2$ perturbations in $D=4$. The equation is of the Heun type, but in $D=4$ the Chandrasekhar relation~\cite{Chandrasekhar:1985kt,1975RSPSA.343..289C} provides a field-redefinition that transforms the Zerilli equation into the Regge--Wheeler equation, which is indeed of the hypergeometric-type in the zero-frequency limit~\cite{1971ApJ...166..197F}. The Chandrasekhar mapping between the Zerilli and Regge--Wheeler equations does not exist for $D\neq 4$, 
but even in the absence of an even-odd duality, it is still possible to find a field redefinition of a similar form that maps the Heun equation of interest to a hypergeometric equation in the zero frequency limit~\cite{Kodama:2003jz,Ishibashi:2003ap}. In the following, we review how this works.

As was noticed in~\cite{Kodama:2003jz,Ishibashi:2003ap}, in the zero-frequency limit~\eqref{heunxw} can be reduced to a hypergeometric equation.  Indeed---setting $r_s=1$ for the moment---the equation
\begin{equation}
f\partial_r (f\partial_r  Y ) -\frac{4 L(L+n-1) (1-r^{1-n})-(2 n-1) r^{2-2n}+2 n r^{1-n}+(n-2) n}{4r^2} Y =0 \, ,
\label{eqYtilde}
\end{equation}
where we have defined $n\equiv D-2$, is equivalent to the equation~\eqref{actioncalV}, provided that $Y$ satisfies
\be
 Y = \frac{f^{1/2}  }{r  f'}  \left(\frac{r Q(r) v'(r)}{4 \left(m+\frac{1}{2} (n+1) n x(r)\right)}-\frac{P(r) v(r)}{16 \left(m+\frac{1}{2} (n+1) n x(r)\right)^2}\right) \, ,
\label{YcalV}
\ee
where we have defined the parameter $m\equiv L (L+n-1)-n$, and where the variable $v$ is related to ${\cal V}$ through
\be
v(r) = \frac{\left(1-r^{1-n}\right) r^{\frac{n}{2}-1}}{2 (L-1) (L+n)+n (n+1) r^{1-n} }  {\cal V}(r) \,.
\ee
In order to condense notation we have defined the following functions in~\eqref{YcalV}
\begin{subequations}
\begin{align}
\nonumber
P(r) & \equiv 2 \left(n^2-1\right) n (4 m-n (n-2) (n+1)) x(r)^2 
\\
&~~+4 m (n-1) n (3 m+(n+1) n) x(r)+(n-1) (n+1)^2 n^3 x(r)^3,
\\
Q(r) & \equiv n (n-1) (n+1) x(r)^2-2 (n-1) (m+(n+1) n) x(r) .
\end{align}
\end{subequations}
The equation~\eqref{eqYtilde} can then be put in hypergeometric form by introducing 
 $x\equiv (r_s/r)^{D-3}$ and by making the redefinition
\be
y(x) =  x^{-\frac{L+D-2}{D-3}} \frac{r^{\frac{n}{2}-1}f'}{f^{1/2}} Y (r(x)) \, .
\ee
After all this, the equation for $y$ is a hypergeometric equation in standard form
\be
x (1-x)y''(x) + (c- (a+b+1)x)y'(x)  - a \, b \,  y(x) =0 \, ,
\label{eq:evenspin2hyper}
\ee
where the parameters appearing are fixed by $\hat L$:
\be
a =  \hat L + 2\, ,
\qquad\quad
b= \hat L +1 \, ,
\qquad\quad
c= 2\hat L +2 \, .
\ee
Note that these parameters
satisfy the relation $a+b-c=1$. Notice that we have traded the original Heun equation for a hypergeometric equation and the ODE~\eqref{YcalV}.

For generic values of $\hat L$,  the linearly independent solutions to~\eqref{eq:evenspin2hyper} are\footnote{For simplicity we will focus on the generic case. Explicit expressions in the degenerate cases can be found in~\cite{Kol:2011vg}, or they can be computed using the equations in the Appendix~\ref{App:hyperG}.}
\be
\def\arraystretch{.8}
y_1(x) = \hypergeom{2}{1}\left[\begin{array}{c}
a,~~b\\
c
\end{array}\Big\rvert \,x\,\right]~~~{\rm and}~~~y_5(x) = x^{1-c }\hypergeom{2}{1}\left[\begin{array}{c}
a-c+1,~~b-c+1\\
2-c
\end{array}\Big\rvert \,x\,\right] .
\ee
In order to find the solution that is regular at the horizon $x=1$, we first use the identity
\be
\def\arraystretch{.8}
 \hypergeom{2}{1}\left[\begin{array}{c}
a,~~b\\
a+b-1
\end{array}\Big\rvert \,x\,\right] =  (1-x)^{-1 }\hypergeom{2}{1}\left[\begin{array}{c}
a-1,~~b-1\\
a+b-1
\end{array}\Big\rvert \,x\,\right] .
\ee
and then use  the expansion~\eqref{app:id1} with $m=1$:
\be
\def\arraystretch{.8}
 \hypergeom{2}{1}\left[\begin{array}{c}
a,~~b\\
a+b-1
\end{array}\Big\rvert \,x\,\right]  \xrightarrow{x\rightarrow 1}
\frac{\Gamma(a+b-1)}{\Gamma(a)\Gamma(b)} \left[ (1-x)^{-1}
+ (a-1)(b-1)\ln(1-x) \right]
+ \text{finite terms} \, .
\ee
From this, we infer than the 
 linear combination that is finite at $x=1$ is
\be
\def\arraystretch{.8}
\begin{aligned}
y_0 (x) = A \Bigg(\frac{\Gamma(3-a-b)}{\Gamma(2-a)\Gamma(2-b)} \hypergeom{2}{1}\left[\begin{array}{c}
a,~~b\\
a+b-1
\end{array}\Big\rvert \,x\,\right] - \frac{\Gamma(a+b-1)}{\Gamma(a)\Gamma(b)}x^{2-a-b }\hypergeom{2}{1}\left[\begin{array}{c}
2-b,~~2-a\\
3-a-b
\end{array}\Big\rvert \,x\,\right] \Bigg) \, ,
\end{aligned}
\ee
up to an overall  constant amplitude, $A$. 
In the small-$x$ limit (near radial infinity) this solution has the expansion
\be
y_0 (x\rightarrow 0 ) \simeq A\left[  \frac{\Gamma(-2\hat L)}{\Gamma(-\hat L)\Gamma(1-\hat L)}
- \frac{\Gamma(2+2\hat L)}{\Gamma(2+\hat L)\Gamma(1+\hat L)} x^{-1-2\hat L } +\cdots\right] \, .
\ee
Going back through the chain of field redefinitions we have done, and using the definition of the Zerilli variable~\eqref{eq:Zvariable}, we can write the expansion of the Zerilli potential near infinity as 
\begin{equation}
\Psi_{\rm Z} (r\rightarrow\infty)\simeq A_{\rm Z}r^\frac{4-D}{2}\left( r^{L+1} + k_{\rm Z} \,r^{-L-D+4} +\cdots\right) \, ,
\label{lambdacalV}
\end{equation}
where $A_{\rm Z}$ is some  overall  constant and $k_{\rm Z}$ is the Love number for the Zerilli potential\footnote{Note that this Love number differs from that in~\cite{Kol:2011vg}, essentially because they computed the Love numbers for the variable $Y$, using the fall-offs of the solutions to~\eqref{eq:evenspin2hyper}. Defining their result as $\lambda^{\text{KS}}$, the relation between the two results is
\be
k^{\rm here}_{\rm Z}	= - \frac{(L+D-3) (L+D-2)^2 }{L (1-L)^2  }k^{\text{KS}} \, ,
\notag
\ee
so that the two Love numbers are normalized differently. In both cases they vanish in $D=4$ though.
}
 \begin{tcolorbox}[colframe=white,arc=0pt,colback=greyish2]
%\vspace{-1pt}
\be
k_{\rm Z}
	= -\frac{1}{4^{2\hat L+1}} \frac{(L+D-3) (L+D-2)^2 }{L (1-L)^2  } \frac{\Gamma(\hat L)\Gamma (\hat{L}+2) }{\Gamma (\hat{L}+\frac{1}{2}) \Gamma (\hat{L}+\frac{3}{2}) } {\tan} (\pi\hat L) \, ,
\label{eq:zerillilovenumber}
\ee
\end{tcolorbox}
\vspace{-5pt}
\noindent
which has units of $r_s^{2L+D-3}$ and where we have used the identities in Appendix~\ref{App:hyperG}.

In order to compare~\eqref{eq:zerillilovenumber} with the numerical result of~\cite{Cardoso:2019vof}, we have to account for the fact that the equation they were solving was for the variable $H_0$, which is defined in~\eqref{eq:hdecomp}. Since this field is auxiliary, relating it to $\Psi_{\rm Z}$ is slightly subtle. From~\eqref{actionevenD} we can obtain the ${\cal H}_1$ equation of motion
\begin{equation}
\frac{2(D-3)}{r}{\cal H}_1 - (D-3) H_2 + (D-3) H_0 - (D-4)H_0 - rH'_0 \approx 0 \, ,
\label{eqHoapprox}
\end{equation}
where we neglected subleading terms in the limit $\omega\rightarrow0$ and in $1/r\rightarrow0$.
Then, expressing ${\cal H}_1 $ and $H_2$ in terms of ${\cal V}$ using~\eqref{eq:Vaux} and~\eqref{eq:H2aux} expanded in the $r\to\infty$ limit
\begin{align}
{\cal H}_1&  \approx {\cal V}  + \frac{(D-2)r}{2L(L+D-3)} H_2 \, ,\\
H_2 & \approx \frac{2L(L+D-3)}{L(L+D-3)-(D-2) } \left( {\cal V}' + \frac{D-3}{r}{\cal V} \right) \, ,
\end{align}
we can simplify eq.~\eqref{eqHoapprox} into the relation
\be
\frac{2(D-3)}{r}\left[ (D-4){\cal V} +r {\cal V}'\right]  - H_0 +r H_0'\approx 0 \, .
\ee
Finally using eqs.~\eqref{eq:Zvariable} and~\eqref{lambdacalV} we can relate $H_0$ to ${\cal V}$. If we then define the Love number
$k_0$ through $H_0(r\rightarrow\infty)\approx A_0(r^L+k_0 r^{-L-D+3})$, we find 
\begin{equation}
k_0 = \frac{L(L-1)}{(L+D-3)(L+D-2)}k_{\rm Z} \, ,
\end{equation}
which agrees with the numerical computation of  \cite{Cardoso:2019vof}.

Though the numerical values of $k_0$ and $k_{\rm Z}$ differ, both capture the same underlying static response of the black hole. In fact, since there is only a single gauge-invariant propagating degree of freedom in the spin-2 scalar sector---and consequently only one type of external field we could apply---once one has computed the metric response for a particular variable, the other can be solved for via constraint equations, as we did above. In order to sidestep this apparent ambiguity, we instead compute the response directly for the Weyl tensor built out of these metric perturbations and match to a point-particle EFT. This has the added benefit of being manifestly gauge invariant. Other reasonable definitions will differ from this one at most by an overall multiplicative constant, reflecting the choice of units used to measure the (dimension-ful) response.  For example, the extraction of a gauge-invariant response was performed in a slightly different (but we expect equivalent) way in~\cite{Cardoso:2019vof}, where Love numbers were defined in terms of the Geroch--Hansen multipole moments of the perturbed spacetime, without recourse to an EFT description.

\paragraph{Summary:} Here we summarize the results for the tidal Love numbers. The tensor-type Love numbers are identical to the scalar case (spin 0), which is summarized in~\eqref{eq:scalarfieldLove}. For the vector-type (odd parity) perturbations, we found
\be
k_{\rm RW} = 
\begin{cases}
\vspace{.15cm}
~  (2\hat L +1)
 \frac{\Gamma(\hat L +2 + \frac{1}{D-3})^2\Gamma(\hat L - \frac{1}{D-3})^2}{ \Gamma(2\hat L +2)^2  } \frac{\sin[\pi (\hat L + \frac{1}{D-3}) ] \sin[\pi (\hat L - \frac{1}{D-3}) ]}{\pi \sin (2\pi \hat L)} &\qquad\text{for generic $\hat L,D$}\,,\\
~
 \frac{ (-1)^{2\hat{L}}(D-3) \Gamma(\hat{L}-\tfrac{1}{D-3})\Gamma(\hat{L}+2+\tfrac{1}{D-3})  }{(2\hat L+1)!(2\hat L)!\Gamma(-\hat{L}-1-\tfrac{1}{D-3})\Gamma(-\hat{L}+1+\tfrac{1}{D-3})}\log\left(\frac{r_0}{r}\right)
 & \qquad\text{for integer $c$}\,,\\
~0&\qquad\text{for integer $a$ or $b$}\,,
\end{cases}
\label{eq:spin2summaryRW}
\ee
where the parameters $a,b,c$ are defined in~\eqref{abcoddspin2}. In the scalar-type (even parity) sector, the generic Love number is given by
\be
k_{\rm Z}
	= -\frac{1}{4^{2\hat L+1}} \frac{(L+D-3) (L+D-2)^2 }{L (1-L)^2  } \frac{\Gamma(\hat L)\Gamma (\hat{L}+2) }{\Gamma (\hat{L}+\frac{1}{2}) \Gamma (\hat{L}+\frac{3}{2}) } {\tan} (\pi\hat L) \, ,\qquad\text{for generic $\hat L, D$}.
\label{eq:zerillilovesummary}
\ee
Specializing to the case of $D=4$, we find that Love numbers of all types vanish (this case can be obtained from~\eqref{eq:zerillilovesummary} by a suitable limit). This implies that all types of black hole response to massless external fields vanish in $D=4$, in contrast to their higher-dimensional counterparts.

\newpage
\section{Matching to point particle effective field theory}
\label{sec:EFT}

Viewed from sufficiently far away, all objects look the same: a point in the distance. If we are able to perform finer and finer measurements, we can systematically correct this point-particle approximation to account for the internal structure of the object. The most familiar example of this logic is the multipole expansion in electrostatics, where the measurement of higher multipoles tells us about the spatial distribution of charge. However, the general philosophy is widely applicable: at distances large compared to the characteristic size of an object, there is an effective description where the object is modeled as a point particle. In this way of thinking about things, corrections due to the object's finite size and its internal structure are encoded in higher-derivative operators in the effective theory.

In this section, we relate the static response coefficients computed in Section~\ref{sec:lovenums} to the coefficients appearing in this point-particle effective field theory (EFT). The motivation for this reorganization is twofold. First, there is some concern in the literature about whether or not black hole Love numbers are well-defined, given that they are computed in some particular choice of coordinates~\cite{Fang:2005qq,Gralla:2017djj}. The definition of black hole static responses as coefficients in a worldline effective action is unambiguous and manifestly gauge invariant.
A second motivation for the translation to effective field theory is that it makes more transparent the relation between various phenomena that are controlled by the same operator coefficient, which may not be so obvious in general relativity calculations.

We first briefly review the point particle effective field theory formalism that we employ, and then proceed to match the coefficients in the worldline theory of a non-spinning black hole to the static response coefficients computed previously.

\subsection{Point particle EFT basics}
We being by setting up the effective theory that we will use to describe the interaction of a black hole with external fields at long distances.\footnote{For a pedagogical introduction, see~\cite{Porto:2016pyg} and these~\href{http://webtheory.sns.it/ggilectures2018/ira/notesCLEFT.pdf}{\tt lecture notes}.} In this approach we model a black hole as a point particle and construct an effective field theory on the black hole's worldline. This type of point-particle effective field theory has wide applicability across physics, ranging from atomic physics~\cite{Burgess:2017mhz}, to superradiance~\cite{Endlich:2016jgc}, to superfluid rotons~\cite{Nicolis:2017eqo}.  In the black hole context, this approach was first developed in~\cite{Goldberger:2004jt,Goldberger:2005cd}, motivated by the modeling of gravitational waves from binary inspirals. In the Love number context, this approach was used in~\cite{Kol:2011vg} to match the black hole response to $L=2$ scalar perturbations to a worldline EFT. Here we extend the analysis to all multipoles and types of applied external field.

The idea is to model a black hole as a point particle, described by the action
\be
S_{\rm ng} = -m\int\rd\tau \sqrt{- \eta_{\mu\nu}\frac{\rd x^\mu}{\rd\tau}\frac{\rd x^\nu}{\rd\tau}},
\label{eq:ngaction}
\ee
where $\tau$ is a coordinate that parameterizes the particles's worldline and $x^\mu(\tau)$ denotes the spacetime position of the particle as a function of this parameter.  This worldline action is invariant under reparametrizations $\tau\mapsto \tilde \tau(\tau)$. In order to couple the point particle to external fields, it is often convenient to go to Polyakov form by introducing the worldline vielbein $E$ (not to be confused with the electric field) as
\be
\rd s^2 = -E^2\, \rd\tau^2
\ee
and coupling the point particle action to worldline gravity. The action~\eqref{eq:ngaction} is equivalent to\footnote{In this formulation, the reparametrization invariance of the action acts on the fields as
\be
\delta\tau = \xi\,,~~~~~~\delta E = \partial_\tau(E\xi)\,,~~~~~~\delta x^\mu = \xi \dot x^\mu,
\ee
where $\xi$ is an infinitesimal parameter.
}
\be
S_{\rm Polyakov} = \frac{1}{2}\int\rd\tau\left(E^{-1} \eta_{\mu\nu}\frac{\rd x^\mu}{\rd\tau}\frac{\rd x^\nu}{\rd\tau} - Em^2
\right).
\label{eq:polyakov}
\ee
To see this, we can integrate out $E$ using its equation of motion,
\be
E^{-2} \dot x^\mu\dot x_\mu + m^2 = 0,
\label{eq:momentumeq}
\ee
where an overdot denotes $\rd/\rd\tau$. This equation is just $p^2+m^2 = 0$ for the point particle, and describes its free propagation. Upon substituting~\eqref{eq:momentumeq} back into the action, we recover~\eqref{eq:ngaction}.

The action~\eqref{eq:polyakov} describes a free point particle. In order to go beyond this approximation, we could introduce particle self-interactions, but for our purposes the relevant interactions will come from coupling this point particle to external fields. The types of couplings relevant for studying static response are those that are quadratic in the external fields. The interpretation is that one of the fields in the interaction serves as a background field that causes a response from the point particle, which manifests as an induced field measured at infinity. This is depicted schematically in Figure~\ref{fig:response}.

\subsection{Coupling to a scalar field}
We first consider the interaction of the point particle with an external scalar field. The effective field theory logic dictates that we should write down all possible couplings to $\phi$ on the worldline that are consistent with the symmetries of the problem. Using the particle's spacetime velocity $v^\mu \equiv \dot x^\mu$, which satisfies $v_\mu v^\mu = -1$ because it is time-like, we can construct the transverse projector
\be
P_\mu^\nu \equiv \delta_\mu^\nu + v_\mu v^\nu.
\label{eq:projector}
\ee
This projector allows us to separate derivatives into temporal and spatial parts. Because of this, we will often directly write operators with spatial indices in the rest frame of the particle, with the understanding that they can be covariantized using the operator~\eqref{eq:projector}.\footnote{We denote these spatial indices by Latin indices from the beginning of the alphabet, e.g, $a,b,c,\cdots$.}
Additionally, since we are interested in the black hole's static response, we will ignore operators with time derivatives of the scalar, and will work to leading order in the velocity of the particle.
With these considerations, the most general action (up to field redefinitions) that we can write down to second order in the bulk scalar field and at leading order in time derivatives is\footnote{The operator $\partial_{a_1}\cdots \partial_{a_L} \phi$ can be covariantized as $E^{(L)}_{\mu_1\cdots\mu_L} \equiv P_{\mu_1}^{\nu_1}\cdots P_{\mu_1}^{\nu_1}\partial_{\nu_1}\cdots\partial_{\nu_L}\phi$.}
\be
S = -\frac{1}{2}\int \rd^D x \,(\partial\phi)^2+ \int \rd \tau E \left[\frac{1}{2}E^{-2} \dot x^\mu\dot x_\mu - \frac{m^2}{2}- g\phi+\sum_{L=1}^{\infty} \frac{\lambda_L }{2L!} \left(  \partial_{(a_1}\cdots \partial_{a_L)_T} \phi \right)^2\right],
\label{eq:pointscalaraction}
\ee
where $(\cdots)_T$ denotes the symmetrized traceless component of the enclosed indices.
Notice that this action has two parts. The first term is the kinetic term for the scalar field, which propagates in spacetime. The second is the worldline action which, in addition to the degrees of freedom describing the position of the particle, is a coupling to (spatial derivatives of) the bulk scalar field. The $\lambda_L$ couplings are the worldline definitions of black hole static response coefficients. It is these coefficients that we want to fix by comparing to the full general relativity calculation.
One of the benefits of the EFT approach is that we could fix these coefficients by matching any processes between the EFT and the full theory. In this particular case, it will be most convenient to match static solutions in the two descriptions. Note that from the EFT perspective everything takes place in flat Minkowski space; the effects of dynamical gravity are included perturbatively.

In addition to the terms that we have written down in~\eqref{eq:pointscalaraction} there are possible terms with time derivatives, which will control the frequency-dependent response of the system---see~\cite{Wong:2019yoc,Kuntz:2019zef} for a discussion of some of them---along with terms involving more powers of fields, which would be relevant for the nonlinear response of the system. However, since we are concerned only with the {\it static} linear response, these are the only relevant operators.

The term $g\phi$ characterizes the charge of the point particle under the scalar field. Indeed, ignoring the other terms for a moment, it would lead to an equation of motion for the scalar field of the form:
\be
\square \phi  = g\int \rd\tau \delta^{(D)}(x-x(\tau)),
\ee
where we have introduced a delta function at the location of the particle in order to write the worldline action as a spacetime integral. We see that the $g\phi$ coupling makes the point particle behave like a charged point source from the perspective of the bulk $\phi$ field. Interestingly, in the black hole context, the no-hair theorems~\cite{Bekenstein:1971hc,Bekenstein:1995un,Hui:2012qt} indicate that this coupling is absent. 

In order to match $\lambda_L$ to the response coefficients computed in Section~\ref{Sec:spin-0}, we imagine that there is some external source (which we do not have to specify explicitly) that causes the solution to the linearized equation of motion $\square\phi =0$ to be
\be
\phi^{(0)} = c_{a_1\cdots a_L}x^{a_1}\cdots x^{a_L},
\ee
where $c_{a_1\cdots a_L}$ is a symmetric trace-free tensor.
In spherical coordinates, this is precisely a tidal field $\phi \sim r^L Y_L^M(\theta)$ at infinity. 
Now, we want to understand the profile for the field induced by the $\lambda_L$ terms in response to this boundary condition. To do so, we formally expand the field as
\be
\phi = \phi^{(0)} + \epsilon\,\phi^{(1)}+\cdots,
\label{eq:fieldexpansion}
\ee
and solve order-by-order in $\epsilon$. In particular, the linear response corresponds to the solution $\phi^{(1)}$, where the effects of $\phi^{(0)}$ cause the $\lambda_L$ terms to act as a source. Inserting the expansion~\eqref{eq:fieldexpansion} into the action leads to the equation of motion 
\be
\square \phi^{(1)} = -\lambda_L(-1)^L\int\rd\tau\,c^{a_1\cdots a_L}\partial_{a_1}\cdots \partial_{a_L}\delta^{(D)}(x-x(\tau)),
\ee
where the right-hand side plays the role of an effective source:
\be
J_{\rm eff}(x) \equiv -\lambda_L(-1)^L\int\rd\tau\,c^{a_1\cdots a_L}\partial_{a_1}\cdots \partial_{a_L}\delta^{(D)}(x-x(\tau)).
\ee
This is a linear equation, so we can solve it by convolving the source with the Green's function for $\square$
\be
\phi^{(1)}(x) = \int\rd^Dy\,G(x-y)J_\mathrm{eff}(y).
\ee
Since we are interested in the static 
response, all Green's functions coincide. It is easiest to work in Fourier space, where the convolution is replaced by multiplication. As a simplifying assumption, we take the point particle to be static at the origin, so that the Fourier transform of the source is
\begin{align}
J_{\rm eff}(\vec p) &= -\lambda_L (-1)^L\int\rd^{d}x \,e^{-i\vec p\cdot \vec x} \int\rd\tau \,c^{a_1\cdots a_L}\partial_{a_1}\cdots \partial_{a_L}\delta^{(D)}(x-x(\tau)) \nonumber\\
&= -\lambda_L\,(-i)^L \,c^{a_1\cdots a_L} p_{a_1}\cdots p_{a_L}.
\label{eq:sourcevertex}
\end{align} 
In Fourier space, the Green's function is just $G(\vec p) = -1/p^2$. so that
\be
\phi^{(1)} (\vec p) = \lambda_L\,(-i)^L \,c^{a_1\cdots a_L}\frac{p_{a_1}\cdots p_{a_L}}{p^2}.
\label{eq:p1soln}
\ee
\begin{figure}[tb]
\centering
      \includegraphics[scale=1.7]{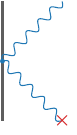}
           \caption{Diagrammatic representation of static response in the EFT. The thick grey line is the worldline of the point particle. One of the external fields (wavy lines) behaves as a background---denoted by the red $\times$---which induces a response that can be read off at infinity. The computation of the static response can be carried out diagrammatically using this Feynman diagram.} 
    \label{fig:response}
\end{figure}
This solution is exactly the response induced by the presence of the quadratic $\lambda_L$ coupling to an external tidal field, as we expect. Of course, the expression~\eqref{eq:p1soln} can equally well be obtained via a Feynman diagram expansion, where the relevant diagram is shown in Figure~\ref{fig:response}. In this approach, the source term~\eqref{eq:sourcevertex} is associated to the four-point vertex with one leg on the background (denoted by the red $\times$), and the Green's function is (the static limit of) the propagator for the external scalar line (wavy line). At this order, the interpretation in terms of Feynman diagrams does not afford a large simplification, but at higher orders it becomes much more economical.

At this point, all that is left is to Fourier transform back to position space and match to the solution we obtained in the full theory calculation. We can do the Fourier transform using the standard Fourier integral
\be
\int\frac{\D^d \vec{p}}{(2\pi)^d}\, e^{i \vec{p}\cdot\vec{x}} \frac{1}{\vec{p}^2} = \frac{\Gamma(\frac{d}{2}-1)}{(4\pi)^{d/2}} \left( \frac{\vec{x}^2}{4} \right)^{1-\frac{d}{2}} \,.
\ee
Of course this integral does not converge for some values of $d$, but in those cases the Fourier transform can be defined in the standard way by taking a limit. In the following we will focus on the cases where $d$ and $L$ are generic, but quantities in special degenerate cases can also be obtained by the limiting procedure.
This is not precisely the integral we need, but we can obtain the integral of interest by differentiating both sides of this formula with respect to $x^a$ $L$ times. In general this would generate a very complicated formula, but since we are ultimately going to contract the free indices with the totally traceless tensor $c_{a_1\cdots a_L}$ the only nonzero contributions will arise when derivatives hit the $\vec x^2$ factor. With this simplification, we have 
\be
(-i)^L \int\frac{\D^d \vec{p}}{(2\pi)^d}\,e^{i \vec{p}\cdot\vec{x}}\,c^{a_1\cdots a_L}\frac{p_{a_1}\cdots p_{a_L}}{\vec p^2} = (-1)^L\frac{\Gamma(\tfrac{d}{2}-1)\Gamma(2-\tfrac{d}{2})}{2^L(4\pi)^{d/2}\Gamma(2-L-\tfrac{d}{2})}\,c_{a_1\cdots a_L}x^{a_1}\cdots x^{a_L}\left( \frac{\vec{x}^2}{4} \right)^{1-\frac{d}{2}-L}.
\ee
We can now combine together the solution $\phi^{(1)}$ with the tidal field background $\phi^{(0)}$ into the full field $\phi$:
\be
\phi(\vec x) = c_{a_1\cdots a_L}x^{a_1}\cdots x^{a_L}\left[1+\lambda_L(-1)^L\frac{2^{L-2}\Gamma(\tfrac{d}{2}-1)\Gamma(2-\tfrac{d}{2})}{\pi^{d/2}\Gamma(2-L-\tfrac{d}{2})}\lvert\vec x\rvert^{-d-2L+2}\right].
\label{eq:eftscalar}
\ee
Notice that the prefactor is (up to an unimportant normalization) $r^L Y_L^M(\theta)$, so we can readily match this solution to the full theory calculation we did in Section~\ref{Sec:spin-0}. Unpacking the definitions there, we find that the solution in the $r\to \infty$ limit is given by
\be
\phi_L(r) \simeq A r^LY_L^M(\theta)\left[1+\cdots+ k_{\rm scalar} \left(\frac{r}{r_s}\right)^{-D-2L+3}+\cdots
\right],
\label{eq:uvscalar}
\ee
where $A$ is an unimportant overall normalization and
where $k_{\rm scalar}$ is the (dimensionless) scalar Love number given by~\eqref{eq:scalarfieldLove}. Here $\cdots$ denotes other subleading terms in the $r\to\infty$ expansion. Recall also that $D = d+1$. Comparing~\eqref{eq:eftscalar} to~\eqref{eq:uvscalar}, we can read off $\lambda_L$:
\begin{tcolorbox}[colframe=white,arc=0pt,colback=greyish2]
%\vspace{-7pt}
\be
\lambda_L = k_{\rm scalar} (-1)^L \frac{\pi^{\frac{D-1}{2}}}{2^{L-2}} \frac{\Gamma(\frac{5-D}{2}-L)}{ \Gamma(\frac{5-D}{2})\Gamma(\frac{D-3}{2})   } r_s^{2L+D-3}    \, ,
\ee
\end{tcolorbox}
\noindent
which relates the EFT coefficient to the quantity computed in the full theory, $k_{\rm scalar}$. For $L=2$ this reproduces the result in~\cite{Kol:2011vg}. The overall numerical prefactor multiplying $k_{\rm scalar}$ is not particularly important. Rather, conceptually, the interesting thing is that when $k_{\rm scalar}$ vanishes---as it does in $D=4$---the corresponding EFT coefficient also vanishes. The vanishing of this EFT coefficient is therefore an unambiguous characterization of what it means for the scalar Love numbers of a black hole to vanish.

\subsection{Coupling to electromagnetism}
We next consider worldline couplings to the electromagnetic field. These are the couplings that encode the electromagnetic susceptibilities of the black hole. In addition to demanding worldline reparametrization invariance, as in the scalar case, we must impose that the worldline couplings are gauge invariant. This indicates that the worldline operators should be built from the electric and magnetic fields:
\begin{align}
E_{a} &\equiv  F_{0a} = \dot A_a - \partial_aA_0,\\
B_{a b} &\equiv F_{ab} = \partial_aA_b-\partial_b A_a,
\end{align}
which can be written in terms of $F_{\mu\nu}$ in a covariant way using the projector~\eqref{eq:projector}. We therefore consider the action
\be
S = S_{\rm pp}-\frac{1}{4}\int \D^D x F_{\mu\nu}F^{\mu\nu} +  \sum^\infty_{L=1} \frac{1}{2L!}\int \D \tau \,E\left[ 
\lambda_L^{(E)}  \left(  \partial_{(a_1}\cdots \partial_{a_{L-1}} E_{a_L)_T} \right)^2 
+ \frac{\lambda_L^{(B)}}{2}  \left(  \partial_{( a_1}\cdots \partial_{a_{L-1}} B_{a_L )_T b} \right)^2  
\right] ,
\ee
where $S_{\rm pp}$ is the free point particle action~\eqref{eq:polyakov}. Note that the letter $E$ appears in this equation in two different ways: as the worldline vielbein, $E$, and as the electric field $E_{a}$; these should not be confused.  There is another possible operator $\sim v^\mu A_\mu$ that we have not written, which just accounts for the possibility that the point particle is charged. Since the black holes of interest are uncharged, we set this coupling to zero.
We now solve the same problem as before: consider a background tidal electric or magnetic field and compute the response induced by the operators proportional to $\lambda_L^{(E)}$ and $\lambda_L^{(B)}$.

\vspace{-.5cm}
\paragraph{Electric response:} In the static case, we can induce a tidal electric field by considering the background field profile
\be
A_0^{(0)}(\vec x) = c^{(E)}_{a_1\cdots a_L}x^{a_1}\cdots x^{a_L},
\ee
so that the effective source appearing in the $A^{(1)}_\mu$ equation of motion is
\be
J^\mu_{\rm eff}(\vec p) = - \delta_0^\mu \lambda_L^{(E)} \, (-i)^L c^{(E)}_{a_1\cdots a_L}p^{a_1}\cdots p^{a_L}.
\ee
It is easiest to solve for $A_\mu$ by multiplying this by the Feynman gauge propagator (at $\omega = 0$):
\be
G_F^{\mu\nu} = -\frac{1}{\vec{p}^{\, 2}} \eta^{\mu\nu}.
\ee
Using this, we find the following solution for $A_\mu^{(1)}$ where only $A_0$ is nonzero
\be
A_0^{(1)}(\vec p) = \lambda_L^{(E)} \,(-i)^Lc^{(E)}_{a_1\cdots a_L} \frac{p^{a_1}\cdots p^{a_L}}{\vec{p}^{\, 2}}.
\ee
Since this is identical to~\eqref{eq:p1soln} we can use the formulas from the previous subsection to go back to real space. Summing up the tidal field with this induced profile we obtain
\be
A_0(\vec x) = \bar c^{(E)} r^L Y_L^M(\theta)\left[1+\lambda^{(E)}_L(-1)^L\frac{2^{L-2}\Gamma(\tfrac{d}{2}-1)\Gamma(2-\tfrac{d}{2})}{\pi^{d/2}\Gamma(2-L-\tfrac{d}{2})}r^{-d-2L+2}\right],
\label{eq:EFTelectric}
\ee
where we have switched to spherical coordinates, and $\bar c$ is an (irrelevant) overall constant.

In order to match the solution~\eqref{eq:EFTelectric} to the quantities computed in Section~\ref{sec:EMresponse}  we have to account for one additional subtlety, which is that the two computations have been carried out in different gauges. One way to deal with this would be to explicitly change gauges in one of the answers and then compare. However, it is much simpler to just compare a gauge-invariant quantity like the radial electric field. In the EFT, the radial electric field is just $E_r = -\partial_rA_0$, while in the full GR computation the radial electric field in the static limit is $E_r = -\partial_r a_0 Y_L^M$, where $a_0$ can be written in terms of $\Psi_S$ using~\eqref{eq:a0intermsofpsi}. Since the radial derivatives are the same in the two cases, we can actually just match $A_0$ and $a_0$. The latter is given by (as $r\to\infty$)
\be
a_0\simeq A r^{L} \left[1+\cdots - \frac{L}{D-3+L}  k_S \left(\frac{r}{r_s}\right)^{-D-2L+3}+\cdots \right]
\ee
where $k_S$ is the electric polarizability~\eqref{eq:spin1Ssummary}. Comparing the two, we find that
\begin{tcolorbox}[colframe=white,arc=0pt,colback=greyish2]
%\vspace{-7pt}
\be
\lambda_L^{(E)} = - k_S (-1)^L  \frac{L}{L+D-3} \frac{\pi^{\frac{D-1}{2}}}{2^{L-2}} \frac{\Gamma(\frac{5-D}{2}-L)}{ \Gamma(\frac{5-D}{2})\Gamma(\frac{D-3}{2})   } r_s^{2L+D-3}   \, ,
\ee
\end{tcolorbox}
\noindent
which relates the Wilson coefficient in the point particle action to the black hole polarizability. Again, we find that the coupling to $E^2$ is absent in the action when the polarizability vanishes.

\vspace{-.5cm}
\paragraph{Magnetic response:} Now we match the magnetic coefficient $\lambda_L^{(B)}$. As in the electric case we consider a tidal magnetic field, which is induced by
\be
A_a^{(0)}(\vec x) = c^{(B)}_{a | b_1\cdots b_L}x^{b_1}\cdots x^{b_L},
\ee
where $c^{(B)}_{a|b_1\cdots b_L}$ has the symmetry type $\scalebox{.7}{\gyoung(_4{L},;)}^{\,T}$. That is, it is traceless and symmetric in its $b$ indices, and if we symmetrize over $a$ additionally the tensor vanishes. This leads to a source in the equation of motion for $A_\mu$ 
\be
J^\mu_{\rm eff}(\vec p) = -\delta_a^\mu\lambda_L^{(B)}(-i)^L c^{(B)}_{a|b_1\cdots b_L} p_{b_1}\cdots p_{b_L}.
\ee
We can, as before, contract this with the Feynman gauge Green's function to solve for the induced field and then combine with the tidal contribution to get the full solution,
\be
A_i(\vec x) = \bar c^{(B)} r^{L} Y_i^{(T)}{}_L^M\left[1+\lambda_L^{(B)}(-1)^L\frac{2^{L-2}\Gamma(\tfrac{d}{2}-1)\Gamma(2-\tfrac{d}{2})}{\pi^{d/2}\Gamma(2-L-\tfrac{d}{2})}r^{-d-2L+2}
\right],
\ee
where we have focused on the angular components so that we can use the fact that (see Appendix~\ref{app:spin1sphericalharmonics})
\be
c^{(B)}_{i|b_1\cdots b_L}x^{b_1}\cdots x^{b_L} = \bar c^{(B)} r^{L+1} Y_i^{(T)}{}_L^M,
\ee
to write things in spherical coordinates.\footnote{Strictly speaking, the translation from Cartesian coordinates to spherical coordinates can have a part proportional to the gradient of a scalar spherical harmonic. However, this piece will get projected out of gauge-invariant quantities because it is pure gauge, so we suppress this (irrelevant) piece.} In order to match to the full GR calculation, we again match to the gauge-invariant magnetic field, which is given by $B_{ab} = \partial_aA_b- \partial_bA_a$. It is easiest to match to the angular components of this tensor, $B_{ij}$ (recall $i,j,k,\cdots$ indices indicate angular directions), so that in the EFT we have
\be
B_{ij}(\vec x) =  2\bar c^{(B)} r^{L+1} \nabla_{[i}Y_{j]}^{(T)}{}_L^M\left[1+\lambda_L^{(B)}(-1)^L\frac{2^{L-2}\Gamma(\tfrac{d}{2}-1)\Gamma(2-\tfrac{d}{2})}{\pi^{d/2}\Gamma(2-L-\tfrac{d}{2})}r^{-d-2L+2}
\right].
\ee
In the full theory, on the other hand, we have that $B_{ij} = r^{(4-D)/2}\Psi_V \nabla_{[i}Y_{j]}^{(T)}{}_L^M$, where $\Psi_V$ can be expanded near infinity as
\be
\Psi_V \simeq A r^{L+1+\tfrac{D-4}{2}}\left[1+\cdots+ k_V  \left(\frac{r}{r_s}\right)^{-D-2L+3}+\cdots\right],
\ee
where $k_V$ is given in \eqref{kVspin1m}.
Putting these together, we find
\begin{tcolorbox}[colframe=white,arc=0pt,colback=greyish2]
%\vspace{-7pt}
\be
\lambda_L^{(B)} =  k_V (-1)^L   \frac{\pi^{\frac{D-1}{2}}}{2^{L-2}} \frac{\Gamma(\frac{5-D}{2}-L)}{ \Gamma(\frac{5-D}{2})\Gamma(\frac{D-3}{2})   } r_s^{2L+D-3}   \, ,
\ee
\end{tcolorbox}
\noindent
which matches $\lambda_L^{(B)}$ in terms of $k_V$. Again, the $B^2$ operators are absent when the corresponding Love numbers vanish in the full theory.

\subsection{Coupling to gravity}
Finally, we consider coupling the black hole point particle to gravity. Strictly speaking, the mere presence of black holes tells us that we should have coupled to gravity from the beginning. However, in the previous sections we were interested in the leading effects near $r\to\infty$, which are insensitive to dynamical gravity. Indeed, this is another advantage of the EFT approach. If we were interested in the subleading behavior of the responses we would have to include gravitational effects even in the scalar and electromagnetic cases, which can be done perturbatively. At the free level, coupling to gravity is very simple: we just promote $\eta_{\mu\nu}\mapsto g_{\mu\nu}$ so that the point particle action is given by
\be
S_{\rm pp} =  \int \rd \tau E \left( -\frac{1}{2}E^{-2} g_{\mu\nu}\dot x^\mu\dot x^\nu - \frac{m^2}{2}\right).
\ee
This coupling causes gravity to respond to the stress tensor of the point particle.
In order to capture the tidal response of the point particle and match to the Love numbers in the full theory, we have to include higher-derivative worldline couplings to the graviton. Much as in the spin-1 case, these couplings are constrained by worldline reparametrization invariance as well as gauge invariance. This implies that these higher-derivative couplings will have to be built from the Weyl tensor, $C_{\mu\nu\rho\sigma}$, which we consider instead of the Riemann tensor because the traces are redundant couplings and can be removed by field redefinitions. The differential Bianchi identity tells us that divergences of the Weyl tensor vanish, and d'Alembertians acting on the Weyl tensor can also be redefined away at quadratic order in $h_{\mu\nu}$. Since we are interested in static solutions, the most general basis of operators at second order in the fields therefore consists of symmetrized traceless derivatives of the Weyl tensor.

There are three types of operators that we can write in the worldline action, coming from the irreducible components of the Weyl tensor under the $D\to d+1$ space/time split. Given the $D$-dimensional Weyl tensor, we can construct its electric part as
\be
E^{(2)}_{ab} \equiv C_{0a0b},
\ee
which can be written covariantly using $v^\mu$ and the projector~\eqref{eq:projector}. This gravito-electric field is a symmetric and traceless tensor. From the Weyl tensor we can also construct the magnetic part
\be
B^{(2)}_{ab|c} \equiv C_{0abc}.
\ee
This tensor is totally traceless and is of mixed symmetry type $\scalebox{.7}{\gyoung(;;,;)}^{\,T}$. It is convenient to work in the convention where it is symmetric under the interchange of its first two indices. If we were to try to symmetrize as well over $c$ the tensor vanishes. In the special case $D=4$ this tensor can be dualized using $\epsilon_{abc}$ to a symmetric-traceless two-index tensor.\footnote{The fact that the gravito-electric and gravito-magnetic tensors have the same number of components in $D=4$ makes the existence of gravitational electric-magnetic duality possible.} Finally there is the lower-dimensional Weyl-like tensor:
\be
C^{(2)}_{ac|bd} \equiv C_{abcd}.
\ee
which has the same symmetries as the Weyl tensor $\scalebox{.65}{\gyoung(;;,;;)}^{\,T}$.
In the special case $D=4$ this tensor identically vanishes, so there are only electric-type and magnetic-type operators that we can add to the worldline action. This is consistent with the fact that the spin-2 tensor degree of freedom only exists for $D\geq 5$, and these Weyl-like operators induce a tidal response in this tensor degree of freedom.

We can then write down the worldline effective action to quadratic order in the graviton fluctuation $g_{\mu\nu} = \eta_{\mu\nu} + 2h_{\mu\nu}/M_{\rm Pl}^{(D-2)/2}$ 
\be
\begin{aligned}
S = S_{\rm pp} +\int\rd^D x \frac{1}{2}h{\cal E}h +\sum_{L=1}^\infty\frac{1}{2L!} \int\rd \tau \Bigg[
&\lambda_L^{(C_E)}\left(\partial_{(a_1}\cdots \partial_{a_{L-2}}E^{(2)}_{a_{L-1}a_L)_T}\right)^2\\
&+\frac{\lambda_L^{(C_B)}}{2}\left(\partial_{(a_1}\cdots \partial_{a_{L-2}}B^{(2)}_{a_{L-1}a_L)_T|b}\right)^2\\
&+\frac{\lambda_L^{(T)}}{4}\left(\partial_{(a_1}\cdots \partial_{a_{L-2}}C^{(2)}_{a_{L-1}a_L)_T|bc}\right)^2\Bigg],
\end{aligned}
\ee
where ${\cal E}$ is the Lichnerowicz operator (see eq.~\eqref{eq:graviton-action} with $\Lambda=0$ and $\nabla=\partial$ for its precise form).
The problem is now the same as in the previous cases: we impose boundary conditions so that the solution to the linearized equation of motion corresponds to a tidal field and then we compute the induced tidal response due to the presence of these operators. It will be convenient to work in de Donder gauge, which is defined by the condition $\partial_\mu \left(h^{\mu\nu} -\tfrac{1}{2}\eta^{\mu\nu}h\right) = 0$. In this gauge the graviton propagator is 
\be
G_{\rm dD}^{\mu\nu \alpha\beta} = -\frac{1}{2}\left(\eta_{\mu\alpha}\eta_{\nu\beta}+\eta_{\mu\beta}\eta_{\nu\alpha} - \frac{2}{D-2}\eta_{\mu\nu}\eta_{\alpha\beta}\right)\frac{1}{p^2}.
\ee
Since we are interested in static solutions ($\dot h_{\mu\nu} = 0$), there are many simplifications in this gauge, we find that $\nabla^2 h_a^a = \nabla^2 h_{00} = \nabla^2 h_{0a} = \partial_a h^{a0} = 0$ and $\partial_a h^{ab} = \tfrac{1}{2} \partial^b h_c^c-\tfrac{1}{2}\partial^b h_{00}$, where $\nabla^2$ is the spatial laplacian.

%\vspace{-.5cm}
\paragraph{Gravito-electric response:} We begin by considering the electric (scalar-type) Love number. In de Donder gauge only the $h_{00}$ component of the metric contributes to the gravito-electric field, so to induce a background field we can take the solution to the linear equation of motion
\be
h^{(0)}_{00} = c^{(E)}_{a_1\cdots a_{L}}x^{a_1}\cdots x^{a_L}.
\ee
Following the same procedure as above, the $E^2$ coupling in the action induces the following response:
\be
h_{00}^{(1)}(\vec p) = \lambda_L^{(C_E)} (-i)^L \frac{D-3}{D-2}\,c^{(E)}_{a_1\cdots a_L} \frac{p^{a_1}\cdots p^{a_L}}{\vec{p}^{\, 2}}.
\ee
Using the standard Fourier transform formulas, it is straightforward to go back to position space and combine with the tidal field to get the full solution for $h_{00}$:
\be
h_{00}(\vec x) = \bar c^{(E)} r^L Y_L^M(\theta)\left[1+\lambda^{(C_E)}_L(-1)^L\frac{d-2}{d-1}\,\frac{2^{L-2}\Gamma(\tfrac{d}{2}-1)\Gamma(2-\tfrac{d}{2})}{\pi^{d/2}\Gamma(2-L-\tfrac{d}{2})}r^{-d-2L+2}\right].
\ee
Now, in order to match to the GR computation, we again have to match gauge-invariant quantities. In this case, the simplest quantity to match is the $rr$ component of the electric tensor, or equivalently the $C_{0r0r}$ component of the Weyl tensor. This is given by $C_{0r0r}= -\partial_r^2 h_{00}$, which evaluates to
\be
C_{0r0r} \propto  r^{L-2} Y_L^M
\left[1+\lambda^{(C_E)}_L(-1)^L\frac{d-2}{d-1}\,\frac{(d+L-2)(d+L-1)}{L(L-1)}\frac{2^{L-2}\Gamma(\tfrac{d}{2}-1)\Gamma(2-\tfrac{d}{2})}{\pi^{d/2}\Gamma(2-L-\tfrac{d}{2})}r^{-d-2L+2}\right].
\ee
We also have to compute the Weyl tensor arising from the potentials computed in Section~\ref{sec:Lovenumbers}. At a general point $r$, this is a somewhat complicated task. However, since we only need to match things at $r\to\infty$ we can compute the Weyl tensor to leading order in this limit, which is considerably simpler. In this limit, the part of the Weyl tensor of interest is given by
\be
C_{0r0r} \xrightarrow{r\to\infty} 2(D-3)\,L(L+D-3) Y_L^M(\theta)r^{-5+\tfrac{D}{2}}\Psi_{\rm Z}
\ee
where $\Psi_{\rm Z}$ is the Zerilli variable, which has the expansion near $r\to\infty$
\be
\Psi_Z = A r^{L+1-\frac{(D-4)}{2}}\left[1+\cdots +  k_{\rm Z} \left(\frac{r}{r_s}\right)^{-D-2L-3}+\cdots \right],
\ee
where $k_{\rm Z}$ is the electric Love number given by~\eqref{eq:zerillilovesummary}. Comparing the two solutions we can match,
\begin{tcolorbox}[colframe=white,arc=0pt,colback=greyish2]
%\vspace{-7pt}
\be
\lambda^{(C_E)}_L =  k_{\rm Z}(-1)^L \frac{(D-2)L(L-1)}{(D-3)(D+L-3)(D+L-2)}   \frac{\pi^{\frac{D-1}{2}}}{2^{L-2}} \frac{\Gamma(\frac{5-D}{2}-L)}{ \Gamma(\frac{5-D}{2})\Gamma(\frac{D-3}{2})   } r_s^{2L+D-3}   \, ,
\ee
\end{tcolorbox}
\noindent
which relates the EFT parameter to the Love number. As expected, the EFT coefficient vanishes whenever the Love number does.

\vspace{-.5cm}
\paragraph{Gravito-magnetic response:} We can repeat the exercise for the gravito-magnetic field. In de Donder gauge, only $h_{0a}$ contributes to the $C_{0abc}$ part of the Weyl tensor, so we can consider
\be
h_{0a}^{(0)}(\vec x) = c^{(B)}_{a | b_1\cdots b_L}x^{b_1}\cdots x^{b_L},
\ee
which induces a tidal gravito-magnetic field with multipole structure $L$.
Following the same procedure, we can contract the corresponding source obtained by perturbing the action and then Fourier transform back to real space to find the field profile
\be
h_{0i}(\vec x) = \bar c^{(B)} r^{L+1} Y_i^{(T)}{}_L^M(\theta)\left[1+\lambda_L^{(C_B)}(-1)^L\frac{L+1}{L}\frac{2^{L-4}\Gamma(\tfrac{d}{2}-1)\Gamma(2-\tfrac{d}{2})}{\pi^{d/2}\Gamma(2-L-\tfrac{d}{2})}r^{-d-2L+2}
\right],
\ee
where we have again focused on the angular component. As is now familiar, we have to match gauge-invariant quantities, so we construct the Weyl tensor $C_{0rij}  = 2r^2\partial_r\left(r^{-2}\nabla_{[i}h_{j]0}\right)$, which is
\be
C_{0rij} = 2(L+1)\bar c^{(B)} r^{L} \nabla_{[i}Y_{j]}^{(T)}{}_L^M(\theta)\left[1-\lambda_L^{(C_B)}(-1)^L\frac{d+L-1}{L -1} \frac{L+1}{L}\frac{2^{L-4}\Gamma(\tfrac{d}{2}-1)\Gamma(2-\tfrac{d}{2})}{\pi^{d/2}\Gamma(2-L-\tfrac{d}{2})}r^{-d-2L+2}
\right].
\label{eq:c0rij}
\ee
In order to match, we require the same Weyl tensor component computed in the full theory, expanded around $r\to\infty$. We find that (up to an irrelevant overall factor)
\be
C_{0rij} \xrightarrow{r\to\infty} \nabla_{[i}Y_{j]}^{(T)}{}_L^M(\theta)r^{\frac{2-D}{2}}\Psi_{\rm RW} .
\ee
Then, using the expansion of the Regge--Wheeler variable near infinity,
\be
\Psi_{\rm RW} = Ar^{L+\frac{D}{2}-1} \left[1+\cdots+k_{\rm RW}\left(\frac{r}{r_s}\right)^{-D-2L+3}+\cdots\right],
\ee
where $k_{\rm RW}$ is the magnetic Love number~\eqref{eq:spin2summaryRW}, and comparing the two solutions, we can read off 
\begin{tcolorbox}[colframe=white,arc=0pt,colback=greyish2]
%\vspace{-7pt}
\be
\lambda^{(C_B)}_L =  -k_{\rm RW}(-1)^L \frac{L-1}{D+L- 2} \frac{L}{L+1}  \frac{\pi^{\frac{D-1}{2}}}{2^{L- 4}} \frac{\Gamma(\frac{5-D}{2}-L)}{ \Gamma(\frac{5-D}{2})\Gamma(\frac{D-3}{2})   } r_s^{2L+D-3}   \, ,
\label{eq:BLovematch}
\ee
\end{tcolorbox}
\noindent
which relates the response coefficient in the EFT to the Love number computed via a GR calculation.

\vspace{-.5cm}
\paragraph{Tensor response:} Finally, we consider the response induced by the $\lambda^{(T)}$ terms. To do this, we consider the growing-mode profile for $h_{ab}$:
\be
h_{ab}^{(0)}(\vec x) = c^{(T)}_{ab | c_1\cdots c_L}x^{c_1}\cdots x^{c_L}
\ee
Through the $\lambda^{(T)}$ operators, this induces the sub-leading falloff (which can be computed by contracting the growing-mode source with the de Donder propagator)
\be
h^{(1)}_{ab} = \lambda_L^{(T)}\, (-i)^Lc^{(T)}_{ab | c_1\cdots c_L}\frac{p^{c_1}\cdots p^{c_L}}{\vec{p}^{\, 2}}.
\ee
Going back to real space and adding back in the tidal field, we have the solution
\be
h_{ab}  = c^{(T)}_{ab | c_1\cdots c_L}x^{c_1}\cdots x^{c_L}\left[1+\lambda^{(T)}_L(-1)^L\frac{2^{L-2}\Gamma(\tfrac{d}{2}-1)\Gamma(2-\tfrac{d}{2})}{\pi^{d/2}\Gamma(2-L-\tfrac{d}{2})}\lvert\vec x\rvert^{-d-2L+2}\right].
\ee
Notice that if we write this expression in spherical coordinates, the transverse-traceless part in the angular directions is actually gauge invariant
\be
h_{ij}^{TT}  = \bar c\,r^{L+2} Y_{ij}^{(TT)}(\theta)\left[1+\lambda^{(T)}_L(-1)^L\frac{2^{L-2}\Gamma(\tfrac{d}{2}-1)\Gamma(2-\tfrac{d}{2})}{\pi^{d/2}\Gamma(2-L-\tfrac{d}{2})}r^{-d-2L+2}\right],
\ee
so we can just match it directly to the tensor component computed in the full theory. This tensor component coefficient has an expansion that is identical to that of the scalar~\eqref{eq:uvscalar} (with an additional overall factor of $r^2$) so we can use that solution to read off
\begin{tcolorbox}[colframe=white,arc=0pt,colback=greyish2]
%\vspace{-7pt}
\be
\lambda^{(T)}_L = k_{\rm scalar} (-1)^L \frac{\pi^{\frac{D-1}{2}}}{2^{L-2}} \frac{\Gamma(\frac{5-D}{2}-L)}{ \Gamma(\frac{5-D}{2})\Gamma(\frac{D-3}{2})   } r_s^{2L+D-3}    \, ,
\ee
\end{tcolorbox}
\noindent
which defines the EFT coefficient in terms of the tensor-type Love numbers (which happen to coincide with the scalar Love numbers). If we had preferred, we could also have done the matching through the Weyl tensor, though it is not necessary in this case. 

In summary, we have derived the mapping between the worldline EFT coefficients for all types of black hole static response, and have confirmed that the worldline couplings also vanish anytime the Love numbers computed directly in GR vanish.

\newpage
\section{Conclusions}
\label{sec:conclude}

%\vspace{-4pt}
We have systematically computed the static response of non-spinning black holes in flat spacetime to external massless perturbations of spin-0, spin-1, and spin-2 in all dimensions. These results confirm known results where applicable, while filling in gaps in the literature.
The final results are reported in eq.~\eqref{eq:scalarfieldLove} for spin-0, in eqs.~\eqref{eq:spin1Ssummary} and~\eqref{eq:spin1Vsummary} for spin-1, and in eqs.~\eqref{eq:spin2summaryRW} and~\eqref{eq:zerillilovesummary} for spin-2. We find that, like the spin-2 tidal Love numbers, the spin-0 scalar response and spin-1 electromagnetic susceptibilities of black holes vanish only in $D=4$. In order to give an unambiguous definition of the static response coefficients, we have connected solutions obtained by a general relativity calculation---which are calculated in a particular coordinate system---to gauge-invariant quantities by matching to the point particle effective action that describes the black hole at long distances.

%\vspace{-2pt}
These results deepen the mystery of the vanishing of black hole Love numbers: all static responses vanish in four dimensions, but all are generically nonzero in other dimensions (apart from special values of the multipole moment). This adds further evidence that there is some underlying explanation for this vanishing of black hole responses in four dimensions.
A natural possibility is that there is some hidden symmetry responsible for this behavior, particularly given these responses' interpretation as Wilson coefficients in the point particle effective theory. One intriguing possibility is that the Geroch group~\cite{Geroch:1972yt,Breitenlohner:1986um} plays some role, because the static sector relevant for computation of Love numbers should have an action of these transformations. In~\cite{uspaper2} we explore this more fully by systematically studying the symmetries of perturbations around non-spinning black holes and their consequences for both Love numbers and time-dependent solutions.

%%\vspace{-2pt}
One reason that it would be interesting to understand {\it why} Love numbers vanish is that it would allow them to be used as powerful tests of gravity. The fact that Love numbers are zero is rather delicate. Generic deviations away from general relativity cause the tidal response to be nonzero~\cite{Cardoso:2017cfl}. Once we understand the minimal requirements for their vanishing, measurements would allow us to constrain broad classes of theories.

%\vspace{-2pt}
Aside from understanding the underlying reason for vanishing of Love numbers, there are other natural directions suggested by this study. It is known that changing the asymptotic boundary conditions~\cite{Emparan:2017qxd}
or adding higher curvature terms~\cite{Cardoso:2018ptl} causes black hole Love numbers to be non-zero, but another natural generalization away from the Schwarzschild case is to consider charged black holes. Aspects of the perturbation theory of charged black holes have been studied~\cite{Moncrief:1975sb,Cardoso:2017cfl,Cardoso:2019mes}, but it is not yet known whether Love numbers vanish in the presence of both gravitational and electromagnetic tidal fields, and it would be interesting to find out. The other natural extension is to consider black holes with spin. In~\cite{LeTiec:2020spy}, Kerr black hole Love numbers were computed to be nonzero. It would be very interesting to verify this by explicitly matching to a worldline effective theory to determine if some non-minimal coupling is required to reproduce the solutions that they found. More generally, it would be interesting to phrase the computation of Love numbers in a more on-shell language~\cite{Cheung:2020sdj,Kalin:2020lmz,Haddad:2020que}, which may shed some light on the underlying structure.
A related---but computationally simpler---example is provided by a scalar field, which has somewhat similar properties~\cite{Wong:2019yoc}, and could serve as a useful test case.
Another possible simplification could be provided by considering the very rapidly spinning case, where the symmetries of the near-horizon metric can be used to organize calculations~\cite{Bardeen:1971eba,Bardeen:1999px}.

%\vspace{-2pt}
We expect that understanding these issues will provide some insights both into the nature of black holes themselves, and into the structure of Einstein gravity.

\vspace{-.3cm}
%=======================================
\paragraph{Acknowledgements} Thanks to Horng Sheng Chia, Kurt Hinterbichler, Dan Kabat, Alberto Nicolis, Robert Penna, Alessandro Podo, Rachel A. Rosen, Ira Rothstein, Mikhail Solon, John Stout, and Sam Wong for helpful conversations, and to Vitor Cardoso, Leonardo Gualtieri, and David Pere\~niguez for useful comments. Special thanks to Emily Hui for technical assistance.
We are especially grateful to Tomer Hadad, Barak Kol, and Michael Smolkin for pointing out an error in the previous version of~\eqref{eq:BLovematch}.
LH is supported by the DOE DE-SC0011941 and a Simons Fellowship in Theoretical Physics. The work of AJ is part of the Delta-ITP consortium. The work of RP is supported in part by the National Science Foundation under Grant No. PHY-1915611. LS is supported by Simons Foundation Award No. 555117. ARS is supported by DOE HEP grants DOE DE-FG02-04ER41338 and FG02-06ER41449 and by the McWilliams Center for Cosmology, Carnegie Mellon University. We thank the participants of the KITP program ``Probing Effective Theories of Gravity in Strong Fields and Cosmology" for stimulating discussions.
This research was supported in part by the National Science Foundation under Grant No. NSF PHY-1748958.

\appendix

\newpage
\section{Spherical harmonics}
\label{app:sphericalharmonics}
In order to isolate the physical degrees of freedom in Section~\ref{sec:actions}, we decomposed fields into SO$(D-1)$ eigenfunctions, namely spherical harmonics. In this Appendix we collect some useful information about spherical harmonics in general dimension. Many useful results can be found in~\cite{Higuchi:1986wu,vanNieuwenhuizen:2012zk}, and particularly in~\cite{Chodos:1983zi}.

\subsection{Scalar spherical harmonics}
\label{sec:scalarsphericalharmonics}

We begin by discussing scalar spherical harmonics on the $n$-sphere. These are the higher-dimensional versions of the familiar harmonics on the 2-sphere.

The basic idea is to embed the $n$-sphere $S^n$ into ${\mathbb R}^{n+1}$. (Hyper)spherical harmonics are then the restriction of homogeneous harmonic polynomials in this ambient space to the sphere.
To see this, consider a homogeneous polynomial of order $L$
\be
P^{(L)}(x) = c_{a_1\cdots a_{L}}x^{a_1}\cdots x^{a_L}.
\ee
This polynomial is harmonic if $c_{a_1\cdots a_{L}}$ is traceless. That is,
\be
\square_{n+1}P^{(L)}(x) =  0 ~~~~~{\rm if}~~~~c_{a~a_3\cdots a_{L}}^{~a} = 0,
\ee
where $\square_{n+1}$ is the ambient space laplacian.
We now consider slicing ${\mathbb R}^{n+1}$ by $n$-spheres:
\be
\rd s^2 = \gamma_{ab}\rd x^a\rd x^b =\rd r^2 + r^2 \rd\Omega_{S^n}^2.
\label{eq:sphericalcoords}
\ee
The $(n+1)$-dimensional laplacian then decomposes as
\be
\square_{n+1}= \frac{1}{r^n}\partial_r r^n\partial_r + \frac{1}{r^2}\Delta_{S^n},
\ee
where the derivatives act on everything to their right.

In the spherical coordinates we have chosen, the radial dependence of the homogeneous polynomial $P^{(L)}$ is very simple:
\be
P^{(L)}(r,\theta)  = r^{L} Y_L(\theta),
\ee
where $\theta = \{ \theta_1, \dots, \theta_n \}$ are the coordinates on $S^n$, which we choose using the recursive definition of the line element:
$
\rd\Omega_{S^n}^2 = \rd \theta_n^2 + \sin^2\theta_n \rd\Omega_{S^{n-1}}^2,
$
where the line element on the circle is just $\rd\Omega_{S^1}^2 = \rd\theta_1^2$.
It is then straightforward to see that the harmonic condition in the ambient space translates to 
\be
\square_{n-1}P^{(L)}(r,\theta) = r^{L-2}\Big(L(L+n-1)+\Delta_{S^n}\Big)Y_L(\theta) = 0.
\ee
This implies that the $Y_L(\theta_a)$ are eigenfunctions of the spherical laplacian with eigenvalue
\be
\Delta_{S^n}Y_L(\theta) = -L(L+n-1)Y_L(\theta).
\ee
These functions provide a representation of the rotation group SO$(n+1)$. To count the dimension of this representation, note that a symmetric $L$-index tensor in $(n+1)$-dimensions has ${n+L}\choose L$ independent components and that the tracelessness condition imposes ${n+L-2}\choose L-2$ conditions. There are therefore
\be
N_L = {{n+L}\choose{L}} - {{n+L-2}\choose{L-2}} = \frac{(L+n-2)!(2L+n-1)}{(n-1)!L!},
\ee
independent harmonics, which gives the dimension of the representation.

It is useful to give a concrete basis for these functions. There are various ways to do this, but probably the most intuitive relies on the observation that there is a sequence of group inclusions
\be
{\rm SO}(n+1) \supset{\rm SO}(n) \supset{\rm SO}(n-1) \supset\cdots\supset {\rm SO}(2).
\ee
This is reflected in the fact that the laplacian on the sphere factorizes nicely:
\be
\Delta_{S^n} = \sin^{1-n}\theta_n\frac{\partial}{\partial \theta_n}\sin^{n-1}\theta_n\frac{\partial}{\partial \theta_n}+\sin^{-2}\theta_n\Delta_{S^{n-1}}.
\ee
It is therefore convenient to label the spherical harmonics by their eigenvalues of the laplacian of each embedded sphere. There are thus $(n-1)$ ``magnetic" quantum numbers, which satisfy
\be
\lvert m_1\rvert\leq m_2\leq \cdots \leq m_{n-1}\leq L\equiv m_n.
\label{eq:Mnumbs}
\ee
We will usually collect these into a multi-index, so that we label the spherical harmonics as $Y^M_L(\theta)$. The magnetic quantum numbers are the angular momentum projections along the various embedded spheres:\footnote{The SO$(2)$ case is treated slightly differently because its eigenfunctions are just $\sim e^{im_1\theta_1}$, and specifying the eigenvalue of the laplacian, $m_1^2$, does not uniquely pick the positive or negative $m_1$ option.}
\begin{align}
\Delta_{S^{n-j}} Y^M_L(\theta) &= -m_{n-j}\left(m_{n-j}+n-j-1\right)Y^M_L(\theta)~~~~{\rm for }~~n-j \geq 2\\
\frac{\partial}{\partial \theta_1} Y^M_L(\theta)&= \pm im_1 Y^M_L(\theta).
\end{align}
These functions provide an orthonormal and complete basis of functions on the $n$-sphere:
\begin{align}
\label{eq:sharmcompleteness}
\int\rd\Omega_{S^n}  Y^M_L(\theta){}^*\, Y^{M'}_{L'}(\theta) &= \delta_{LL'}\delta^{MM'}\\
\sum_{L,M} Y^M_L(\theta){}^*\, Y^{M}_{L}(\tilde\theta) &= \frac{1}{\sqrt{\gamma}}\delta^{(n)}(\theta-\tilde\theta)
\label{eq:sharmcompleteness2}
\end{align}
Using separation of variables, we can construct explicit formulae for these functions~\cite{Higuchi:1986wu}:
\begin{align}
\label{eq:harmonicdef}
Y_L^{m_1\cdots\,m_n}(\theta) &= \frac{1}{\sqrt{2\pi}}e^{im_1\theta_1}\prod_{i=2}^n {}_i\bar P_{m_i}^{m_{i-1}}(\theta_i)\\
{}_k\bar P_{i}^{j}(\theta) &= \sqrt\frac{(2i+k-1)(i+j+k-2)!}{2(i-j)!}\sin^\frac{2-k}{2}(\theta) P^{-\frac{2j+k-2}{2}}_{\frac{2i+k-2}{2}}(\cos\theta)\\
P^{-a}_b(x) &=\frac{1}{\Gamma[1+a]}\left(\frac{1-x}{1+x}\right)^\frac{a}{2} 
\def\arraystretch{.75}
\hypergeom{2}{1}\left[\begin{array}{c}
-b,~~~~b+1\\
1+a
\end{array}\Big\rvert \,\frac{1-x}{2}\,\right]
\end{align}
The functions $P^{-a}_b$ are also known as associated Legendre functions, and $\hypergeom{2}{1}$ is the (Gauss) hypergeometric function, discussed more in Appendix~\ref{App:hyperG}. Note that the ${}_k\bar P_\ell^m$ are each eigenfunctions of the $k$-sphere:
\be
\Delta_{S^k} \,{}_k\bar P^m_\ell = -\ell(\ell+k-1) {}_k\bar P^m_\ell.
\ee
While it is useful for some applications to have explicit formulas for the spherical harmonics, we actually do not require these expressions in the main text. Instead we only use their orthogonality properties~\eqref{eq:sharmcompleteness} and~\eqref{eq:sharmcompleteness2}.

\paragraph{Complex conjugation:} 
In general the harmonics defined by~\eqref{eq:harmonicdef} are not real-valued because of the $e^{im_1\theta_1}$ phase factor. Under complex conjugation they transform as
\be
Y_L^{m_1\cdots\,m_n}(\theta)^* = (-1)^{m_1} Y_L^{-m_1 \,m_2\cdots\,m_n}(\theta) .
\ee
This implies in particular that the $m_1 = 0$ harmonics are real, a fact that we utilize in the main text. For other values of $m_1$ it is straightforward to construct real harmonics by taking linear combinations.

\paragraph{Parity:} 
In some cases it is useful to know
 how the spherical harmonics transform under parity. The action of parity on the $n$-sphere is to send a point to its antipodal point:
\begin{align}
\theta_i &\mapsto \pi - \theta_i ~~~~{\rm for}~i\neq 1\\
\theta_1 &\mapsto \theta_1+\pi
\end{align}
Under this transformation the scalar spherical harmonics pick up a factor of $(-1)^L$:
\be
{\cal P}\, Y_L^M(\theta) = (-1)^L Y_L^M(\theta).
\ee
Functions with this transformation property are said to be parity even.

\subsection{Spin-1 spherical harmonics}
\label{app:spin1sphericalharmonics}
In order to decompose vector fields on the sphere, we also require spin-1 spherical harmonics, which are an extension of the usual scalar spherical harmonics to have an SO$(n+1)$ index. These harmonics can
be constructed in a similar way to the scalar case~\cite{Chodos:1983zi}. We consider instead a harmonic ambient space vector function
\be
P^{(L)}_a(x) = c_{a \rvert b_1\cdots b_L} x^{b_1}\cdots x^{b_L},
\label{eq:vectharmonic}
\ee
where the tensor $c_{a \rvert b_1\cdots b_L}$ is of mixed symmetry type
\be
c_{a\rvert b_1\cdots b_L}  \,\in\,\raisebox{1ex}{\gyoung(_4{L},;)}^{\,T} ,
\ee
and is traceless, which ensures that~\eqref{eq:vectharmonic} is harmonic. Note that spin-1 harmonics exist only for $ L\geq 1$.
When we write~\eqref{eq:vectharmonic} in the spherical coordinates~\eqref{eq:sphericalcoords}, we focus on the $i$ component, which can be written as
\be
P_i^{(L)} = r^{L+1} Y_{i}^{(T)}{}_L^M(\theta)+r^{L+1}\nabla_i\sigma(\theta),
\ee
where $\sigma$ is a scalar and $\nabla^iY_{i}^{(T)}{}_L^M = 0$.
In spherical coordinates, the harmonic condition in the ambient space leads to the eigenvalue equation for the vector spherical harmonics
\be
\Delta_{S^n} Y_{i}^{(T)}{}_L^M = -\left(L(L+n-1)-1\right)Y_{i}^{(T)}{}_L^M
\ee
We see that the spin-1 spherical harmonics have a shifted eigenvalue spectrum compared to the scalar spherical harmonics. Note that the dimension of the transverse vector harmonic representation is different from that of the scalar harmonics (there are more vector harmonics, see~\cite{Chodos:1983zi} for an explicit counting), so the basis used above in~\eqref{eq:Mnumbs} cannot be used here, and the magnetic quantum numbers range over different values. Since we only require the completeness properties of the vector harmonics, we do not enumerate such a basis.

Another way to
construct a spin-1 spherical harmonic is to take a gradient of a scalar spherical harmonic:
\be
Y_{i}^{(L)}{}_L^M(\theta) \equiv \frac{1}{\sqrt{L(L+n-1)}}\nabla_i Y_{L}^M(\theta),
\ee
Commuting the spherical laplacian past the gradient, we find
that the eigenvalue equation is instead
\be
\Delta_{S^n} Y_{i}^{(L)}{}_L^M(\theta)  = -\left(L^2+(L-1)(n-1)\right)Y_{i}^{(L)}{}_L^M(\theta) .
\ee
The fact that the transverse and longitudinal spin-1 spherical harmonics have different eigenvalues makes it manifest that they will decouple in general dimension.
In the special case $n=2$ both the transverse and longitudinal vector spherical harmonics can be constructed by taking derivatives of scalar spherical harmonics.\footnote{On the two sphere, $n=2$, the transverse vector spherical harmonic can be written in terms of the Levi--Civita symbol as $Y_{i}^{(T)}{}_L^M(\theta) = \epsilon_{ij}\nabla^j Y_{L}^M(\theta)/\sqrt{L(L+1)}$.}
In this case, it is the fact that they have different parity eigenvalues that guarantee that they decouple.

Vector spherical harmonics obey similar orthogonality and completeness relations to scalar spherical harmonics
\begin{align}
\int\rd\Omega_{S^d}  Y_{i\,L}^M(\theta){}^*\, Y_{L'}^{i\,M'}(\theta) &= \delta_{LL'}\delta^{MM'},\\
\sum_{L,M} Y_{i}^{(T)}{}_L^M(\theta){}^*\, Y_{j}^{(T)}{}_L^{M}(\tilde\theta) +Y_{i}^{(L)}{}_L^M(\theta){}^*\, Y_{j}^{(L)}{}_L^{M}(\tilde\theta) &= \frac{1}{\sqrt{\gamma}}\delta_{ij}\delta^{(d)}(\theta-\tilde\theta).
\end{align}
Though, note that because $Y_{i}^{(T)}{}_L^M$ is transverse, it is orthogonal to $Y_{i}^{(L)}{}_L^M$. In the main text, we use a slightly different normalization of the gradient vector harmonic. Rather than writing $Y_{i}^{(L)}{}_L^M$ explicitly, we write things directly as $\nabla_i Y_L^M$, which has a different normalization factor.

The reality properties of the vector harmonics are essentially the same as the scalar harmonics: the $m_1=0$ harmonics are real, which we utilize in the main text.
The two types of vector spherical harmonics have different transformation properties under parity. The gradient of a scalar transforms like a vector:
\be
{\cal P}\,Y_{i}^{(L)}{}_L^M(\theta) = (-1)^{L+1}Y_{i}^{(L)}{}_L^M(\theta).
\ee
while the transverse vector spherical harmonic transforms as
\be
{\cal P}\,Y_{i}^{(T)}{}_L^M(\theta) = (-1)^{L}Y_{i}^{(T)}{}_L^M(\theta),
\ee
and so transforms like a pseudo-vector.

\subsection{Spin-2 spherical harmonics}

Finally we consider spin-2 spherical harmonics, which are necessary to decompose spin-2 fields on the sphere. 
To construct them, we consider a harmonic ambient space tensor function~\cite{Chodos:1983zi}
\be
P^{(L)}_{ab}(x) = c_{ab\rvert c_1\cdots c_L} x^{c_1}\cdots x^{c_L},
\ee
where $c_{ab\rvert c_1\cdots c_L}$ has the symmetry type
\be
c_{ab\rvert c_1\cdots c_L}  \,\in\,  \raisebox{1ex}{\gyoung(_4{L},;;)}^{\,T}. \ee
Much like the spin-1 case, these harmonics only exist for $L\geq 2$.
When we write $P^{(L)}$ in spherical coordinates~\eqref{eq:sphericalcoords}, we can split it up into $rr, ri, ij$ components. We focus on the $ij$ component:
\be
P_{ij}^{(L)} = r^{L+2} Y_{ij}^{(TT)}{}_L^M(\theta)+\cdots,
\ee
where the $\cdots$ are the other components of the decomposition---a transverse vector and two scalars (we don't need the details of this decomposition, but they can be found in, e.g., Appendix B of~\cite{Goon:2018fyu}). Note that $Y_{ij}^{(TT)}$ is both transverse and traceless.

In spherical coordinates, the harmonic condition in the ambient space translates into the following eigenvalue equation for $Y_{ij}^{(TT)}$~\cite{Chodos:1983zi}:
\be
\Delta_{S^n} Y_{ij}^{(TT)}{}_L^M = -\left(L(L+n-1)-2\right)Y_{ij}^{(TT)}{}_L^M,
\ee
so we see that the spin-2 spherical harmonics' eigenvalues are shifted by 2 compared to scalar harmonics. As in the vector case, the $M$ label ranges over different values, filling out the representation space (a counting of its dimension can be found in~\cite{Chodos:1983zi}).

The harmonics $Y_{ij}^{(TT)}{}_L^M$ are ``true" spin-2 spherical harmonics, but there
are three other ways to construct a tensor spherical harmonic with the right transformation properties:

\begin{itemize}

\item
 We can multiply a scalar spherical harmonic by the metric on the $n$-sphere: $\gamma_{ij}$:
\be
Y_{ij}^{(\tr)}{}_L^M \equiv \frac{1}{\sqrt n}\gamma_{ij} Y_L^M.
\ee
It is then clear that this object will have the same eigenvalue as a scalar spherical harmonic:
\be
\Delta_{S^n}Y_{ij}^{(\tr)}{}_L^M = -L(L+n-1)Y_{ij}^{(\tr)}{}_L^M.
\ee
This is the only tensor harmonic that has a trace. 

\item
We can also construct a tensor spherical harmonic by taking derivatives of a scalar spherical harmonic:
\be
Y_{ij}^{(S)}{}_L^M \equiv \frac{1}{\sqrt{L(L+d-1)}}\nabla_{(i}\nabla_{j)_T}Y_L^M.
\ee
This object has the following eigenvalue~\cite{Chodos:1983zi}
\be
\Delta_{S^n}Y_{ij}^{(S)}{}_L^M = -\left(L(L+n-1)-2n\right)Y_{ij}^{(S)}{}_L^M.
\ee

\item
Finally we can construct a tensor spherical harmonic by taking a derivative of a vector spherical harmonic:
\be
Y_{ij}^{(T)}{}_L^M \equiv \frac{1}{\sqrt{L(L+n-1)-1}}\nabla_{(i}Y_{j)}^{(T)}{}_L^M,
\ee
which is traceless because the trace is carried by $Y^{(\tr)}$.
The laplacian eigenvalue is~\cite{Chodos:1983zi}
\be
\Delta_{S^n}Y_{ij}^{(T)}{}_L^M = -\left(L(L+n-1)-(n+2)\right)Y_{ij}^{(T)}{}_L^M.
\ee
\end{itemize}
These harmonics are all orthonormal
\begin{align}
\int\rd\Omega_{S^d}  Y_{ij\,L}^M(\theta){}^*\, Y_{L'}^{ij\,M'}(\theta_a) &= \delta_{LL'}\delta^{MM'},\\
\sum_{I,L,M} Y^{(I)}_{ij}{}_L^M(\theta){}^*\, Y^{(I)}_{kl}{}_L^{M}(\tilde\theta) &= \frac{1}{\sqrt{\gamma}}\mathbbm{1}_{ij,kl}\delta^{(d)}(\theta-\tilde\theta),
\end{align}
where $\mathbbm{1}_{ij,kl}$ is the identity on symmetric $2$-index tensors and
where the index $I$ runs over all the types of spin-2 harmonics we have introduced.
In general all of these harmonics are necessary to decompose a spin-2 field. In the main text, in order to avoid proliferation of notation, we only introduce the transverse-traceless spherical harmonic, and denote the other harmonics by explicitly derivatives acting on scalar harmonics.

The reality properties of the spin-2 harmonics are the same as the other cases: the $m_1=0$ harmonics are real.
We can also figure out the transformation of each of the different tensor spherical harmonics under parity:
\begin{align}
{\cal P}Y_{ij}^{(TT)}{}_L^M &=(-1)^L Y_{ij}^{(TT)}{}_L^M \\
{\cal P}Y_{ij}^{(\tr)}{}_L^M &= (-1)^LY_{ij}^{(\tr)}{}_L^M  \\
{\cal P}Y_{ij}^{(S)}{}_L^M &= (-1)^LY_{ij}^{(S)}{}_L^M \\
{\cal P}Y_{ij}^{(T)}{}_L^M &= (-1)^{L+1}Y_{ij}^{(T)}{}_L^M 
\end{align}
We see that $Y^{(TT)}, T^{(\tr)}, Y^{(S)}$ transform like tensors under parity, while $Y^{(T)}$ transforms like a pseudo-tensor.

\newpage
\section{The hypergeometric equation}
\label{App:hyperG}

Aspects of the theory of hypergeometric functions play an important role in the computation of black hole Love numbers described in Section~\ref{sec:lovenums}, so in this Appendix we provide a brief review of the relevant facts for the convenience of the reader. For more details, some useful references are~\cite{slavjanov2000special,Bateman:100233,Beukers2007,beals_wong_2010,Kristensson}.\footnote{\href{https://webspace.science.uu.nl/~beuke106/HypergeometricFunctions/HGFcourse2009.pdf}{\tt These} notes also have additional useful information and the~\href{https://dlmf.nist.gov/15}{\tt Digital Library of Mathematical Functions} is a helpful resource with many formulas.}

\subsection{Generalities of Fuchsian equations}

The hypergeometric differential equation is a second-order differential equation of the Fuchsian type, possessing three regular singular points. Fuchsian equations are differential equations for complexified functions of the form
\be
u^{(n)}(z)+p_1(z)u^{(n-1)}(z)+\cdots +p_{n-1}(z)u'(z)+p_n(z)u(z) = 0,
\label{eq:fuchseq}
\ee
where all points in the complex plane are regular or regular singular points. In this equation $u^{(n)}$ denotes the $n^{\rm th}$ derivative with respect to $z$. Points are said to be regular if the coefficient functions, $p_a(z)$, are finite for all $a$. The coefficients are allowed to have singularities as long as the singularities are sufficiently mild. In particular a {\it regular singular point}, $z_\star$, is one for which $\lim_{z\to z_\star} (z-z_\star)^a p_a(z_\star)$ exists. In other words, $p_a(z)$ can have a pole at most of order $a$ at a regular singular point. In general there is the possibility that the point at infinity is a singular point; it turns out that the point at infinity is a regular singular point (or is just regular) provided that $\lim_{z\to\infty} z^a p_a(z_\star)$ exists for all $a$.\footnote{One way to understand this is to map $z\mapsto 1/t$ and study the singularities of the resulting differential equation. The point at infinity in the $z$ variable is now mapped to $t=0$ and can be treated normally.}

The singularities of Fuchsian differential equations in large part control the behavior of the solutions and dictate that they must have a particular form in the vicinity of a singular point. Near a singular point, it makes sense to change variables to study the local behavior of solutions by defining $t \equiv z-z_\star$. We then study the equation near $t=0$ by multiplying~\eqref{eq:fuchseq} by $t^n$ and taking the limit $t\to 0$, so that the equation takes the form:\footnote{If we are interested in the point at infinity, we define $t\equiv 1/z$ so that the analogous equation to~\eqref{eq:indicialdifeq} is
\be
(-1)^n{\cal D}({\cal D}+1)\cdots({\cal D}+n-1)u+ (-1)^{n-1}c_1{\cal D}({\cal D}+1)\cdots({\cal D}+n-2)u+\cdots+(-1)c_{n-1}{\cal D}u+c_n u = 0,
\label{eq:indicialeqinf}
\ee
which can be treated in a similar manner.
}
\be
{\cal D}({\cal D}-1)\cdots({\cal D}-n+1)u+ c_1{\cal D}({\cal D}-1)\cdots({\cal D}-n+2)u+\cdots+c_{n-1}{\cal D}u+c_n u = 0,
\label{eq:indicialdifeq}
\ee
where the (constant) coefficients $c_a \equiv \lim_{z\to z_\star}(z-z_\star)^aP_a(z)$ are the limits of the coefficient functions as we approach the singularity, and where we have defined the dilation operator ${\cal D} \equiv t\frac{\rd}{\rd t}$, which acts on everything to its right. Notice that ${\cal D}$ commutes with the differential operator acting on $u$ in~\eqref{eq:indicialdifeq}, so we see that the differential equation develops a scaling symmetry in the vicinity of its singular points. It is therefore useful to distinguish solutions based on their behavior as they approach these singularities; in particular, it makes sense to consider their ${\cal D}$ eigenvalue. If we make the ansatz that $\lim_{t\to0} u\sim t^\alpha$, then the equation~\eqref{eq:indicialdifeq} becomes algebraic:
\be
\alpha(\alpha-1)\cdots(\alpha-n+1)u+ c_1\alpha(\alpha-1)\cdots(\alpha-n+2)u+\cdots+c_{n-1}\alpha u+c_n u = 0.
\label{eq:indicialeq} 
\ee
This equation is called the {\it indicial equation} and its $n$ roots are called {\it local exponents}. Note that there is an indicial equation---and correspondingly a set of local exponents---for each regular singular point.\footnote{The local exponents are not totally independent. Their sum is fixed by the so-called {\it Fuchs' relation}, which is a global constraint on their values.} These local exponents determine the fall-off of a solution as we approach the singular points.
One of the goals of the theory of special functions is to understand how solutions with a given scaling near one singularity are related to solutions expanded near another singularity.

\subsection{The hypergeometric differential equation}
We now turn to the equation of principal interest, the hypergeometric equation. It is a second-order differential equation of the form
\be
 L_{a,b,c} u(z) = z(1-z) u''(z)+ \big[c - (a+b+1)z\big] u'(z) - a \,  b \,  u(z) =0 \, ,
\label{app:hyperEq}
\ee
where for later convenience we have defined the differential operator $L_{a,b,c}$.
Dividing through by $z(1-z)$ we can put it in the standard form~\eqref{eq:fuchseq}:
\be
u''(z) + \frac{c - (a+b+1)z}{ z(1-z) } u'(z) - \frac{a \,  b}{ z(1-z) }  u(z) =0 \, ,
\ee
from which we deduce that it has three regular singular points: at $z=0$, $z=1$, and $z=\infty$. In fact, it can be shown that any second order differential equation with three regular singular points can be cast in the standard hypergeometric form by means of M\"obius transformations of the coordinate $z$~\cite{slavjanov2000special,Bateman:100233,beals_wong_2010,Kristensson}. These transformations move the singular points of the differential equation, and there is enough freedom to place them at $0,1,\infty$. Then, redefinitions of the dependent variable can be used to write the equation in the form~\eqref{app:hyperEq}. Consequently the solutions to any such equation can be obtained from solutions to the hypergeometric equation. Many of the differential equations appearing in physics are of this type, motivating the systematic study of the hypergeometric equation.

\bgroup
\def\arraystretch{1.35}% 
\begin{table}
\centering
{\small
\begin{tabular}{c |c | c | c }
Singular point: &~0~ &  ~1~  & ~$\infty$~\\\hline
~& 0&0& $a$\\ 
~&$1-c$& $c-a-b$& $b$
\end{tabular}
}
\caption[Table caption text]{\small Local exponents near each of the singularities of the hypergeometric differential equation~\eqref{app:hyperEq}.}
\label{table:hyperindices}
\end{table}
\egroup

Around each of the singular points we can derive the indicial equation~\eqref{eq:indicialeq} and compute the local exponents, each singularity has a pair.  The results are collected in Table~\ref{table:hyperindices}. We see that in the standard form, near each of the singularities at $0$ and $1$, one of the two solutions just goes to a constant (though not the {\it same} solution). In fact the standard solution to the hypergeometric equation is precisely the one that goes to a constant as $z\to 0$.

The differential equation~\eqref{app:hyperEq} is solved by the series:\footnote{The hypergeometric function is written in some references as
\begin{equation*}
\def\arraystretch{.7}
{}_2F_1(a,b;c;z)\equiv \hypergeom{2}{1}\left[\begin{array}{c}
a,~~b\\
c
\end{array}\Big\rvert \,z\,\right] ,
\end{equation*}
which is entirely equivalent. We have chosen this notation because it manifests the symmetry under interchanging $a$ and $b$.
}
\begin{equation}
\def\arraystretch{.8}
u_1(z) = \hypergeom{2}{1}\left[\begin{array}{c}
a,~~b\\
c
\end{array}\Big\rvert \,z\,\right] \equiv \sum_{n=0}^\infty \frac{(a)_n(b)_n}{(c)_n n!}z^n
= \frac{\Gamma(c)}{\Gamma(a) \Gamma(b)} \sum_{n=0}^\infty 
\frac{\Gamma(a+n)\Gamma(b+n)}{\Gamma(c+n)n!} z^n ,
\label{eq:solhype1}
\end{equation}
where $(\cdot)_n$ is the Pochhammer symbol. This series defines the {\it hypergeometric function};  it converges for $\lvert z\rvert < 1$, and is normalized so that $u_1(0) = 1$. The numbers $a,b,c$ are typically called the parameters of the hypergeometric function, while $z$ is the argument. From the definition it is clear that everything is symmetric under the interchange of $a$ and $b$. For generic values of $a,b,c$, the singularity causing the breakdown of convergence of the Taylor expansion is a branch point at $z=1$. The definition of the hypergeometric function can be extended to $\lvert z\rvert > 1$ via analytic continuation, where the usual convention is to take the branch cut to run along the real line from $z=1$ to $\infty$.\footnote{For example, the hypergeometric function can be analytically continued for $c>b>0$ using the following integral representation~\cite{Bateman:100233}:
\be
\def\arraystretch{.65}
\hypergeom{2}{1}\left[\begin{array}{c}
a,~~b\\
c
\end{array}\Big\rvert \,z\,\right] = \frac{\Gamma(c)}{\Gamma(b)\Gamma(c-b)}\int_0^1\rd t\, t^{b-1}(1-t)^{c-b-1}(1-tz)^{-a}.
\ee
The restriction on the parameters is required for the integral to converge, but can be removed by choosing a more general contour~\cite{Beukers2007}.
}
For special values of parameters, the hypergeometric function reduces to more familiar elementary functions, and correspondingly the hypergeometric differential equation simplifies.

A second solution to the hypergeometric equation is given by\footnote{We label the solutions in the same way as~\cite{Bateman:100233}.}
\be
\def\arraystretch{.8}
u_5(z) = z^{1-c }\hypergeom{2}{1}\left[\begin{array}{c}
a-c+1,~~b-c+1\\
2-c
\end{array}\Big\rvert \,z\,\right] .
\label{eq:solhype2}
\ee
For generic values of the parameters $a,b,c$, this solution is linearly independent from~\eqref{eq:solhype1}. Near $z=0$ it scales as $\sim z^{1-c}$, as expected from the analysis of the local exponents near this singularity. In the following section we will see how to generate this solution from~\eqref{eq:solhype1}, along with many other representations.

In order to efficiently describe the relations between different solutions to the hypergeometric equation, it is helpful to introduce the {\it Riemann $P$-symbol}:
\be
\def\arraystretch{.9}
u(z) = P\left\{\begin{array}{ccc}
0 & 1 & \infty\\
0 & 0 & a\\
1-c & c-1-b & b
\end{array}\,\Bigg\rvert\,z\,\right\},
\ee
which contains the same information as Table~\ref{table:hyperindices}, keeping track of the singular points of the differential equation and their local exponents. The $P$-symbol abstractly denotes the space of solutions to~\eqref{app:hyperEq}. There are many equivalent ways of writing hypergeometric functions and the symbol allows us to express the relations between them in a simple way.

\subsubsection{Symmetries of the hypergeometric equation}
The hypergeometric equation has a large degree of symmetry, leading to many equivalent ways of writing its elementary solutions, and leading to many identities satisfied by hypergeometric functions. These identities are particularly important for particular parameter choices, when the natural solutions~\eqref{eq:solhype1} and~\eqref{eq:solhype2} can become linearly dependent, and we need to find alternative expressions for the other solution. This situation arises in the computation of black hole Love numbers, so we describe briefly how these different representations are related.

There are two conceptually distinct transformations we can perform on the hypergeometric equation. The first is to change coordinates, while the other is to redefine the dependent function, $u$. Of course, general such transformations will not preserve the hypergeometric form, but a subset will. We consider each of them in turn.

\vspace{-.5cm}
\paragraph{Coordinate transformations:}
Recall that there is an action of M\"obius transformations on $z$. Most of this symmetry is used to fix the singularities of a general linear equation with three regular singular points to lie at $0,1,\infty$. However, there is a residual discrete subgroup that serves to permute these singular points. This symmetry group is just the symmetric group on three letters, $S_3$. It is generated by the two transformations
\be
z\mapsto 1-z~~~~~~~{\rm and}~~~~~~~z\mapsto \frac{1}{z}~.
\label{eq:permutations}
\ee
These transformations preserve the form of the hypergeometric equation, but change the parameters. For example if $u(z)$ satisfies~\eqref{app:hyperEq} then defining $y = 1-z$, the function $v(y) = v(1-z) = u(z)$ will satisfy the differential equation
\be
 L_{a,b,1+a+b-c}v(y)=y(1-y)v''(y)+\big[1+a+b-c-(a+b+1)y\big]v'(y) -a\,b\,v(y) = 0,
\ee
which is a hypergeometric equation, but with a shifted value of $c$. From this, we 
see for example the solution $u_1(z)$~\eqref{eq:solhype1} gets mapped to the solution
\be
\def\arraystretch{.8}
u_2(z) =  \hypergeom{2}{1}\left[\begin{array}{c}
a,~~~~b\\
1+a+b-c
\end{array}\Big\rvert \,1-z\,\right].
\label{eq:hypernearz11}
\ee
Since the transformation we have done has interchanged the singularities at $1$ and $0$, this is now the solution to the hypergeometric equation that is normalized as $u_2(1) = 1$. The linearly independent solution~\eqref{eq:solhype2} gets mapped to
\be
\def\arraystretch{.8}
u_6(z) = (1-z)^{c-a-b } \hypergeom{2}{1}\left[\begin{array}{c}
c-b,~~c-a\\
1+c-a-b
\end{array}\Big\rvert \,1-z\,\right],
\label{eq:hypernearz12}
\ee
which has the expected fall-off near $z = 1$ for the solution linearly independent of~\eqref{eq:hypernearz11}.

We can also consider the change of variable that interchanges the points at $0$ and $\infty$. Defining $y = 1/z$ we can find a differential equation for $v(y) = v(1/z) = u(z)$ which is:
\be
-y^2(1-y)v''(y)+y\big[(a+b-1)-(c-2)y\big]v'(y)-a\,b\,v(y) = 0.
\ee
This equation is not of the hypergeometric form, but if we define $v(y) = (-y)^{a}\tilde v(y)$ it implies the following hypergeometric equation for $\tilde v$:
\be
L_{a,1+a-c,1+a-b}\tilde v(y) =  y(1-y) \tilde v''(y)+ \big[1+a-c - (2+2a-c)y\big] \tilde v'(y) - a (1+a-c)  \tilde v(y) =0.
\ee
Unpacking the definitions, this implies that the following is a solution to~\eqref{app:hyperEq}
\be
\def\arraystretch{.8}
u_3(z) = (-z)^{-a}\hypergeom{2}{1}\left[\begin{array}{c}
a,~~1+a-c\\
1+a-b
\end{array}\Big\rvert \,\frac{1}{z}\,\right].
\ee
Notice that this solution falls off as $z^{-a}$ near $z=\infty$, as expected from the local exponents there. In the previous manipulations the choice to extract $y^a$ instead of $y^b$ was arbitrary, so we can interchange $a$ and $b$ to obtain another solution:
\be
\def\arraystretch{.8}
u_4(z) = (-z)^{-b}\hypergeom{2}{1}\left[\begin{array}{c}
1+b-c,~~b\\
1+b-a
\end{array}\Big\rvert \,\frac{1}{z}\,\right],
\ee
which is the one that falls off like $z^{-b}$ near infinity.

By composing the transformations~\eqref{eq:permutations} we can generate all possible permutations of the points $0,1,\infty$, and correspondingly generate other equivalent solutions. This does not exhaust all the symmetries of the hypergeometric equation; we now turn to considering field redefinitions that map us between solutions.

\vspace{-.5cm}
\paragraph{Variable redefinitions:}
In addition to coordinate transformations that permute the singular points, there is a set of redefinitions of the variable $u$ that preserve the form of the hypergeometric equation. Conceptually these transformations permute the two solutions at a given singular point, and form a ${\mathbb Z}_2\times{\mathbb Z}_2$ group. These transformations are generated by the following two redefinitions, which can be considered separately
\begin{align}
\label{eq:permute0}
u(z) &\mapsto z^{1-c}v(z),\\
u(z) &\mapsto (z-1)^{c-1-b}v(z).
\label{eq:permute1}
\end{align}
These transformations preserve the locations of the singular points, but change the indicial equation at these points, and therefore map a solution to a different solution to the hypergeometric equation. 

It is easiest to express the action of these redefinitions in terms of the $P$-symbol. In this language, the transformation~\eqref{eq:permute0} leads to the identity
\be
\def\arraystretch{.9}
P\left\{\begin{array}{ccc}
0 & 1 & \infty\\
0 & 0 & a\\
1-c & c-1-b & b
\end{array}\,\Bigg\rvert\,z\,\right\}
=
z^{1-c} P\left\{\begin{array}{ccc}
0 & 1 & \infty\\
c-1 & 0 & a+1-c\\
0 & c-a-b & b+1-c
\end{array}\,\Bigg\rvert\,z\,\right\}.
\ee
What this means is that the solutions to the hypergeometric equation with the standard set of local exponents are equivalent to the set of solutions with a shifted set of local exponents, but multiplied by a factor of $z^{1-c}$. This transformation serves to permute the two solutions that have a single fall-off near $z=0$: the solutions $u_1$ and $u_5$ get swapped.

We can similarly ask how the transformation~\eqref{eq:permute1} acts on the space of solutions. It implies
\be
\def\arraystretch{.9}
P\left\{\begin{array}{ccc}
0 & 1 & \infty\\
0 & 0 & a\\
1-c & c-1-b & b
\end{array}\,\Bigg\rvert\,z\,\right\}
=
(z-1)^{c-a-b} P\left\{\begin{array}{ccc}
0 & 1 & \infty\\
0 & a+b-c & c-b\\
1-c & 0 & c-a
\end{array}\,\Bigg\rvert\,z\,\right\},
\ee
and in particular permutes the two solutions with definite scaling near $z=1$.

We can now describe the full set of solutions related by symmetry transformations. Given a solution that has a definite fall-off near one of the singular points---say the solution that is normalized as $u_1(0) = 1$ at $z=0$---we can act with the transformation that permutes the two solutions at $z=1$, which generates a new solution. We can also compose this with the transformation that interchanges the singularities at $z=1$ and $z=\infty$. Together, these transformations generate four different ways of writing the same solution to the hypergeometric equation. There are similarly four ways of writing the solution that scales as $z^{1-c}$ at $z=0$. There are an analogous $8$ solutions at each of the other two singular points, leading to a total of $24$ solutions related by symmetries of the hypergeometric equation. These are the famous {\it Kummer's 24 solutions}. The corresponding symmetry group is the symmetric group on four letters, $S_4$, which can be thought of as being comprised of the $S_3$ symmetric group that permutes the singularities extended by the ${\mathbb Z}_2\times{\mathbb Z}_2$ group that swaps the fall-offs at $z=0$ and $z=1$.

A complementary way to understand the relations between these solutions is to note that for generic values of the parameters one can show that all these solutions can be obtained from a single solution $u_1$ and its analytic continuation along a closed line that encircles at least one of the branch points $0$~and~$1$~\cite{slavjanov2000special,Bateman:100233,beals_wong_2010}.

\subsubsection{Connection formulas and degeneracies}
Since the hypergeometric equation is second order it only has two linearly independent solutions, so there is a linear relation between any three of the $24$ solutions described above. These relations are captured by so-called connection formulas, which describe how to write a solution with a given set of fall-offs near one of the singular points in terms of solutions with definite fall-offs near other singularities. This is precisely the question of interest in the computation of black hole Love numbers. We will not tabulate all the connection formulas here, but they can be found in many places, for example in~\cite{Bateman:100233} and in the~\href{https://dlmf.nist.gov/15.10#ii}{\tt DLMF}.

A phenomenon that often arises in the computation of Love numbers is that solutions that are linearly independent for generic parameter values become accidentally degenerate for particular parameter choices. This happens if at least one of the numbers $a$, $b$, $c-a$ or $c-b$ is an integer. In these cases, one should use one of the other $24$ solutions that has the desired fall-off conditions and is linearly independent. For the derivation of the two linearly independent solutions in these degenerate cases and their characterization we refer,  e.g.,~to \cite{Bateman:100233,beals_wong_2010}.

\subsection{Formulas for the computation of static solutions}
After these generalities about the hypergeometric function, we now list some properties and formulas particularly useful in the computation of static solutions for linearized perturbations in a Schwarzschild spacetime. In many cases the parameters of interest correspond to situations where the naively linearly independent solutions to the hypergeometric equation degenerate, and we need to find a set of linearly independent solutions in order to solve the boundary value problem.

In general, the problem we are interested in is to find the solution that is regular at the black hole horizon (which sits at $z=1$) and then expand this solution around $r\to\infty$ (which sits at $z=0$) and read off the ratio of the two fall-offs, which encodes the linear response to an external perturbation.
An interesting feature of the $D=4$ case is that the solution that has a single fall-off near the $z=1$ singularity (corresponding to the black hole horizon) {\it also} has a single fall-off near the $z=0$ singularity (corresponding to $r=\infty$). This is somewhat non-generic and ultimately, this is the underlying mathematical fact behind the vanishing of black hole Love numbers in four dimensions.

We now review the main properties of the hypergeometric function and its asymptotic expansions in the cases  relevant for the computation of static solutions. We will leave  $a$, $b$ and $c$ generic, with the only assumptions that $a,b,c>0$ and that $c=a+b$. We then split the computation into four cases, distinguishing between degenerate and non-degenerate cases, depending on whether the parameters take integer or non-integer values.

\subsubsection{Case 1: $a$, $b$ and $c$ are all non-integers}
Let us start by assuming that $a$, $b$ and $c$ are non-integers. Since $c=a+b$, this means that none of the  numbers $a$, $b$, $c-a$, $c-b$ and $c$   is integer. In this case,  the two linearly independent solutions to the hypergeometric equation \eqref{app:hyperEq} around $x=0$ are \cite{Bateman:100233}
\be
\def\arraystretch{.8}
u_1(z) = \hypergeom{2}{1}\left[\begin{array}{c}
a,~~b\\
c
\end{array}\Big\rvert \,z\,\right]~~~{\rm and}~~~u_5(z) = z^{1-c }\hypergeom{2}{1}\left[\begin{array}{c}
a-c+1,~~b-c+1\\
2-c
\end{array}\Big\rvert \,z\,\right] .
\label{hypergcase1}
\ee
Notice that, since $c=a+b$, it is not hard to find the linear combination of \eqref{hypergcase1} that is regular at $z=1$. To this end, we can start from the general  identity
\be
\begin{aligned}
\def\arraystretch{.8}
\hypergeom{2}{1}\left[\begin{array}{c}
a,~~b\\
a+b+m
\end{array}\Big\rvert \,z\,\right]
= &~\frac{\Gamma(a+b+m)}{\Gamma(a+m)\Gamma(b+m)}\sum_{k=0}^{m-1}\frac{(a)_k(b)_k}{k!}(m-k-1)!(z-1)^k
\\
&- \frac{\Gamma(a+b+m)}{\Gamma(a)\Gamma(b)}(z-1)^m\sum_{k=0}^\infty  \frac{(a+m)_k(b+m)_k}{k!(k+m)!}(1-z)^k
\\
&\times \Big[\log(1-z) - \psi(k+1) -\psi(k+m+1) +\psi(a+k+m) + \psi(b+k+m)  \Big] ,
\end{aligned}
\label{app:id1}
\ee
 where $\psi(k)$ is the digamma function, $\psi(k)=\frac{\Gamma'(k)}{\Gamma(k)}$ and where the sum $\sum_{k=0}^{m-1}$ should be replaced with zero if $m=0$. 
In this case---which is the one we are interested in, since $c=a+b$---eq.~\eqref{app:id1} in the limit $z\rightarrow1$  becomes
\begin{equation}
\def\arraystretch{.8}
\hypergeom{2}{1}\left[\begin{array}{c}
a,~~b\\
a+b
\end{array}\Big\rvert \,z\,\right] \xrightarrow{z\rightarrow 1} 
- \frac{\Gamma(a+b)}{\Gamma(a)\Gamma(b)}\log(1-z) + \text{finite terms} \, .
\label{app:id1bis}
\end{equation}
In order to cancel the logarithmic divergence, one should therefore take the following linear combination in \eqref{hypergcase1}:
\begin{equation}
\def\arraystretch{.8}
u(z) = \frac{\Gamma(2-c)}{\Gamma(a-c+1)\Gamma(b-c+1)}\hypergeom{2}{1}\left[\begin{array}{c}
a,~~b\\
c
\end{array}\Big\rvert \,z\,\right]
- \frac{\Gamma(a+b)}{\Gamma(a)\Gamma(b)} z^{1-c }\hypergeom{2}{1}\left[\begin{array}{c}
a-c+1,~~b-c+1\\
2-c
\end{array}\Big\rvert \,z\,\right] \, ,
\label{hypergcase1-regular}
\end{equation}
which is regular at $z=1$. We can then ask about the behavior of this solution near $z=0$.
In the small-$z$ limit the leading terms of~\eqref{hypergcase1-regular} are
\begin{equation}
 u(z) \overset{z\to 0}{\sim} \frac{\Gamma(2-c)}{\Gamma(a-c+1)\Gamma(b-c+1)}
- \frac{\Gamma(a+b)}{\Gamma(a)\Gamma(b)} z^{1-c } + \mathcal{O}(z) \, ,
\label{hypergcase1-regular-smallx}
\end{equation}
where the two terms correspond to the two terms in~\eqref{hypergcase1-regular}.

Note that it is crucial here that $c$ is non-integer. This guarantees that the none of the powers of the Taylor expansion of the first term in \eqref{hypergcase1-regular} appears in the expansion of the second term around $x=0$, and vice versa. This allows us to cleanly separate the two expansions and  keep only the terms that we wrote in \eqref{hypergcase1-regular-smallx} in order to compute the ratio of solutions and extract Love numbers. 
Note also that, in the explicit  cases discussed  in the main text,  $z$ roughly corresponds to some inverse power of  the radial coordinate. Thus,  eq.~\eqref{hypergcase1-regular-smallx} effectively encodes  the large-distance behavior of the solution that is regular at the horizon. Then, we can define the (dimensionless) Love numbers associated with the non-degenerate equation \eqref{app:hyperEq}  as  the  ratio between the two asymptotics,
\begin{equation}
k =- \frac{\Gamma(2-c)}{\Gamma(1-b)\Gamma(1-a)}
 \frac{\Gamma(a)\Gamma(b)}{ \Gamma(a+b)  }  \, .
\label{app:defLove0}
\end{equation}
If none of the numbers $a$, $b$ and $c$ is integer, one could  use the identities in Sec.~\ref{app:id} below and  $c=a+b$  to rewrite
\begin{tcolorbox}[colframe=white,arc=0pt,colback=greyish2]
\begin{equation}
k = (a+b-1)
 \frac{\Gamma(a)^2\Gamma(b)^2}{ \Gamma(a+b)^2  } \frac{\sin(\pi a ) \sin(\pi b)}{\pi \sin [\pi(a+b)]} \, ,
\label{generallove}
\end{equation}
\end{tcolorbox}
\noindent
which is the formula that we will use in the main text.

\subsubsection{Case 2: $a$ and $b$  are non-integers,  $c$ is an integer}
Next we consider the case where $a$, $b$, $c-a$ and $c-b$ are not integer but $c$ is a positive integer ($c\neq 0,-1,-2,\cdots$). In this case  the two independent solutions are~\cite{Bateman:100233}
\begin{equation}
\def\arraystretch{.8}
u_1(z) = \hypergeom{2}{1}\left[\begin{array}{c}
a,~~b\\
c
\end{array}\Big\rvert \,z\,\right]~~~{\rm and}~~~
u_2(z) =  \hypergeom{2}{1}\left[\begin{array}{c}
a,~~~~b\\
1+a+b-c
\end{array}\Big\rvert \,1-z\,\right]\,,
\label{halfinteger1}
\end{equation}
if $1+a+b-c \neq 0,-1,-2,\cdots$, and 
\begin{equation}
\def\arraystretch{.8}
u_1(z) = \hypergeom{2}{1}\left[\begin{array}{c}
a,~~b\\
c
\end{array}\Big\rvert \,z\,\right]~~~{\rm and}~~~
u_6(z) = (1-z)^{c-a-b } \hypergeom{2}{1}\left[\begin{array}{c}
c-b,~~c-a\\
1+c-a-b
\end{array}\Big\rvert \,1-z\,\right],
\end{equation}
if instead $1+a+b-c = 0,-1,-2,\cdots$
Note that all the equations discussed in the main text belong to the first case and, therefore, the independent solutions are those enumerated in~\eqref{halfinteger1}. In this particular case, only the second solution in~\eqref{halfinteger1} is regular at $z=1$. To extract the behavior near the other singular point $z=0$, one can use the following series formula for the hypergeometric function (valid for this choice of parameters)
\be
\def\arraystretch{.8}
\begin{aligned}
\hypergeom{2}{1}\left[\begin{array}{c}
a,~~~~b\\
1+a+b-c
\end{array}\Big\rvert \,1-z\,\right]&=\hypergeom{2}{1}\left[\begin{array}{c}
a,~~b\\
c
\end{array}\Big\rvert \,z\,\right]\log z - \sum_{n=1}^{c-1} \frac{(c-1)!(n-1)!}{(c-n-1)!(1-a)_n(1-b)_n}(-z)^{-n}
\\
&~~~+ \sum_{n=0}^{\infty}\frac{(a)_n(b)_n}{(c)_n n!} \Big[\psi(a+n) + \psi(b+n) - \psi(1+n) - \psi(c+n)\Big]  z^n \, ,
\label{half-integer2}
\end{aligned}
\ee
where  $(k)_n = \frac{\Gamma(k+n)}{\Gamma(k)}$ if $k\neq 0,-1,-2,\cdots$, and where $\psi(z) \equiv \Gamma'(z)/\Gamma(z)$ is the digamma function.  Notice that the appearance  of the $\log z$ in \eqref{half-integer2} makes manifest that this second solution is linearly independent from the first one in  \eqref{halfinteger1}, which is instead regular around $z=0$. 
Again, one can focus on the terms with the same powers of those considered in \eqref{hypergcase1-regular-smallx},
\begin{equation}
 u_2(z) \overset{z\to 0}{\sim}\log z + \gamma_\text{E}  + \psi(a) + \psi(b)  - \psi(c) 
+ (-1)^c (c-1)!(c-2)! \frac{\Gamma(1-a)}{\Gamma(c-a)}\frac{\Gamma(1-b)}{\Gamma(c-b)}z^{1-c} \, ,
\label{eq:expdeg}
\end{equation}
where $\gamma_\text{E}$  is the Euler--Mascheroni constant. The expansion \eqref{eq:expdeg} is substantially different from the one in \eqref{hypergcase1-regular-smallx}, which we obtained in the non-degenerate case, in particular because of the $\log (z)$ term. In \cite{Kol:2011vg} this term was interpreted as a logarithmic running of the Love numbers. Note that the way the calculation in the degenerate cases is organized in \cite{Kol:2011vg} is slightly different from what we discussed here. In \cite{Kol:2011vg}, the Love numbers corresponding to cases in which $c$ is integer were defined starting from the general expression~\eqref{generallove} and then taking the limit in which $c=a+b$ becomes integer. Equation~\eqref{generallove} is clearly singular in  this limit, therefore one needs to introduce a regularization procedure that gets rid of the divergence. As a byproduct, one ends up with a logarithmic term, whose coefficient can be interpreted as a beta-function in a RG-flow sense~\cite{Kol:2011vg}. In our case the logarithm appears from finding two linearly independent solutions to the hypergeometric equation. One can think of this as a different regularization procedure from~\cite{Kol:2011vg}. 
Because of the logarithm, the Love numbers are ambiguous, and change with the distance at which one measures the response. However, the coefficient of the logarithm is universal and unambiguous, and indeed this coefficient matches between our computation and~\cite{Kol:2011vg}, as it must.

\subsubsection{Case 3: $a$, $b$ and $c$ are all integers}
\label{app:abcints}

This case is  degenerate. We will  assume that $a\geq1$ and that $b-a \equiv l$ is a non-negative integer (see the main text for explicit cases). The two linearly independent solutions are (recall $a+b=c$)\footnote{This case corresponds to line 20 of the table in Sec.~2.2.2 of \cite{Bateman:100233}, with $m=n$. The two independent solutions can be found in eqs.~2.9(1) and 2.9(13) of \cite{Bateman:100233}. Note that there is a typo in the case 20 of the table in Sec.~2.2.2: the ``$u_2$'' should be instead ``$u_4$''. }
\begin{equation}
\def\arraystretch{.8}
u_1(z) = \hypergeom{2}{1}\left[\begin{array}{c}
a,~~b\\
c
\end{array}\Big\rvert \,z\,\right]~~~{\rm and}~~~
u_4(z)=(-z)^{-b}\hypergeom{2}{1}\left[\begin{array}{c}
1-a,~~b\\
l+1
\end{array}\Big\rvert \,\frac{1}{z}\,\right]\, .
\label{case2indsols}
\end{equation}

Since the first argument of $u_4$ is a non-positive integer, $1-a=0,-1,-2,\cdots$, and the third argument $l+1$ is a positive number, we can use eq.~\eqref{ana1} to rewrite  the second solution in \eqref{case2indsols}  as
\begin{equation}
\def\arraystretch{.8}
(-z)^{-b}\hypergeom{2}{1}\left[\begin{array}{c}
1-a,~~b\\
l+1
\end{array}\Big\rvert \,\frac{1}{z}\,\right] = (-z)^{-b }\sum_{n=0}^{a-1} \frac{(1-a)_n(b)_n}{(l+1)_n n!}z^{-n} \, .
\label{case3appreg}
\end{equation}
Notice that only this second solution is regular at $z=1$ (the first one in \eqref{case2indsols} contains a logarithmic divergence). This is therefore the solution that we would keep to describe physical perturbations around a Schwarzschild black hole (recall that, in the notation of the main text, the horizon is located at $z=1$). 
Interestingly~\eqref{case3appreg} is a polynomial with only positive powers of $r$ (remember that $z$ goes like  inverse powers of $r$), namely it does not retain the fall-off behavior \eqref{hypergcase1-regular-smallx} of the non-degenerate regular solution \eqref{hypergcase1-regular}. Another way of saying this is that it only has growing mode behavior near $z=0$, without any response. This indeed corresponds to a vanishing of the Love number \eqref{generallove}. This result can be equivalently understood from eq.~\eqref{generallove} by replacing $a\mapsto a+\epsilon$ and $b\mapsto b+\epsilon$, and then sending $\epsilon\rightarrow 0$~\cite{Kol:2011vg}.

\subsubsection{Case 4: either $a$ or  $b$ is integer, while  $c$ is non-integer} 
\label{app:aorbintcnonint}

Finally, 
let us assume that $a$ is integer, while both $b$ and $c$ are non-integers. This case is also degenerate and the two independent solutions are:\footnote{This degenerate case is discussed e.g.,~in \cite{Bateman:100233}, see line 8 of the table in Sec.~2.2.2, with $l=0$. The two independent solutions can be read off from eqs.~2.9(1) and 2.9(18) of \cite{Bateman:100233}.}
\begin{equation}
\def\arraystretch{.8}
u_1(z) = \hypergeom{2}{1}\left[\begin{array}{c}
a,~~b\\
c
\end{array}\Big\rvert \,z\,\right]~~~{\rm and}~~~
u_5(z) = z^{1-c }\hypergeom{2}{1}\left[\begin{array}{c}
1-a,~~1-b\\
2-c
\end{array}\Big\rvert \,z\,\right] .
\end{equation}
Only the second solution $u_5$ is regular at $z=1$. Since its first argument is negative and $c$ is not a non-positive integer, we can still use the formula \eqref{ana1} and rewrite it as
\begin{equation}
\def\arraystretch{.8}
z^{1-c }\hypergeom{2}{1}\left[\begin{array}{c}
1-a,~~1-b\\
2-c
\end{array}\Big\rvert \,z\,\right]
=  z^{1-c } \sum_{n=0}^{a-1} \frac{(1-a)_n(1-b)_n}{(2-c)_n n!}z^n \, .
\label{case4ex}
\end{equation}
Comparing with \eqref{hypergcase1-regular-smallx}, it is clear that \eqref{case4ex} contains  the $z^{1-c }$ term but not the constant one. As a result, the Love numbers defined as in \eqref{app:defLove0} vanish. Notice that, if $b>0$, as in all particular cases discussed in the main text, then \eqref{case4ex} is a polynomial with only negative powers of $z$.

\subsection{Useful identities}
\label{app:id}
Here we collect some useful formulas used in the previous section and in the main text.

If $a=-m$ with $m$  a non-negative integer and if $c\neq 0,-1,-2,\cdots$, then the hypergeometric function $\hypergeom{2}{1} \left(a,b;c;x\right)$ takes the form of a polynomial:
\begin{equation}
\def\arraystretch{.8}
\hypergeom{2}{1}\left[\begin{array}{c}
-m,~~b\\
c
\end{array}\Big\rvert \,z\,\right]
 = \sum_{n=0}^m \frac{(-m)_n(b)_n}{(c)_n n!}z^n \, .
\label{ana1}
\end{equation}
The hypergeometric function satisfies the following difference equations---see e.g.~\cite{slavjanov2000special}:
\begin{equation}
\def\arraystretch{.8}
\begin{aligned}
z\hypergeom{2}{1}\left[\begin{array}{c}
a,~~b\\
c
\end{array}\Big\rvert \,z\,\right]& =
\frac{a(b-c)}{(a-b)(a-b+1)} \hypergeom{2}{1}\left[\begin{array}{c}
a+1,~~b-1\\
c
\end{array}\Big\rvert \,z\,\right]
\\
& \quad
+\frac{c(a+b-1)-2ab}{(a-b+1)(a-b-1)}\hypergeom{2}{1}\left[\begin{array}{c}
a,~~b\\
c
\end{array}\Big\rvert \,z\,\right]
\\
& \quad
+\frac{b(a-c)}{(a-b)(a-b-1)} \hypergeom{2}{1}\left[\begin{array}{c}
a-1,~~b+1\\
c
\end{array}\Big\rvert \,z\,\right] \, ,
\end{aligned}
\label{diffeq1}
\end{equation}
and
\begin{equation}
\def\arraystretch{.8}
\begin{aligned}
z(1-z) \frac{\D}{\D z}\hypergeom{2}{1}\left[\begin{array}{c}
a,~~b\\
c
\end{array}\Big\rvert \,z\,\right]& =
\frac{ab(b-c)}{(a-b)(a-b+1)}\hypergeom{2}{1}\left[\begin{array}{c}
a+1,~~b-1\\
c
\end{array}\Big\rvert \,z\,\right]
\\
& \quad
+\frac{ab(2c-a-b-1)}{(a-b+1)(a-b-1)}\hypergeom{2}{1}\left[\begin{array}{c}
a,~~b\\
c
\end{array}\Big\rvert \,z\,\right]
\\
& \quad
+\frac{ab(a-c)}{(a-b)(a-b-1)} \hypergeom{2}{1}\left[\begin{array}{c}
a-1,~~b+1\\
c
\end{array}\Big\rvert \,z\,\right]\, .
\end{aligned}
\label{diffeq2}
\end{equation}
In the previous sections, when manipulating gamma functions we made use of the  Legendre duplication formula,
\begin{equation}
\Gamma(2z) = \frac{2^{2z-1}}{\sqrt{\pi}}\Gamma(z)\Gamma\left(z+\tfrac{1}{2}\right)
\qquad
\text{if } 2z\neq 0,-1,-2,\cdots
\label{Legendredup}
\end{equation}
and Euler's reflection formula,
\begin{equation}
\Gamma(1-z)\Gamma(z) = -z \Gamma(-z)\Gamma(z) = \frac{\pi}{\sin(\pi z)} \qquad \text{if } ~~z \, \notin \, \mathbb{Z} \, ,
\label{Eulerf}
\end{equation}
which are useful for simplifying the various expressions that appear.

\newpage
\renewcommand{\em}{}
\bibliographystyle{utphys}
\addcontentsline{toc}{section}{References}
\bibliography{notesbib}

\providecommand{\href}[2]{#2}\begingroup\raggedright\begin{thebibliography}{100}

\bibitem{Kerr:1963ud}
R.~P. Kerr, ``{Gravitational field of a spinning mass as an example of
  algebraically special metrics},''
  \href{http://dx.doi.org/10.1103/PhysRevLett.11.237}{{\em Phys. Rev. Lett.}
  {\bf 11} (1963)  237--238}.

\bibitem{Newman:1965my}
E.~T. Newman, R.~Couch, K.~Chinnapared, A.~Exton, A.~Prakash, and R.~Torrence,
  ``{Metric of a Rotating, Charged Mass},''
  \href{http://dx.doi.org/10.1063/1.1704351}{{\em J. Math. Phys.} {\bf 6}
  (1965)  918--919}.

\bibitem{Israel:1967wq}
W.~Israel, ``{Event horizons in static vacuum space-times},''
  \href{http://dx.doi.org/10.1103/PhysRev.164.1776}{{\em Phys. Rev.} {\bf 164}
  (1967)  1776--1779}.

\bibitem{Carter:1971zc}
B.~Carter, ``{Axisymmetric Black Hole Has Only Two Degrees of Freedom},''
  \href{http://dx.doi.org/10.1103/PhysRevLett.26.331}{{\em Phys. Rev. Lett.}
  {\bf 26} (1971)  331--333}.

\bibitem{Bekenstein:1971hc}
J.~D. Bekenstein, ``{Nonexistence of baryon number for static black holes},''
  \href{http://dx.doi.org/10.1103/PhysRevD.5.1239}{{\em Phys. Rev. D} {\bf 5}
  (1972)  1239--1246}.

\bibitem{Bekenstein:1995un}
J.~Bekenstein, ``{Novel ``no-scalar-hair" theorem for black holes},''
  \href{http://dx.doi.org/10.1103/PhysRevD.51.R6608}{{\em Phys. Rev. D} {\bf
  51} (1995) no.~12, 6608}.

\bibitem{Hui:2012qt}
L.~Hui and A.~Nicolis, ``{No-Hair Theorem for the Galileon},''
  \href{http://dx.doi.org/10.1103/PhysRevLett.110.241104}{{\em Phys. Rev.
  Lett.} {\bf 110} (2013)  241104}, \href{http://arxiv.org/abs/1202.1296}{{\tt
  arXiv:1202.1296 [hep-th]}}.

\bibitem{Jacobson:1999vr}
T.~Jacobson, ``{Primordial black hole evolution in tensor scalar cosmology},''
  \href{http://dx.doi.org/10.1103/PhysRevLett.83.2699}{{\em Phys. Rev. Lett.}
  {\bf 83} (1999)  2699--2702},
  \href{http://arxiv.org/abs/astro-ph/9905303}{{\tt arXiv:astro-ph/9905303}}.

\bibitem{Horbatsch:2011ye}
M.~Horbatsch and C.~Burgess, ``{Cosmic Black-Hole Hair Growth and Quasar
  OJ287},'' \href{http://dx.doi.org/10.1088/1475-7516/2012/05/010}{{\em JCAP}
  {\bf 05} (2012)  010}, \href{http://arxiv.org/abs/1111.4009}{{\tt
  arXiv:1111.4009 [gr-qc]}}.

\bibitem{Hui:2019aqm}
L.~Hui, D.~Kabat, X.~Li, L.~Santoni, and S.~S. Wong, ``{Black Hole Hair from
  Scalar Dark Matter},''
  \href{http://dx.doi.org/10.1088/1475-7516/2019/06/038}{{\em JCAP} {\bf 06}
  (2019)  038}, \href{http://arxiv.org/abs/1904.12803}{{\tt arXiv:1904.12803
  [gr-qc]}}.

\bibitem{Clough:2019jpm}
K.~Clough, P.~G. Ferreira, and M.~Lagos, ``{Growth of massive scalar hair
  around a Schwarzschild black hole},''
  \href{http://dx.doi.org/10.1103/PhysRevD.100.063014}{{\em Phys. Rev. D} {\bf
  100} (2019) no.~6, 063014}, \href{http://arxiv.org/abs/1904.12783}{{\tt
  arXiv:1904.12783 [gr-qc]}}.

\bibitem{Penrose:1969pc}
R.~Penrose, ``{Gravitational collapse: The role of general relativity},''
  \href{http://dx.doi.org/10.1023/A:1016578408204}{{\em Riv. Nuovo Cim.} {\bf
  1} (1969)  252--276}.

\bibitem{1971JETPL..14..180Z}
Y.~B. {Zel'Dovich}, ``{Generation of Waves by a Rotating Body},''{\em Soviet
  Journal of Experimental and Theoretical Physics Letters} {\bf 14} (Aug.,
  1971)  180.

\bibitem{1972JETP...35.1085Z}
Y.~B. {Zel'Dovich}, ``{Amplification of Cylindrical Electromagnetic Waves
  Reflected from a Rotating Body},''{\em Soviet Journal of Experimental and
  Theoretical Physics} {\bf 35} (Jan., 1972)  1085.

\bibitem{1972BAPS...17..472M}
C.~{Misner}, ``{Stability of Kerr black holes against scalar perturbations},''
  in {\em Bulletin of the American Physical Society}, vol.~17, p.~472.
\newblock Dec., 1972.

\bibitem{Bardeen:1972fi}
J.~M. Bardeen, W.~H. Press, and S.~A. Teukolsky, ``{Rotating black holes:
  Locally nonrotating frames, energy extraction, and scalar synchrotron
  radiation},'' \href{http://dx.doi.org/10.1086/151796}{{\em Astrophys. J.}
  {\bf 178} (1972)  347}.

\bibitem{Press:1972zz}
W.~H. Press and S.~A. Teukolsky, ``{Floating Orbits, Superradiant Scattering
  and the Black-hole Bomb},'' \href{http://dx.doi.org/10.1038/238211a0}{{\em
  Nature} {\bf 238} (1972)  211--212}.

\bibitem{Starobinsky:1973aij}
A.~Starobinsky, ``{Amplification of waves reflected from a rotating "black
  hole".},'' {\em Sov. Phys. JETP} {\bf 37} (1973) no.~1, 28--32.

\bibitem{Teukolsky:1974yv}
S.~Teukolsky and W.~Press, ``{Perturbations of a rotating black hole. III -
  Interaction of the hole with gravitational and electromagnet ic radiation},''
  \href{http://dx.doi.org/10.1086/153180}{{\em Astrophys. J.} {\bf 193} (1974)
  443--461}.

\bibitem{Arvanitaki:2010sy}
A.~Arvanitaki and S.~Dubovsky, ``{Exploring the String Axiverse with Precision
  Black Hole Physics},''
  \href{http://dx.doi.org/10.1103/PhysRevD.83.044026}{{\em Phys. Rev. D} {\bf
  83} (2011)  044026}, \href{http://arxiv.org/abs/1004.3558}{{\tt
  arXiv:1004.3558 [hep-th]}}.

\bibitem{Endlich:2016jgc}
S.~Endlich and R.~Penco, ``{A Modern Approach to Superradiance},''
  \href{http://dx.doi.org/10.1007/JHEP05(2017)052}{{\em JHEP} {\bf 05} (2017)
  052}, \href{http://arxiv.org/abs/1609.06723}{{\tt arXiv:1609.06723
  [hep-th]}}.

\bibitem{Baumann:2019eav}
D.~Baumann, H.~S. Chia, J.~Stout, and L.~ter Haar, ``{The Spectra of
  Gravitational Atoms},''
  \href{http://dx.doi.org/10.1088/1475-7516/2019/12/006}{{\em JCAP} {\bf 12}
  (2019)  006}, \href{http://arxiv.org/abs/1908.10370}{{\tt arXiv:1908.10370
  [gr-qc]}}.

\bibitem{Cardoso:2017cfl}
V.~Cardoso, E.~Franzin, A.~Maselli, P.~Pani, and G.~Raposo, ``{Testing
  strong-field gravity with tidal Love numbers},''
  \href{http://dx.doi.org/10.1103/PhysRevD.95.084014}{{\em Phys. Rev. D} {\bf
  95} (2017) no.~8, 084014}, \href{http://arxiv.org/abs/1701.01116}{{\tt
  arXiv:1701.01116 [gr-qc]}}. [Addendum: Phys.Rev.D 95, 089901 (2017)].

\bibitem{Regge:1957td}
T.~Regge and J.~A. Wheeler, ``{Stability of a Schwarzschild singularity},''
\href{http://dx.doi.org/10.1103/PhysRev.108.1063}{{\em Phys. Rev.} {\bf 108}
  (1957)  1063--1069}.
%%CITATION = PHRVA,108,1063;%%.

\bibitem{Zerilli:1970se}
F.~J. Zerilli, ``{Effective potential for even parity Regge-Wheeler
  gravitational perturbation equations},''
\href{http://dx.doi.org/10.1103/PhysRevLett.24.737}{{\em Phys. Rev. Lett.} {\bf
  24} (1970)  737--738}.
%%CITATION = PRLTA,24,737;%%.

\bibitem{Zerilli:1971wd}
F.~J. Zerilli, ``{Gravitational field of a particle falling in a schwarzschild
  geometry analyzed in tensor harmonics},''
\href{http://dx.doi.org/10.1103/PhysRevD.2.2141}{{\em Phys. Rev.} {\bf D2}
  (1970)  2141--2160}.
%%CITATION = PHRVA,D2,2141;%%.

\bibitem{Teukolsky:1972my}
S.~Teukolsky, ``{Rotating black holes - separable wave equations for
  gravitational and electromagnetic perturbations},''
  \href{http://dx.doi.org/10.1103/PhysRevLett.29.1114}{{\em Phys. Rev. Lett.}
  {\bf 29} (1972)  1114--1118}.

\bibitem{Teukolsky:1973ha}
S.~A. Teukolsky, ``{Perturbations of a rotating black hole. 1. Fundamental
  equations for gravitational electromagnetic and neutrino field
  perturbations},'' \href{http://dx.doi.org/10.1086/152444}{{\em Astrophys. J.}
  {\bf 185} (1973)  635--647}.

\bibitem{Chandrasekhar:1985kt}
S.~Chandrasekhar, ``{The mathematical theory of black holes},'' in {\em
  {Oxford, UK: Clarendon (1992) 646 p., Oxford, UK: Clarendon (1985) 646 P.}}
\newblock
1985.
\newblock
%%CITATION = INSPIRE-224457;%%.

\bibitem{Krtous:2018bvk}
P.~Krtou\v{s}, V.~P. Frolov, and D.~Kubiz\v{n}\'ak, ``{Separation of Maxwell
  equations in Kerr\textendash{}NUT\textendash{}(A)dS spacetimes},''
  \href{http://dx.doi.org/10.1016/j.nuclphysb.2018.06.019}{{\em Nucl. Phys. B}
  {\bf 934} (2018)  7--38}, \href{http://arxiv.org/abs/1803.02485}{{\tt
  arXiv:1803.02485 [hep-th]}}.

\bibitem{Dolan:2018dqv}
S.~R. Dolan, ``{Instability of the Proca field on Kerr spacetime},''
  \href{http://dx.doi.org/10.1103/PhysRevD.98.104006}{{\em Phys. Rev. D} {\bf
  98} (2018) no.~10, 104006}, \href{http://arxiv.org/abs/1806.01604}{{\tt
  arXiv:1806.01604 [gr-qc]}}.

\bibitem{Chandrasekhar:1975zza}
S.~Chandrasekhar and S.~L. Detweiler, ``{The quasi-normal modes of the
  Schwarzschild black hole},''
  \href{http://dx.doi.org/10.1098/rspa.1975.0112}{{\em Proc. Roy. Soc. Lond. A}
  {\bf 344} (1975)  441--452}.

\bibitem{Leaver:1985ax}
E.~Leaver, ``{An Analytic representation for the quasi normal modes of Kerr
  black holes},'' \href{http://dx.doi.org/10.1098/rspa.1985.0119}{{\em Proc.
  Roy. Soc. Lond. A} {\bf 402} (1985)  285--298}.

\bibitem{Nollert:1999ji}
H.-P. Nollert, ``{Topical Review: Quasinormal modes: the characteristic `sound'
  of black holes and neutron stars},''
\href{http://dx.doi.org/10.1088/0264-9381/16/12/201}{{\em Class. Quant. Grav.}
  {\bf 16} (1999)  R159--R216}.
%%CITATION = CQGRD,16,R159;%%.

\bibitem{Kokkotas:1999bd}
K.~D. Kokkotas and B.~G. Schmidt, ``{Quasinormal modes of stars and black
  holes},'' \href{http://dx.doi.org/10.12942/lrr-1999-2}{{\em Living Rev. Rel.}
  {\bf 2} (1999)  2}, \href{http://arxiv.org/abs/gr-qc/9909058}{{\tt
  arXiv:gr-qc/9909058}}.

\bibitem{Berti:2009kk}
E.~Berti, V.~Cardoso, and A.~O. Starinets, ``{Quasinormal modes of black holes
  and black branes},''
  \href{http://dx.doi.org/10.1088/0264-9381/26/16/163001}{{\em Class. Quant.
  Grav.} {\bf 26} (2009)  163001},
\href{http://arxiv.org/abs/0905.2975}{{\tt arXiv:0905.2975 [gr-qc]}}.
%%CITATION = ARXIV:0905.2975;%%.

\bibitem{Love}
A.~E.~H. Love, ``The yielding of the earth to disturbing forces,''
  \href{http://dx.doi.org/10.1098/rspa.1909.0008}{{\em Proceedings of the Royal
  Society of London. Series A, Containing Papers of a Mathematical and Physical
  Character} {\bf 82} (1909) no.~551, 73--88}.

\bibitem{Flanagan:2007ix}
E.~E. Flanagan and T.~Hinderer, ``{Constraining neutron star tidal Love numbers
  with gravitational wave detectors},''
  \href{http://dx.doi.org/10.1103/PhysRevD.77.021502}{{\em Phys. Rev. D} {\bf
  77} (2008)  021502}, \href{http://arxiv.org/abs/0709.1915}{{\tt
  arXiv:0709.1915 [astro-ph]}}.

\bibitem{Hinderer:2007mb}
T.~Hinderer, ``{Tidal Love numbers of neutron stars},''
  \href{http://dx.doi.org/10.1086/533487}{{\em Astrophys. J.} {\bf 677} (2008)
  1216--1220}, \href{http://arxiv.org/abs/0711.2420}{{\tt arXiv:0711.2420
  [astro-ph]}}.

\bibitem{Chirenti:2020bas}
C.~Chirenti, C.~Posada, and V.~Guedes, ``{Where is Love? Tidal deformability in
  the black hole compactness limit},''
  \href{http://dx.doi.org/10.1088/1361-6382/abb07a}{{\em Class. Quant. Grav.}
  {\bf 37} (2020) no.~19, 195017}, \href{http://arxiv.org/abs/2005.10794}{{\tt
  arXiv:2005.10794 [gr-qc]}}.

\bibitem{Brustein:2020tpg}
R.~Brustein and Y.~Sherf, ``{Quantum Love},''
  \href{http://arxiv.org/abs/2008.02738}{{\tt arXiv:2008.02738 [gr-qc]}}.

\bibitem{Damour:2009vw}
T.~Damour and A.~Nagar, ``{Relativistic tidal properties of neutron stars},''
  \href{http://dx.doi.org/10.1103/PhysRevD.80.084035}{{\em Phys. Rev. D} {\bf
  80} (2009)  084035}, \href{http://arxiv.org/abs/0906.0096}{{\tt
  arXiv:0906.0096 [gr-qc]}}.

\bibitem{Binnington:2009bb}
T.~Binnington and E.~Poisson, ``{Relativistic theory of tidal Love numbers},''
  \href{http://dx.doi.org/10.1103/PhysRevD.80.084018}{{\em Phys. Rev.} {\bf
  D80} (2009)  084018},
\href{http://arxiv.org/abs/0906.1366}{{\tt arXiv:0906.1366 [gr-qc]}}.
%%CITATION = ARXIV:0906.1366;%%.

\bibitem{Fang:2005qq}
H.~Fang and G.~Lovelace, ``{Tidal coupling of a Schwarzschild black hole and
  circularly orbiting moon},''
  \href{http://dx.doi.org/10.1103/PhysRevD.72.124016}{{\em Phys. Rev. D} {\bf
  72} (2005)  124016}, \href{http://arxiv.org/abs/gr-qc/0505156}{{\tt
  arXiv:gr-qc/0505156}}.

\bibitem{Kol:2011vg}
B.~Kol and M.~Smolkin, ``{Black hole stereotyping: Induced gravito-static
  polarization},'' \href{http://dx.doi.org/10.1007/JHEP02(2012)010}{{\em JHEP}
  {\bf 02} (2012)  010},
\href{http://arxiv.org/abs/1110.3764}{{\tt arXiv:1110.3764 [hep-th]}}.
%%CITATION = ARXIV:1110.3764;%%.

\bibitem{Chakrabarti:2013lua}
S.~Chakrabarti, T.~Delsate, and J.~Steinhoff, ``{New perspectives on neutron
  star and black hole spectroscopy and dynamic tides},''
  \href{http://arxiv.org/abs/1304.2228}{{\tt arXiv:1304.2228 [gr-qc]}}.

\bibitem{Gurlebeck:2015xpa}
N.~G\"urlebeck, ``{No-hair theorem for Black Holes in Astrophysical
  Environments},'' \href{http://dx.doi.org/10.1103/PhysRevLett.114.151102}{{\em
  Phys. Rev. Lett.} {\bf 114} (2015) no.~15, 151102},
  \href{http://arxiv.org/abs/1503.03240}{{\tt arXiv:1503.03240 [gr-qc]}}.

\bibitem{Cardoso:2019vof}
V.~Cardoso, L.~Gualtieri, and C.~J. Moore, ``{Gravitational waves and higher
  dimensions: Love numbers and Kaluza-Klein excitations},''
  \href{http://dx.doi.org/10.1103/PhysRevD.100.124037}{{\em Phys. Rev.} {\bf
  D100} (2019) no.~12, 124037},
\href{http://arxiv.org/abs/1910.09557}{{\tt arXiv:1910.09557 [gr-qc]}}.
%%CITATION = ARXIV:1910.09557;%%.

\bibitem{Emparan:2017qxd}
R.~Emparan, A.~Fernandez-Pique, and R.~Luna, ``{Geometric polarization of
  plasmas and Love numbers of AdS black branes},''
  \href{http://dx.doi.org/10.1007/JHEP09(2017)150}{{\em JHEP} {\bf 09} (2017)
  150}, \href{http://arxiv.org/abs/1707.02777}{{\tt arXiv:1707.02777
  [hep-th]}}.

\bibitem{Cardoso:2018ptl}
V.~Cardoso, M.~Kimura, A.~Maselli, and L.~Senatore, ``{Black Holes in an
  Effective Field Theory Extension of General Relativity},''
  \href{http://dx.doi.org/10.1103/PhysRevLett.121.251105}{{\em Phys. Rev.
  Lett.} {\bf 121} (2018) no.~25, 251105},
  \href{http://arxiv.org/abs/1808.08962}{{\tt arXiv:1808.08962 [gr-qc]}}.

\bibitem{Pani:2015hfa}
P.~Pani, L.~Gualtieri, A.~Maselli, and V.~Ferrari, ``{Tidal deformations of a
  spinning compact object},''
  \href{http://dx.doi.org/10.1103/PhysRevD.92.024010}{{\em Phys. Rev. D} {\bf
  92} (2015) no.~2, 024010}, \href{http://arxiv.org/abs/1503.07365}{{\tt
  arXiv:1503.07365 [gr-qc]}}.

\bibitem{Pani:2015nua}
P.~Pani, L.~Gualtieri, and V.~Ferrari, ``{Tidal Love numbers of a slowly
  spinning neutron star},''
  \href{http://dx.doi.org/10.1103/PhysRevD.92.124003}{{\em Phys. Rev. D} {\bf
  92} (2015) no.~12, 124003}, \href{http://arxiv.org/abs/1509.02171}{{\tt
  arXiv:1509.02171 [gr-qc]}}.

\bibitem{Landry:2015zfa}
P.~Landry and E.~Poisson, ``{Tidal deformation of a slowly rotating material
  body. External metric},''
  \href{http://dx.doi.org/10.1103/PhysRevD.91.104018}{{\em Phys. Rev. D} {\bf
  91} (2015)  104018}, \href{http://arxiv.org/abs/1503.07366}{{\tt
  arXiv:1503.07366 [gr-qc]}}.

\bibitem{Landry:2015cva}
P.~Landry and E.~Poisson, ``{Gravitomagnetic response of an irrotational body
  to an applied tidal field},''
  \href{http://dx.doi.org/10.1103/PhysRevD.91.104026}{{\em Phys. Rev. D} {\bf
  91} (2015) no.~10, 104026}, \href{http://arxiv.org/abs/1504.06606}{{\tt
  arXiv:1504.06606 [gr-qc]}}.

\bibitem{Landry:2017piv}
P.~Landry, ``{Tidal deformation of a slowly rotating material body: Interior
  metric and Love numbers},''
  \href{http://dx.doi.org/10.1103/PhysRevD.95.124058}{{\em Phys. Rev. D} {\bf
  95} (2017) no.~12, 124058}, \href{http://arxiv.org/abs/1703.08168}{{\tt
  arXiv:1703.08168 [gr-qc]}}.

\bibitem{Poisson:2020mdi}
E.~Poisson, ``{Gravitomagnetic Love tensor of a slowly rotating body:
  post-Newtonian theory},'' \href{http://arxiv.org/abs/2007.01678}{{\tt
  arXiv:2007.01678 [gr-qc]}}.

\bibitem{LeTiec:2020spy}
A.~Le~Tiec and M.~Casals, ``{Spinning Black Holes Fall in Love},''
  \href{http://arxiv.org/abs/2007.00214}{{\tt arXiv:2007.00214 [gr-qc]}}.

\bibitem{LeTiec:2020bos}
A.~Le~Tiec, M.~Casals, and E.~Franzin, ``{Tidal Love Numbers of Kerr Black
  Holes},'' \href{http://arxiv.org/abs/2010.15795}{{\tt arXiv:2010.15795
  [gr-qc]}}.

\bibitem{Chia:2020yla}
H.~S. Chia, ``{Tidal Deformation and Dissipation of Rotating Black Holes},''
  \href{http://arxiv.org/abs/2010.07300}{{\tt arXiv:2010.07300 [gr-qc]}}.

\bibitem{Goldberger:2020fot}
W.~D. Goldberger, J.~Li, and I.~Z. Rothstein, ``{Non-conservative effects on
  Spinning Black Holes from World-Line Effective Field Theory},''
  \href{http://arxiv.org/abs/2012.14869}{{\tt arXiv:2012.14869 [hep-th]}}.

\bibitem{Charalambous:2021mea}
P.~Charalambous, S.~Dubovsky, and M.~M. Ivanov, ``{On the Vanishing of Love
  Numbers for Kerr Black Holes},'' \href{http://arxiv.org/abs/2102.08917}{{\tt
  arXiv:2102.08917 [hep-th]}}.

\bibitem{Goldberger:2004jt}
W.~D. Goldberger and I.~Z. Rothstein, ``{An Effective field theory of gravity
  for extended objects},''
  \href{http://dx.doi.org/10.1103/PhysRevD.73.104029}{{\em Phys. Rev. D} {\bf
  73} (2006)  104029}, \href{http://arxiv.org/abs/hep-th/0409156}{{\tt
  arXiv:hep-th/0409156}}.

\bibitem{Goldberger:2005cd}
W.~D. Goldberger and I.~Z. Rothstein, ``{Dissipative effects in the worldline
  approach to black hole dynamics},''
  \href{http://dx.doi.org/10.1103/PhysRevD.73.104030}{{\em Phys. Rev. D} {\bf
  73} (2006)  104030}, \href{http://arxiv.org/abs/hep-th/0511133}{{\tt
  arXiv:hep-th/0511133}}.

\bibitem{Porto:2016zng}
R.~A. Porto, ``{The Tune of Love and the Nature(ness) of Spacetime},''
  \href{http://dx.doi.org/10.1002/prop.201600064}{{\em Fortsch. Phys.} {\bf 64}
  (2016) no.~10, 723--729}, \href{http://arxiv.org/abs/1606.08895}{{\tt
  arXiv:1606.08895 [gr-qc]}}.

\bibitem{LopezOrtega:2006vn}
A.~Lopez-Ortega, ``{Electromagnetic quasinormal modes of D-dimensional black
  holes},'' \href{http://dx.doi.org/10.1007/s10714-006-0358-2}{{\em Gen. Rel.
  Grav.} {\bf 38} (2006)  1747--1770},
\href{http://arxiv.org/abs/gr-qc/0605034}{{\tt arXiv:gr-qc/0605034 [gr-qc]}}.
%%CITATION = GR-QC/0605034;%%.

\bibitem{Rosa:2011my}
J.~G. Rosa and S.~R. Dolan, ``{Massive vector fields on the Schwarzschild
  spacetime: quasi-normal modes and bound states},''
  \href{http://dx.doi.org/10.1103/PhysRevD.85.044043}{{\em Phys. Rev. D} {\bf
  85} (2012)  044043}, \href{http://arxiv.org/abs/1110.4494}{{\tt
  arXiv:1110.4494 [hep-th]}}.

\bibitem{Avery:2016zce}
S.~G. Avery and B.~U.~W. Schwab, ``{Soft Black Hole Absorption Rates as
  Conservation Laws},'' \href{http://dx.doi.org/10.1007/JHEP04(2017)053}{{\em
  JHEP} {\bf 04} (2017)  053},
\href{http://arxiv.org/abs/1609.04397}{{\tt arXiv:1609.04397 [hep-th]}}.
%%CITATION = ARXIV:1609.04397;%%.

\bibitem{Kodama:2000fa}
H.~Kodama, A.~Ishibashi, and O.~Seto, ``{Brane world cosmology: Gauge invariant
  formalism for perturbation},''
  \href{http://dx.doi.org/10.1103/PhysRevD.62.064022}{{\em Phys. Rev. D} {\bf
  62} (2000)  064022}, \href{http://arxiv.org/abs/hep-th/0004160}{{\tt
  arXiv:hep-th/0004160}}.

\bibitem{Kodama:2003jz}
H.~Kodama and A.~Ishibashi, ``{A Master equation for gravitational
  perturbations of maximally symmetric black holes in higher dimensions},''
  \href{http://dx.doi.org/10.1143/PTP.110.701}{{\em Prog. Theor. Phys.} {\bf
  110} (2003)  701--722},
\href{http://arxiv.org/abs/hep-th/0305147}{{\tt arXiv:hep-th/0305147
  [hep-th]}}.
%%CITATION = HEP-TH/0305147;%%.

\bibitem{Ishibashi:2003ap}
A.~Ishibashi and H.~Kodama, ``{Stability of higher dimensional Schwarzschild
  black holes},'' \href{http://dx.doi.org/10.1143/PTP.110.901}{{\em Prog.
  Theor. Phys.} {\bf 110} (2003)  901--919},
\href{http://arxiv.org/abs/hep-th/0305185}{{\tt arXiv:hep-th/0305185
  [hep-th]}}.
%%CITATION = HEP-TH/0305185;%%.

\bibitem{uspaper2}
L.~Hui, A.~Joyce, R.~Penco, L.~Santoni, and A.~Solomon, ``{Symmetries of black
  hole perturbations},'' {\em To appear}  .

\bibitem{Penna:2018gfx}
R.~F. Penna, ``{Near-horizon Carroll symmetry and black hole Love numbers},''
  \href{http://arxiv.org/abs/1812.05643}{{\tt arXiv:1812.05643 [hep-th]}}.

\bibitem{Damour:2009va}
T.~Damour and O.~M. Lecian, ``{On the gravitational polarizability of black
  holes},'' \href{http://dx.doi.org/10.1103/PhysRevD.80.044017}{{\em Phys.
  Rev.} {\bf D80} (2009)  044017},
\href{http://arxiv.org/abs/0906.3003}{{\tt arXiv:0906.3003 [gr-qc]}}.
%%CITATION = ARXIV:0906.3003;%%.

\bibitem{dos2016relativistic}
G.~M. dos Santos~Raposo, ``Relativistic tidal love numbers: Tests of
  strong-field gravity,''.

\bibitem{1950SRToh..34..160N}
H.~{Nariai}, ``{On some static solutions of Einstein's gravitational field
  equations in a spherically symmetric case},''{\em Sci. Rep. Tohoku Univ.
  Eighth Ser.} {\bf 34} (Jan., 1950)  160.

\bibitem{10026018884}
H.~Nariai, ``On a new cosmological solution of einstein's field equations of
  gravitation,'' {\em Sci. Rep. Tohoku Univ. Ser. I} {\bf 35} (1951)  62.
  \url{https://ci.nii.ac.jp/naid/10026018884/en/}.

\bibitem{Tangherlini:1963bw}
F.~Tangherlini, ``{Schwarzschild field in n dimensions and the dimensionality
  of space problem},'' \href{http://dx.doi.org/10.1007/BF02784569}{{\em Nuovo
  Cim.} {\bf 27} (1963)  636--651}.

\bibitem{Cardoso:2004uz}
V.~Cardoso, O.~J. Dias, and J.~P. Lemos, ``{Nariai, Bertotti-Robinson and
  anti-Nariai solutions in higher dimensions},''
  \href{http://dx.doi.org/10.1103/PhysRevD.70.024002}{{\em Phys. Rev. D} {\bf
  70} (2004)  024002}, \href{http://arxiv.org/abs/hep-th/0401192}{{\tt
  arXiv:hep-th/0401192}}.

\bibitem{Thorne:1980ru}
K.~Thorne, ``{Multipole Expansions of Gravitational Radiation},''
  \href{http://dx.doi.org/10.1103/RevModPhys.52.299}{{\em Rev. Mod. Phys.} {\bf
  52} (1980)  299--339}.

\bibitem{Lagos:2013aua}
M.~Lagos, M.~Bañados, P.~G. Ferreira, and S.~García-Sáenz, ``{Noether
  Identities and Gauge-Fixing the Action for Cosmological Perturbations},''
  \href{http://dx.doi.org/10.1103/PhysRevD.89.024034}{{\em Phys. Rev.} {\bf
  D89} (2014)  024034},
\href{http://arxiv.org/abs/1311.3828}{{\tt arXiv:1311.3828 [gr-qc]}}.
%%CITATION = ARXIV:1311.3828;%%.

\bibitem{Motohashi:2016prk}
H.~Motohashi, T.~Suyama, and K.~Takahashi, ``{Fundamental theorem on gauge
  fixing at the action level},''
  \href{http://dx.doi.org/10.1103/PhysRevD.94.124021}{{\em Phys. Rev.} {\bf
  D94} (2016) no.~12, 124021},
\href{http://arxiv.org/abs/1608.00071}{{\tt arXiv:1608.00071 [gr-qc]}}.
%%CITATION = ARXIV:1608.00071;%%.

\bibitem{Deser:1976iy}
S.~Deser and C.~Teitelboim, ``{Duality Transformations of Abelian and
  Nonabelian Gauge Fields},''
  \href{http://dx.doi.org/10.1103/PhysRevD.13.1592}{{\em Phys. Rev. D} {\bf 13}
  (1976)  1592--1597}.

\bibitem{Moncrief:1974am}
V.~Moncrief, ``{Gravitational perturbations of spherically symmetric systems.
  I. The exterior problem.},''
  \href{http://dx.doi.org/10.1016/0003-4916(74)90173-0}{{\em Annals Phys.} {\bf
  88} (1974)  323--342}.

\bibitem{Cunningham:1978zfa}
C.~Cunningham, R.~Price, and V.~Moncrief, ``{Radiation from collapsing
  relativistic stars. I - Linearized odd-parity radiation},''
  \href{http://dx.doi.org/10.1086/156413}{{\em Astrophys. J.} {\bf 224} (1978)
  643}.

\bibitem{Cunningham:1979px}
C.~Cunningham, R.~Price, and V.~Moncrief, ``{Radiation From Collapsing
  Relativistic Stars. II. Linearized Even Parity Radiation},''
  \href{http://dx.doi.org/10.1086/157147}{{\em Astrophys. J.} {\bf 230} (1979)
  870--892}.

\bibitem{Rosen:2020crj}
R.~A. Rosen and L.~Santoni, ``{Black hole perturbations of massive and
  partially massless spin-2 fields in (anti) de Sitter spacetime},''
  \href{http://arxiv.org/abs/2010.00595}{{\tt arXiv:2010.00595 [hep-th]}}.

\bibitem{Franciolini:2018uyq}
G.~Franciolini, L.~Hui, R.~Penco, L.~Santoni, and E.~Trincherini, ``{Effective
  Field Theory of Black Hole Quasinormal Modes in Scalar-Tensor Theories},''
  \href{http://dx.doi.org/10.1007/JHEP02(2019)127}{{\em JHEP} {\bf 02} (2019)
  127}, \href{http://arxiv.org/abs/1810.07706}{{\tt arXiv:1810.07706
  [hep-th]}}.

\bibitem{Kobayashi:2012kh}
T.~Kobayashi, H.~Motohashi, and T.~Suyama, ``{Black hole perturbation in the
  most general scalar-tensor theory with second-order field equations I: the
  odd-parity sector},''
  \href{http://dx.doi.org/10.1103/PhysRevD.85.084025}{{\em Phys. Rev. D} {\bf
  85} (2012)  084025}, \href{http://arxiv.org/abs/1202.4893}{{\tt
  arXiv:1202.4893 [gr-qc]}}. [Erratum: Phys.Rev.D 96, 109903 (2017)].

\bibitem{Martel:2005ir}
K.~Martel and E.~Poisson, ``{Gravitational perturbations of the Schwarzschild
  spacetime: A Practical covariant and gauge-invariant formalism},''
  \href{http://dx.doi.org/10.1103/PhysRevD.71.104003}{{\em Phys. Rev. D} {\bf
  71} (2005)  104003}, \href{http://arxiv.org/abs/gr-qc/0502028}{{\tt
  arXiv:gr-qc/0502028}}.

\bibitem{Gibbons:2002pq}
G.~Gibbons and S.~A. Hartnoll, ``{A Gravitational instability in higher
  dimensions},'' \href{http://dx.doi.org/10.1103/PhysRevD.66.064024}{{\em Phys.
  Rev.} {\bf D66} (2002)  064024},
\href{http://arxiv.org/abs/hep-th/0206202}{{\tt arXiv:hep-th/0206202
  [hep-th]}}.
%%CITATION = HEP-TH/0206202;%%.

\bibitem{DeFelice:2011ka}
A.~De~Felice, T.~Suyama, and T.~Tanaka, ``{Stability of Schwarzschild-like
  solutions in f(R,G) gravity models},''
  \href{http://dx.doi.org/10.1103/PhysRevD.83.104035}{{\em Phys. Rev.} {\bf
  D83} (2011)  104035},
\href{http://arxiv.org/abs/1102.1521}{{\tt arXiv:1102.1521 [gr-qc]}}.
%%CITATION = ARXIV:1102.1521;%%.

\bibitem{Franciolini:2018aad}
G.~Franciolini, L.~Hui, R.~Penco, L.~Santoni, and E.~Trincherini, ``{Stable
  wormholes in scalar-tensor theories},''
  \href{http://dx.doi.org/10.1007/JHEP01(2019)221}{{\em JHEP} {\bf 01} (2019)
  221}, \href{http://arxiv.org/abs/1811.05481}{{\tt arXiv:1811.05481
  [hep-th]}}.

\bibitem{1975RSPSA.343..289C}
S.~Chandrasekhar, ``{On the Equations Governing the Perturbations of the
  Schwarzschild Black Hole},''
  \href{http://dx.doi.org/10.1098/rspa.1975.0066}{{\em Proceedings of the Royal
  Society of London Series A} {\bf 343} (1975)  289--298}.

\bibitem{Cooper:1994eh}
F.~Cooper, A.~Khare, and U.~Sukhatme, ``{Supersymmetry and quantum
  mechanics},'' \href{http://dx.doi.org/10.1016/0370-1573(94)00080-M}{{\em
  Phys. Rept.} {\bf 251} (1995)  267--385},
\href{http://arxiv.org/abs/hep-th/9405029}{{\tt arXiv:hep-th/9405029
  [hep-th]}}.
%%CITATION = HEP-TH/9405029;%%.

\bibitem{Glampedakis:2017rar}
K.~Glampedakis, A.~D. Johnson, and D.~Kennefick, ``{Darboux transformation in
  black hole perturbation theory},''
  \href{http://dx.doi.org/10.1103/PhysRevD.96.024036}{{\em Phys. Rev.} {\bf
  D96} (2017) no.~2, 024036},
\href{http://arxiv.org/abs/1702.06459}{{\tt arXiv:1702.06459 [gr-qc]}}.
%%CITATION = ARXIV:1702.06459;%%.

\bibitem{slavjanov2000special}
S.~Slavjanov and L.~Wolfgang, {\em Special Functions: A Unified Theory Based on
  Singularities}.
\newblock Oxford Science Publications. Oxford University Press, 2000.

\bibitem{Bateman:100233}
H.~Bateman and A.~Erdélyi, {\em {Higher transcendental functions}}.
\newblock Calif. Inst. Technol. Bateman Manuscr. Project. McGraw-Hill, New
  York, NY, 1955.
\newblock \url{https://cds.cern.ch/record/100233}.

\bibitem{beals_wong_2010}
R.~Beals and R.~Wong, \href{http://dx.doi.org/10.1017/CBO9780511762543}{{\em
  Special Functions: A Graduate Text}}.
\newblock Cambridge Studies in Advanced Mathematics. Cambridge University
  Press, 2010.

\bibitem{1971ApJ...166..197F}
E.~D. Fackerell, ``{Solutions of Zerilli's Equation for Even-Parity
  Gravitational Perturbations},'' \href{http://dx.doi.org/10.1086/150949}{{\em
  Astrophysical Journal} {\bf 166} (1971)  197}.

\bibitem{MR1392976}
A.~Ronveaux, ed., {\em Heun's differential equations}.
\newblock Oxford Science Publications. The Clarendon Press, Oxford University
  Press, New York, 1995.
\newblock With contributions by F. M. Arscott, S. Yu. Slavyanov, D. Schmidt, G.
  Wolf, P. Maroni and A. Duval.

\bibitem{Gralla:2017djj}
S.~E. Gralla, ``{On the Ambiguity in Relativistic Tidal Deformability},''
  \href{http://dx.doi.org/10.1088/1361-6382/aab186}{{\em Class. Quant. Grav.}
  {\bf 35} (2018) no.~8, 085002}, \href{http://arxiv.org/abs/1710.11096}{{\tt
  arXiv:1710.11096 [gr-qc]}}.

\bibitem{Porto:2016pyg}
R.~A. Porto, ``{The effective field theorist\textquoteright{}s approach to
  gravitational dynamics},''
  \href{http://dx.doi.org/10.1016/j.physrep.2016.04.003}{{\em Phys. Rept.} {\bf
  633} (2016)  1--104}, \href{http://arxiv.org/abs/1601.04914}{{\tt
  arXiv:1601.04914 [hep-th]}}.

\bibitem{Burgess:2017mhz}
C.~Burgess, P.~Hayman, M.~Rummel, and L.~Zalavari, ``{Reduced theoretical error
  for $^4He^+$ spectroscopy},''
  \href{http://dx.doi.org/10.1103/PhysRevA.98.052510}{{\em Phys. Rev. A} {\bf
  98} (2018) no.~5, 052510}, \href{http://arxiv.org/abs/1708.09768}{{\tt
  arXiv:1708.09768 [hep-ph]}}.

\bibitem{Nicolis:2017eqo}
A.~Nicolis and R.~Penco, ``{Mutual Interactions of Phonons, Rotons, and
  Gravity},'' \href{http://dx.doi.org/10.1103/PhysRevB.97.134516}{{\em Phys.
  Rev. B} {\bf 97} (2018) no.~13, 134516},
  \href{http://arxiv.org/abs/1705.08914}{{\tt arXiv:1705.08914 [hep-th]}}.

\bibitem{Wong:2019yoc}
L.~K. Wong, A.-C. Davis, and R.~Gregory, ``{Effective field theory for black
  holes with induced scalar charges},''
  \href{http://dx.doi.org/10.1103/PhysRevD.100.024010}{{\em Phys. Rev. D} {\bf
  100} (2019) no.~2, 024010}, \href{http://arxiv.org/abs/1903.07080}{{\tt
  arXiv:1903.07080 [hep-th]}}.

\bibitem{Kuntz:2019zef}
A.~Kuntz, F.~Piazza, and F.~Vernizzi, ``{Effective field theory for
  gravitational radiation in scalar-tensor gravity},''
  \href{http://dx.doi.org/10.1088/1475-7516/2019/05/052}{{\em JCAP} {\bf 05}
  (2019)  052}, \href{http://arxiv.org/abs/1902.04941}{{\tt arXiv:1902.04941
  [gr-qc]}}.

\bibitem{Geroch:1972yt}
R.~P. Geroch, ``{A Method for generating new solutions of Einstein's equation.
  2},'' \href{http://dx.doi.org/10.1063/1.1665990}{{\em J. Math. Phys.} {\bf
  13} (1972)  394--404}.

\bibitem{Breitenlohner:1986um}
P.~Breitenlohner and D.~Maison, ``{On the Geroch Group},'' {\em Ann. Inst. H.
  Poincare Phys. Theor.} {\bf 46} (1987)  215.

\bibitem{Moncrief:1975sb}
V.~Moncrief, ``{Gauge-invariant perturbations of Reissner-Nordstrom black
  holes},'' \href{http://dx.doi.org/10.1103/PhysRevD.12.1526}{{\em Phys. Rev.
  D} {\bf 12} (1975)  1526--1537}.

\bibitem{Cardoso:2019mes}
V.~Cardoso, T.~Igata, A.~Ishibashi, and K.~Ueda, ``{Massive tensor field
  perturbations on extremal and near-extremal static black holes},''
  \href{http://dx.doi.org/10.1103/PhysRevD.100.044013}{{\em Phys. Rev. D} {\bf
  100} (2019) no.~4, 044013}, \href{http://arxiv.org/abs/1904.05109}{{\tt
  arXiv:1904.05109 [gr-qc]}}.

\bibitem{Cheung:2020sdj}
C.~Cheung and M.~P. Solon, ``{Tidal Effects in the Post-Minkowskian
  Expansion},'' \href{http://arxiv.org/abs/2006.06665}{{\tt arXiv:2006.06665
  [hep-th]}}.

\bibitem{Kalin:2020lmz}
G.~K\"alin, Z.~Liu, and R.~A. Porto, ``{Conservative Tidal Effects in Compact
  Binary Systems to Next-to-Leading Post-Minkowskian Order},''
  \href{http://arxiv.org/abs/2008.06047}{{\tt arXiv:2008.06047 [hep-th]}}.

\bibitem{Haddad:2020que}
K.~Haddad and A.~Helset, ``{Tidal effects in quantum field theory},''
  \href{http://dx.doi.org/10.1007/JHEP12(2020)024}{{\em JHEP} {\bf 12} (2020)
  024}, \href{http://arxiv.org/abs/2008.04920}{{\tt arXiv:2008.04920
  [hep-th]}}.

\bibitem{Bardeen:1971eba}
J.~Bardeen and R.~Wagoner, ``{Relativistic Disks. I. Uniform Rotation},''
  \href{http://dx.doi.org/10.1086/151039}{{\em Astrophys. J.} {\bf 167} (1971)
  359--423}.

\bibitem{Bardeen:1999px}
J.~M. Bardeen and G.~T. Horowitz, ``{The Extreme Kerr throat geometry: A Vacuum
  analog of AdS(2) x S**2},''
  \href{http://dx.doi.org/10.1103/PhysRevD.60.104030}{{\em Phys. Rev. D} {\bf
  60} (1999)  104030}, \href{http://arxiv.org/abs/hep-th/9905099}{{\tt
  arXiv:hep-th/9905099}}.

\bibitem{Higuchi:1986wu}
A.~Higuchi, ``{Symmetric Tensor Spherical Harmonics on the $N$ Sphere and Their
  Application to the De Sitter Group SO($N$,1)},''
  \href{http://dx.doi.org/10.1063/1.527513}{{\em J. Math. Phys.} {\bf 28}
  (1987)  1553}.
[Erratum: J. Math. Phys.43,6385(2002)].
%%CITATION = JMAPA,28,1553;%%.

\bibitem{vanNieuwenhuizen:2012zk}
P.~van Nieuwenhuizen,
  \href{http://dx.doi.org/10.1142/9789814412551_0005}{``{The compactification
  of IIB supergravity on $S_5$ revisted},''} in {\em Strings, gauge fields, and
  the geometry behind: The legacy of Maximilian Kreuzer}, A.~Rebhan,
  L.~Katzarkov, J.~Knapp, R.~Rashkov, and E.~Scheidegger, eds., pp.~133--157.
\newblock 2012.
\newblock
\href{http://arxiv.org/abs/1206.2667}{{\tt arXiv:1206.2667 [hep-th]}}.
\newblock
%%CITATION = ARXIV:1206.2667;%%.

\bibitem{Chodos:1983zi}
A.~Chodos and E.~Myers, ``{Gravitational Contribution to the Casimir Energy in
  Kaluza-Klein Theories},''
\href{http://dx.doi.org/10.1016/0003-4916(84)90039-3}{{\em Annals Phys.} {\bf
  156} (1984)  412}.
%%CITATION = APNYA,156,412;%%.

\bibitem{Goon:2018fyu}
G.~Goon, K.~Hinterbichler, A.~Joyce, and M.~Trodden, ``{Shapes of gravity:
  Tensor non-Gaussianity and massive spin-2 fields},''
\href{http://arxiv.org/abs/1812.07571}{{\tt arXiv:1812.07571 [hep-th]}}.
%%CITATION = ARXIV:1812.07571;%%.

\bibitem{Beukers2007}
F.~Beukers, {\em Gauss' Hypergeometric Function},
  \href{http://dx.doi.org/10.1007/978-3-7643-8284-1_2}{pp.~23--42}.
\newblock Birkh{\"a}user Basel, Basel, 2007.
\newblock \url{https://doi.org/10.1007/978-3-7643-8284-1_2}.

\bibitem{Kristensson}
G.~Kristensson, \href{http://dx.doi.org/10.1007/978-1-4419-7020-6}{{\em Second
  Order Differential Equations}}.
\newblock Springer New York.

\end{thebibliography}\endgroup

\end{document}